# The European Solar Telescope

C. Quintero Noda[1,2⋆], R. Schlichenmaier[3], L. R. Bellot Rubio[4], M. G. Löfdahl[6], E. Khomenko[1,2], J. Jurčák[5], J. Leenaarts[6], C. Kuckein[1,2], S. J. González Manrique[1,2,7,3], S. Gunár[5], C. J. Nelson[8,9], J. de la Cruz Rodríguez[6], K. Tziotziou[10], G. Tsiropoula[10], G. Aulanier[11,12], J. Aboudarham[13], D. Allegri[14], E. Alsina Ballester[1,2], J. P. Amans[15], A. Asensio Ramos[1,2], F. J. Bailén[4], M. Balaguer[4], V. Baldini[16], H. Balthasar[17], T. Barata[18], K. Barczynski[19,20,13], M. Barreto Cabrera[1], A. Baur[21], C. Béchet[22], C. Beck[23], M. Belío-Asín[1], N. Bello-González[3], L. Belluzzi[24,3,25], R. D. Bentley[26], S. V. Berdyugina[3], D. Berghmans[27], A. Berlicki[28,29,5], F. Berrilli[30], T. Berkefeld[3], F. Bettonvil[31], M. Bianda[24,25], J. Bienes Pérez[1], S. Bonaque-González[1], R. Brajša[32], V. Bommier[13], P.-A. Bourdin[33], J. Burgos Martín[1], D. Calchetti[30,34], A. Calcines[35], J. Calvo Tovar[1], R. J. Campbell[8], Y. Carballo-Martín[1], V. Carbone[36], E. S. Carlin[1,2], M. Carlsson[12,37], J. Castro López[38], L. Cavaller[38], F. Cavallini[39], G. Cauzzi[23,39], M. Cecconi[40], H. M. Chulani[1], R. Cirami[16], G. Consolini[41], I. Coretti[16], R. Cosentino[40], J. Cózar-Castellano[1], K. Dalmasse[42], S. Danilovic[6], M. De Juan Ovelar[43], D. Del Moro[30], T. del Pino Alemán[1,2], J. C. del Toro Iniesta[4], C. Denker[17], S. K. Dhara[30], P. Di Marcantonio[16], C. J. Díaz Baso[6], A. Diercke[23], E. Dineva[17], J. J. Díaz-García[1], H.-P. Doerr[34], G. Doyle[44], R. Erdelyi[45,46,47], I. Ermolli[48], A. Escobar Rodríguez[1], S. Esteban Pozuelo[1,2], M. Faurobert[49], T. Felipe[1,2], A. Feller[34], N. Feijoo Amoedo[1], B. Femenía Castellá[1,2], J. Fernandes[50], I. Ferro Rodríguez[1], I. Figueroa[51], L. Fletcher[52,12], A. Franco Ordovas[1], R. Gafeira[18], R. Gardenghi[14], B. Gelly[53], F. Giorgi[48], D. Gisler[24], L. Giovannelli[30], F. González[1], J. B. González[54], J. M. González-Cava[1], M. González García[4], P. Gömöry[7], F. Gracia[1], B. Grauf[34], V. Greco[55], C. Grivel[40], N. Guerreiro[24,25], S. L. Guglielmino[56], R. Hammerschlag[31,57], A. Hanslmeier[33], V. Hansteen[12,37], P. Heinzel[5], A. Hernández-Delgado[1], E. Hernández Suárez[1], S. L. Hidalgo[1], F. Hill[23], J. Hizberger[34], S. Hofmeister[17], A. Jägers[57], G. Janett[24,25], R. Jarolim[33], D. Jess[8], D. Jiménez Mejías[1], L. Jolissaint[21], R. Kamlah[17], J. Kapitán[58], J., Kašparová[5], C. U. Keller[31], T. Kentischer[3], D. Kiselman[6], L. Kleint[59], M. Klvana[5], I. Kontogiannis[17], N. Krishnappa[34], A. Kučera[7], N. Labrosse[52], A. Lagg[34,60], E. Landi Degl'Innocenti[61], M. Langlois[62], M. Lafon[42], D. Laforgue[53], C. Le Men[53], B. Lepori[63], F. Lepreti[36], B. Lindberg[6], P. B. Lilje[37], A. López Ariste[42], V. A. López Fernández[4], A. C. López Jiménez[4], R. López López[1], R. Manso Sainz[34], A. Marassi[16], J. Marco de la Rosa[1], J. Marino[23], J. Marrero[38], A. Martín[38], A. Martín Gálvez[1], Y. Martín Hernando[1], E. Masciadri[39], M. Martínez González[1,2], A. Matta-Gómez[1], A. Mato[1], M. Mathioudakis[8], S. Matthews[26], P. Mein[13], F. Merlos García[1], J. Moity[13], I. Montilla[1], M. Molinaro[16], G. Molodij[64], L. M. Montoya[1], M. Munari[56], M. Murabito[48], M. Núñez Cagigal[1], M. Oliviero[65], D. Orozco Suárez[4], A. Ortiz[12,37,66], C. Padilla-Hernández[1], E. Paéz Mañá[1], F. Paletou[42], J. Pancorbo[38], A. Pastor Cañedo[4], A. Pastor Yabar[6], A. W. Peat[52], F. Pedichini[48], N. Peixinho[18], J. Peñate[1], A. Pérez de Taoro[1], H. Peter[34], K. Petrovay[46], R. Piazzesi[48], E. Pietropaolo[67], O. Pleier[34], S. Poedts[68,69], W. Pötzi[33], T. Podladchikova[70], G. Prieto[38], J. Quintero Nehrkorn[1], R. Ramelli[24], Y. Ramos Sapena[1], J. L. Rasilla[1], K. Reardon[23], R. Rebolo[1,2], S. Regalado Olivares[1], M. Reyes García-Talavera[1], T. L. Riethmüller[34], T. Rimmele[23], H. Rodríguez Delgado[1], N. Rodríguez González[1], J. A. Rodríguez-Losada[1], L. F. Rodríguez Ramos[1], P. Romano[56], M. Roth[3,71], L. Rouppe van der Voort[12,37], P. Rudawy[29], C. Ruiz de Galarreta[1], J. Rybák[7], A. Salvade[14], J. Sánchez-Capuchino[1], M. L. Sánchez Rodríguez[1], M. Sangiorgi[1], F. Sayède[15], G. Scharmer[6], T. Scheiffelen[3], W. Schmidt[3], B. Schmieder[13,68,52], C. Scirè[56], S. Scuderi[72], B. Siegel[38], M. Sigwarth[3], P. J. A. Simões[73,52], F. Snik[31], G. Sliepen[6], M. Sobotka[5], H. Socas-Navarro[1,2], P. Sola La Serna[1,2], S. K. Solanki[34,74], M. Soler Trujillo[1], D. Soltau[3], A. Sordini[55], A. Sosa Méndez[1], M. Stangalini[75], O. Steiner[24,3], J. O. Stenflo[76], J. Štěpán[5], K. G., Strassmeier[17], D. Sudar[32], Y. Suematsu[77], P. Sütterlin[6], M. Tallon[62], M. Temmer[33], F. Tenegi[1], A. Tritschler[23], J. Trujillo Bueno[1,2,78], A. Turchi[39], D. Utz[33], G. van Harten[31], M. van Noort[34], T. van Werkhoven[31], R. Vansintjan[27], J. J. Vaz Cedillo[38], N. Vega Reyes[1], M. Verma[17], A. M. Veronig[33], G. Viavattene[48], N. Vitas[1,2], A. Vögler[34], O. von der Lühe[3], R. Volkmer[3], T. A. Waldmann[3], D. Walton[26], A. Wisniewska[17], J. Zeman[5], F. Zeuner[24], L. Q. Zhang[79,80], F. Zuccarello[81,56], and M. Collados[1,2]

*(Affiliations can be found after the references)*

## ABSTRACT

## 1. Abstract

The European Solar Telescope (EST) is a project aimed at studying the magnetic connectivity of the solar atmosphere, from the deep photosphere to the upper chromosphere. Its design combines the knowledge and expertise gathered by the European solar physics community during the construction and operation of state-of-the-art solar telescopes operating in visible and near-infrared wavelengths: the Swedish 1m Solar Telescope (SST), the German Vacuum Tower Telescope (VTT) and GREGOR, the French Télescope Héliographique pour l'Étude du Magnétisme et des Instabilités Solaires (THÉMIS), and the Dutch Open Telescope (DOT). With its 4.2 m primary mirror and an open configuration, EST will become the most powerful European ground-based facility to study the Sun in the coming decades in the visible and near-infrared bands. EST uses the most innovative technological advances: the first adaptive secondary mirror ever used in a solar telescope, a complex multi-conjugate adaptive optics with deformable mirrors that form part of the optical design in a natural way, a polarimetrically compensated telescope design that eliminates the complex temporal variation and wavelength dependence of the telescope Mueller matrix, and an instrument suite containing several (etalon-based) tunable imaging spectropolarimeters and several integral field unit spectropolarimeters. This publication summarises some fundamental science questions that can be addressed with the telescope, together with a complete description of its major subsystems.

## 2. Introduction

The European Solar Telescope (EST) is an initiative to construct and operate a ground-based large-aperture (4-metre class) solar telescope for the visible and near-infrared. The project is promoted by the European Association for Solar Telescopes (EAST[1]), which comprises research institutions from 18 European countries. The main goal of the project is to study the magnetic connectivity of the solar atmosphere, from the deep photosphere to the upper chromosphere, with high spatial and temporal resolutions and high magnetic sensitivity. Despite the thinness of these two layers (a few Megametrea combined), the physics that governs the layers is intrinsically very different. Therefore, the structuring and dynamics observed in the photosphere and chromosphere are also very different. The main parameter that determines the plasma behaviour is the ratio between the gas and magnetic pressures and is known as the $\beta$ parameter. In the photosphere and solar interior, $\beta$ is larger than one and magnetic energy is stored there by the effect of convective motions. Fields rising due to buoyancy from the deep interior also interact with the plasma and are capable of modifying these convective motions and forming sunspots and active regions. Somewhere in between the photosphere and the chromosphere, the plasma and magnetic forces change balance and lead to a scenario where $\beta$ is smaller than one, giving rise to a plethora of magnetic structures and phenomena that can be observed in the chromosphere. Despite the apparently different spatial distributions and temporal evolutions of the photosphere and chromosphere, these two layers are linked by the continuity of the magnetic field lines and give rise to different manifestations of the same phenomena.

EST is mainly focused on the determination of this connectivity by the magnetic field at various photospheric and chromospheric heights and on establishing its relation with the thermal and dynamic behaviour of the plasma. The top-level science questions that drive EST can be summarised as: how the magnetic field emerges to the surface and evolves; how the energy is transported from the photosphere to the chromosphere; how the energy is released and deposited in the upper atmosphere; why the Sun has a hot chromosphere; how waves propagate from the photosphere to the chromosphere; and what the dynamics of large-scale magnetic structures are. Examples of important candidate targets for EST are phenomena such as quiet Sun magnetism and its impact on the chromospheric energy balance, magneto-acoustic and Alfvén waves, spicules, swirls and tornadoes, chromospheric heating, localised reconnection events, flares, filament eruptions, prominence-corona instabilities, non-ideal magnetohydrodynamics (MHD) effects, and so on.

Large solar telescopes with multiple complex and upgradable instruments have been built on the ground, with current examples including the Vacuum Tower Telescope (VTT; von der Lühe 1998), the Swedish 1m Solar Telescope (SST; Scharmer et al. 2003a), the Dunn Solar Telescope (DST; Dunn 1969), the GREGOR solar telescope (Schmidt et al. 2012), the New Vacuum Solar Telescope (Liu et al. 2014), the Télescope Héliographique pour l'Étude du Magnétisme et des Instabilités Solaires (THÉMIS; Gelly et al. 2016), the Goode Solar Telescope (Goode & Cao 2012), and the Chinese Large Solar Telescope (Rao et al. 2020), all of which have apertures close to 1 m. These ground-based telescopes have routinely performed spectropolarimetric measurements of the photosphere for several decades. The determination of the chromospheric magnetic field is more challenging because of the weak polarisation signals, and its measurement with current instrumentation is only possible in particular cases. The difficulty in determining the chromospheric magnetic field lies partly in the fact that current telescopes have not been designed to minimise the problems linked to the intrinsic weakness of the polarisation signals coming from these layers. EST is designed to overcome this inconvenience and is optimised for accurate and simultaneous multi-wavelength polarimetry with narrow-band (NB) tunable filters and integral field spectrographs. An on-axis design is preferred to minimise the impact of instrumental polarisation. In addition, high-order multi-conjugate adaptive optics (MCAO) is naturally integrated into the telescope light path, minimising the number of optical surfaces.

EST inherits the best qualities of previous European facilities, including but not limited to: excellent imaging capabilities, as, for instance, the 1m SST, by having a simple design; an open design, similar to that of the Dutch Open Telescope (DOT; Hammerschlag & Bettonvil 1998) and GREGOR (Schmidt et al. 2012), to exploit the favourable winds at the Canary Islands; a robust and user-friendly adaptive optics (AO) system such as that operating at the German VTT (van der Luehe et al. 2003), the SST (Scharmer et al. 2003b, 2019), and on GREGOR (Kleint et al. 2020); multi-line spectroscopy, as, for instance, in THÉMIS (e.g. Gelly et al. 2016) and the VTT; multi-wavelength simultaneous spectropolarimetry capabilities similar to combinations such as the Visible Imaging Polarimeter (VIP; Beck et al. 2010) with the Tenerife Infrared Polarimeter (TIP; Collados et al. 2007) or the polarimetric Littrow Spectrograph (POLIS, Schmidt et al. 2003) with TIP at the VTT, as well as THÉMIS in multi-line spectropolarimetric mode (e.g. López Ariste et al. 2000; Paletou & Molodij 2001) and in the multichannel subtractive double pass (MSDP) mode (Mein et al. 2021); efficient NB tunable filters, as, for example, the Interferometric BIdimensional Spectropolarimeter (IBIS; Cavallini 2006), the CRisp Imaging SpectroPolarimeter (CRISP; Scharmer et al. 2008), or the CHROMospheric Imaging Spectrometer (CHROMIS; Scharmer 2017) instruments; fast-modulation high-precision spectropolarimetry as with, for instance, the Zurich Imaging Polarimeter (ZIMPOL; Gandorfer et al. 2004; Ramelli et al. 2010); and simultaneous control of multiple polarimetric imaging and spectrograph instruments as, for instance, at the VTT.

⋆ Corresponding author. e-mail: carlos.quintero@iac.es

[1] https://est-east.eu/east

Special mention is reserved here for the Daniel K. Inouye Solar Telescope (DKIST; Rimmele et al. 2020), the first member of the new 4m class generation of ground-based solar telescopes. This telescope is the successor to some of the 1m class telescopes mentioned above and, with operations starting this year, promises to represent a massive leap forward in our capabilities for sampling the lower part of the solar atmosphere, as well as certain parts of the off-limb solar corona. EST and DKIST, separated by an almost 180º difference in longitude, will provide the opportunity to observe slowly evolving phenomena (e.g. active region or filament formation) in a continuous way and increase the chance of detecting events that happen over short timescales and are difficult to predict (e.g. flares or filament eruptions).

Chromospheric heating is one of the most persistent problems in solar physics. The basic explanation is that physical processes, other than radiation, must be constantly releasing energy in order to compensate for the radiative losses. The exact mechanisms that are causing this behaviour are still unknown. A possible cause could be the continuous reconnection of magnetic field lines on tiny scales, releasing energy every time the magnetic field topology is modified. Other explanations are based on the propagation of MHD waves from the lower atmosphere to the upper atmosphere, transporting and releasing energy as they travel. Most likely, both arguments are valid and are applicable to different situations. In both cases, the inference of the spatial distribution and the temporal evolution of the magnetic field seems mandatory to fully explain the observed phenomena, complemented with estimations of the thermodynamics properties of the solar atmosphere at multiple layers, with high spatial, temporal, and spectral resolutions.

Prominent spectral lines sensitive to the photosphere and chromosphere are mainly found in the visible and near-infrared parts of the spectrum. Ideally, these should be complemented with the measurement of spectral lines sensitive to the corona, to have a complete picture of the solar atmosphere. Coronal lines predominantly fall in the UV and X-ray parts of the solar spectrum and are only accessible from space as the Earth's atmosphere blocks that radiation. This is a major limitation of ground-based solar telescopes, and space missions are required to fill that gap. For instance, the Interface Region Imaging Spectrograph (IRIS; De Pontieu et al. 2014b), the Solar-C EUV High-Throughput Spectroscopic Telescope (EUVST; Shimizu et al. 2020), and the Multi-slit Solar Explorer (MUSE; De Pontieu et al. 2020) cover the UV and extreme-ultraviolet (EUV) wavelengths with high precision spectroscopy, which allows researchers to understand the thermodynamics of the upper chromosphere and corona. In addition, large-scale radio and microwave astrophysical observatories are starting to provide high-resolution observations from the ground of the upper chromosphere and corona (e.g. ALMA; Wootten & Thompson 2009) and complement those of large-aperture optical ground-based telescopes.

Though EST is not a space weather facility, we cannot forget that another long-standing problem in solar physics relates to identifying the driving mechanisms of space weather on the solar surface. In general, the Sun interacts with the Earth continuously through both radiation and particles (solar wind). In addition to these constant interactions, the magnetic activity in the solar atmosphere is sometimes so high that eruptive events occur, launching large quantities of magnetised plasma, known as coronal mass ejections (CMEs), out into the Solar System towards the interplanetary medium. If a CME happens to be directed towards Earth, then satellites, telecommunication grids, power stations, and other technology can be seriously damaged or even destroyed. Hence, it is of great importance to learn the origin of these eruptive events and, to the extent possible, predict their occurrence with enough notice. EST will undoubtedly help us better understand the physical processes that constitute these eruptive events, which may lead to better flare and CME forecasting.

In the following sections, the EST science goals, telescope design, and instrument requirements are explained in detail. Figure 1 shows a rendering of the building and telescope.

## 3. Timeline and present status

The Conceptual Design Study of the EST project began in February 2008 and finished in July 2011. It was funded by the European Commission (FP7) and involved 29 European partners, plus 9 collaborating institutions, from 15 different countries. This was followed by Getting Ready for EST (GREST; 2015-2018), a project intended to take EST to the next level of development by undertaking crucial activities required to improve the performance of current state-of-the-art instrumentation. The project was funded by the European Commission (H2020) and involved 13 European partners from 6 different countries. Additionally, the (FP7 and H2020) EU SOLARNET projects provided open access to first-class infrastructures (through, for example, observing time at telescopes), student funding, early-career researcher mobility schemes, networking opportunities, as well as joint research and development activities. With both SOLARNET projects, the solar community has strengthened its scientific and technological capacity for the next generation of solar observations with EST.

An important milestone for EST was achieved in 2016 when the project was included in the European Roadmap for Research Infrastructures (ESFRI). The EST Preparatory Phase (PRE-EST) started after that milestone in April 2017. The main aim of PRE-EST, funded under the H2020 Framework, is to provide both the EST International Consortium and relevant funding agencies with a detailed plan for the construction of EST.

## 4. Science objectives

This section is divided into two main parts. Firstly, in Section 4.1 we summarise the eight overarching science topics included in the EST science requirement document (SRD; Schlichenmaier et al. 2019). Then, in Section 4.2 several specific sub-topics have been selected from these eight and described in detail. As a quick note, the content of the main SRD topics in Section 4.1 has been kept as short as possible, although we do slightly extend those topics that are less thoroughly described in Section 4.2. The aim is to cover all the EST science drivers in a balanced way while keeping a reasonable length for the paper. Throughout, we try to highlight current observational limitations and describe how the new capabilities EST can overcome those limitations.

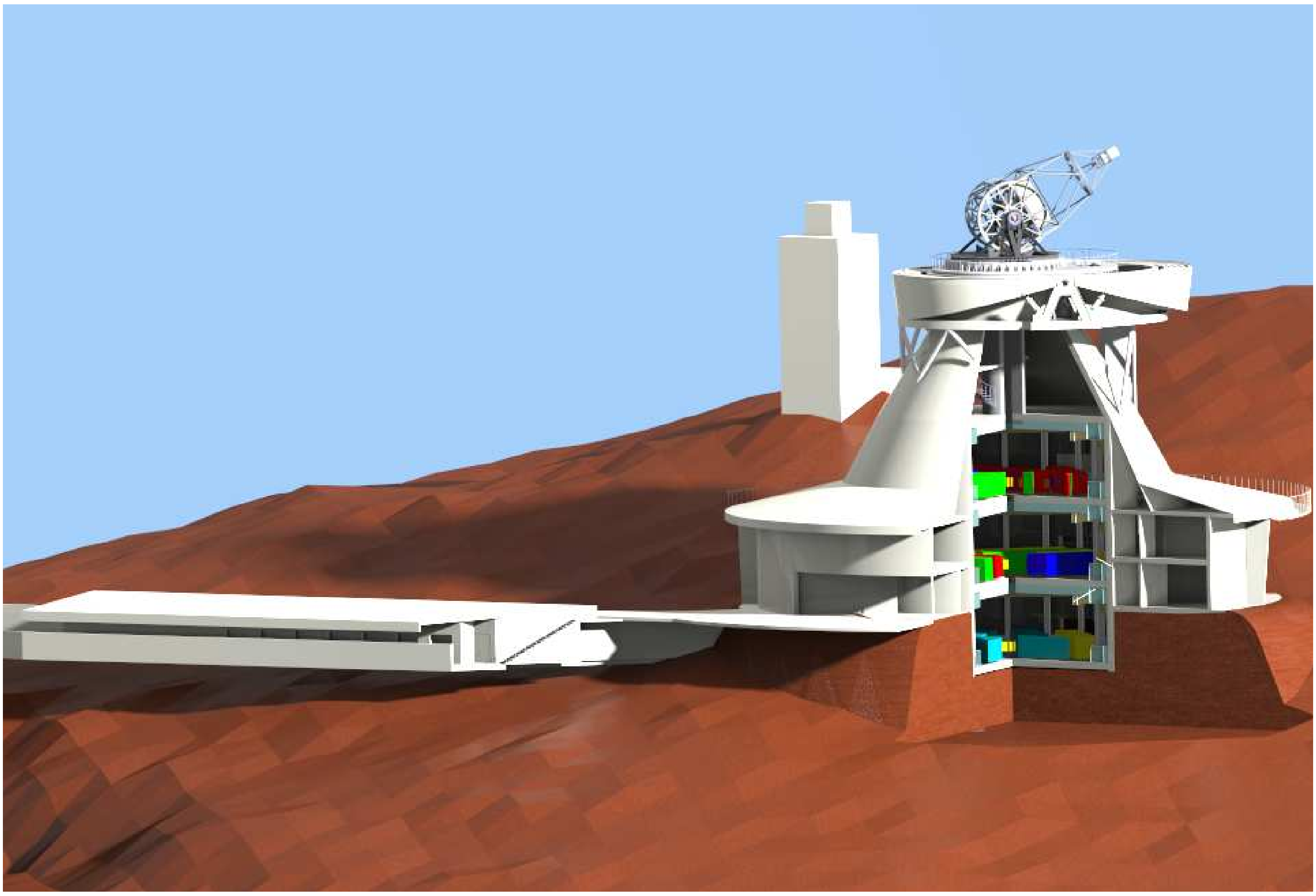

**Fig. 1.** Rendering of the telescope preliminary design on the proposed site at the Roque de Los Muchachos Observatory in La Palma, Spain. The rightmost facility corresponds to the auxiliary building, while the tower in the rightmost part of the figure shows the telescope building and pier. The structure in the background represents the SST.

### *4.1. Eight top level science topics*

In the second edition of the SRD (Schlichenmaier et al. 2019), the top-level science goals were structured into eight sections. In the following, a summary of each section is given:

#### 4.1.1. Structure and evolution of photospheric magnetic flux

The photosphere is the deepest visible layer of the solar atmosphere (see e.g. Stix 1989, for an introduction). It constitutes the upper boundary of the optically thick solar convection zone and transitions from being convectively unstable to stable, by effects of stratification (Stein & Nordlund 1989). Magneto-convective dynamo processes that act in the convection zone add, recycle, and remove magnetic flux, namely, magnetic flux emerges, evolves, and disappears in the photosphere (see, for instance, Nordlund et al. 2009). The current state of our understanding is that the energy source for large-scale events in the outer atmosphere comes from below the photosphere (e.g. Solanki et al. 2006). The interaction of magnetic field and granular convection leads to a zoo of magnetic features that exhibit many magnetic topologies, flow systems, and wave properties (Solanki 1993). To further improve our knowledge of these ubiquitous processes, it is mandatory to characterise their magnetic field through high spatial and high signal-to-noise spectropolarimetric observations of multiple spectral lines.

#### 4.1.2. Wave coupling

Waves are omnipresent in the solar atmosphere and interact in various ways with the plasma medium surrounding them. Thus, they can play an essential role in the energy balance of the different atmospheric layers (see, among others, De Pontieu et al. 2007c; Jess et al. 2009; McIntosh et al. 2011; De Moortel & Nakariakov 2012; Mathioudakis et al. 2013). Those waves can be of multiple kinds, for example acoustic, magneto-acoustic, gravity, or Alfvénic, in the lower atmosphere (e.g. Stein & Leibacher 1974). In addition, as they travel through the magnetic layers of the atmosphere, they may experience mode conversion (Cally & Schunker 2006). Observational evidence for mode conversion in the magnetic network includes transverse waves transferring power to longitudinal waves at twice the original frequency (McAteer et al. 2003). Theoretical considerations suggest that when the plasma $\beta = 1$, longitudinal to transverse (and transverse to longitudinal) mode coupling can also occur (e.g. De Moortel et al. 2004). The properties of the waves depend on whether their direction of propagation is along or transverse to the magnetic field lines. Spectropolarimetric observations of spectral lines sensitive to the atmospheric parameters at both layers are required to track the wave propagation through the photosphere and chromosphere into outer layers, (see, for example, de la Cruz Rodríguez et al. 2013; Quintero Noda et al. 2017; Felipe 2021). Importantly, not only are of interest the background properties of the magnetic field structures that serve as guides for the travelling waves but also the small-scale perturbations to the magnetic field vector that waves induce (see, Khomenko & Collados 2015, as a general reference). The latter implies a challenging requirement for the polarimetric signal-to-noise levels.

#### 4.1.3. Chromospheric dynamics, magnetism, and heating

The chromosphere is the interface between the collisionally dominated plasma of the photosphere and the almost collisionless plasma in the corona (e.g. Mihalas 1978; Stix 1989). In the chromosphere, the physical regime changes relative to the photosphere: magnetic forces dominate over gas pressure forces, the assumption of local thermodynamic equilibrium is no longer valid, and the plasma changes from being mainly neutral to mainly ionised (see, for instance, Dunham 1932; Carlsson et al. 2019). Consequently, the approximations of MHD are no longer accurate and non-thermal and non-local processes need to be considered to describe the physics. Inferring the properties of magnetic field is crucial to understand the chromosphere (e.g. Harvey 2009), and estimating it is challenging (Lagg et al. 2017). Chromospheric features evolve faster than in the photosphere, and the magnetic field is weaker than in the lower atmosphere. It requires going beyond the polarimetric sensitivity of around 1-3$\times10^{-3}$ of the continuum intensity, which is customary in observations of photospheric lines (e.g. de la Cruz Rodríguez et al. 2012; Quintero Noda et al. 2016, for the chromospheric Ca II 8542 Å line). For chromospheric field estimations, the sensitivity must be improved by almost one order of magnitude to $\sim5\times10^{-4}$ of the continuum intensity, while keeping the integration times short enough to track the fast-evolving chromospheric phenomena; for example. filament eruptions can reach velocities up to 100-200 km/s (Sterling & Moore 2005; Penn 2000; Kuckein et al. 2020).

#### 4.1.4. Large-scale magnetic structures

Sunspots and filaments are large-scale magnetic objects that can cover a significant fraction of the solar disk. Sunspots are the primary manifestation of active regions and carry information about how magnetic fields are generated in the solar interior. Their formation, evolution, and decay are relevant for our general understanding of the solar dynamo and magneto-convection. They are the surface manifestation of magnetic flux ropes that originate in the solar interior and reach into the solar atmosphere (see e.g. Solanki 2003; Borrero & Ichimoto 2011; Hinode Review Team et al. 2019). These flux ropes create a wealth of fundamental magnetic processes such as umbral dots, light bridges, or penumbral filaments. These elements are the starting points for umbral flashes, running penumbral waves, and penumbral micro-jets seen higher up. Magnetic field lines that connect active regions arch into the outer atmosphere and manage to confine relatively cool, dense, partially ionised plasma in the form of prominences at heights up to 10 Mm above the visible surface. These prominences have a dynamic fine-scale structure, which fundamentally challenges our understanding of how magnetic fields interact with partially ionised plasma in the outer solar atmosphere (see e.g. Parenti 2014; Hinode Review Team et al. 2019).

#### 4.1.5. Coronal science

Coronal heating is a long-standing unresolved problem in solar physics. Multiple works postulate that the source of coronal heating lies in the photosphere. Many mechanisms have been proposed to transport energy in various forms from the photosphere to the corona, including emerging magnetic flux (e.g. Wu et al. 1994; Schmieder et al. 2004), upward travelling acoustic and magneto-acoustic waves (e.g. Grant et al. 2018), or convective lateral shuffling (braiding) of photospheric magnetic field lines producing magnetic dissipation and reconnection in the corona (for instance, Parker 1983; Bourdin et al. 2014; Viall et al. 2021). In the case of ground-based observatories, it is challenging to have access to spectral lines sensitive to the coronal atmospheric parameters (e.g. Landi et al. 2016). Among large solar telescopes, only DKIST, because of its off-axis design, can perform that task with high spatial resolution observations in specific scenarios (Del Zanna & DeLuca 2018). In particular, the telescope will perform off-limb observations of the coronal magnetic field through occultation techniques. EST on-axis design does not allow for such observations. However, EST will provide both the high resolution observations needed to trace energy from the photosphere to the upper chromosphere and new constraints for coronal magnetic field measurements. These observations will be highly complementary to the coronal phenomena observed by other facilities, such as ground-based radio observatories such as ALMA (Wootten & Thompson 2009) and spacecraft that observe in the X-ray and EUV regions of the spectrum. Coronal measurements are planned to be performed by, among others, Solar Orbiter (Müller et al. 2020), Solar-C EUVST, (Shimizu et al. 2020), and MUSE (De Pontieu et al. 2020).

#### 4.1.6. Flares and eruptive events

The solar atmosphere is a dynamic system featuring flares and eruptive events (see Priest & Forbes 2002; Fletcher et al. 2011; Shibata & Magara 2011, as a reference). Slow differential magnetic field line motions on the solar surface can cause fast reconfiguration of current-carrying coronal magnetic fields through magnetic reconnection, which leads to flares and eruptions. These result in the sudden conversion of magnetic energy into thermal and radiation energy, acceleration of particle beams, and bulk kinetic energy of the plasma when a jet or a CME accompanies a flare. Many of these large-scale developments have local manifestations in the lower layers of the solar atmosphere, particularly along polarity inversion lines, within moving flare ribbons, cooling flare-loops, and at the footpoints of erupting filaments. In order to characterise the physical processes at work in a three-dimensional magnetic configuration before, during, and after such large-scale events, and in order to relate to their corresponding small-scale and time-varying signatures seen in multi-wavelength spectral imaging, it is necessary to measure the magnetic fields, and electric currents both in the photospheric and the chromospheric layers (see, for instance, Kleint 2017; Libbrecht et al. 2019; Vissers et al. 2021; Yadav et al. 2021). In particular, accurate electric current estimations require combining high polarimetric precision with high spatial resolution at multi-height measurements simultaneously. Such measurements can provide the characterisation of key processes of solar flares and eruptions that are still poorly understood, such as the link between chromospheric kernels and the three-dimensional nature of reconnecting flare-loops (e.g. Aulanier et al. 2007; Schmieder & Aulanier 2018; Lörinčík et al. 2019), how the photospheric magnetic field responds to the movement of chromospheric flare ribbons (e.g. Liu et al. 2016; Aulanier 2016; Barczynski et al. 2019), and how much CMEs untwist and stay anchored to their site of origin (e.g. Aulanier & Dudík 2019; Barczynski et al. 2020). In addition to the expected discoveries about the nature of the couplings between various layers of the magnetised solar atmosphere during flares and eruptions, these novel measurements will open a new window to the critical physical processes that contribute to the driving and the evolution of large-scale solar phenomena that form the dangerous aspects of space weather (see, for instance, Georgoulis et al. 2021).

#### 4.1.7. Partially ionised plasma in the photosphere and chromosphere

Most theoretical studies of photospheric and chromospheric plasma dynamics use MHD models (Priest 1982) as the primary tool for successfully understanding the complex structure and dynamical processes of these solar atmospheric layers (among many references, Khomenko & Collados 2006; Cheung et al. 2008; Rempel et al. 2009; Cheung et al. 2010; Khomenko & Collados 2012; Carlsson et al. 2016; Threlfall et al. 2016). Yet, the solar photosphere and chromosphere are only partially ionised with an ionisation fraction below $10^{-3}$ in the photosphere as described in Vernazza et al. (1981), see also Ballester et al. (2018) for a review on this topic and, for example, Pastor Yabar et al. (2021). Therefore, a suitable alternative to the MHD approach is a multi-fluid approach where the plasma species are considered separate fluids interacting by collisions (see, Khomenko et al. 2014, and references within). A multi-fluid treatment is essential for the low collisionally coupled chromosphere because the relevant energy transport and conversion processes happen at spatial and temporal scales similar to ion-neutral collisional scales. With observations of effects of partial ionisation, these theoretical predictions can be tested. To that aim, it is necessary to scan multiple spectral lines from different atoms and ionisation stages, strictly simultaneously and with high spectral resolution and spectral coverage, while also probing the magnetic field topology. Through the analysis of inversion codes (del Toro Iniesta & Ruiz Cobo 2016; de la Cruz Rodríguez & van Noort 2017), one could investigate whether two decoupled atmospheric components are required when reproducing many simultaneously observed spectral lines from the radiation emerging from the model atmosphere. There is still some work be done to characterise the 'response' (Landi Degl'Innocenti & Landi Degl'Innocenti 1977) of candidate spectral lines to changes in the atmospheric parameters to create a comprehensive selection of candidate transitions with different ionisation stages that would be optimised to study these phenomena (see, for example, Socas-Navarro et al. 2008; Demidov & Balthasar 2012; Kuckein et al. 2021; Trelles Arjona et al. 2021).

#### 4.1.8. Scattering physics and Hanle diagnostics

The linearly polarised spectrum of the solar radiation coming from quiet regions close to the solar limb, generally referred to as the second solar spectrum (Stenflo & Keller 1997; Gandorfer 2000, 2002, 2005), is one of the clearest manifestations of anisotropic scattering processes. This spectrum is rich in signals and spectral details and is of double scientific interest. On the one hand, it encodes a wealth of information about the solar atmosphere. Its sensitivity to the magnetic field through the combined action of the Hanle, Zeeman, and magneto-optical effects allows the investigation of the magnetism of the solar atmosphere in domains that are not accessible through standard techniques (e.g. Bommier 1997a,b; Faurobert et al. 2001; Trujillo Bueno et al. 2002; Berdyugina et al. 2002; Trujillo Bueno et al. 2004; Berdyugina & Fluri 2004; Kleint et al. 2011; Ramelli et al. 2019). At the same time, its sensitivity to the anisotropy of the radiation field can be exploited to infer information on the structure and geometrical complexity of the atmospheric plasma (e.g. Trujillo Bueno et al. 2018). On the other hand, many of these signals contain signatures (often not reproducible in laboratory plasmas) of subtle physical mechanisms acting at the atomic scale. These signals are of invaluable interest for improving our understanding of the fundamental physics of matter-radiation interaction in the presence of polarisation phenomena in both atomic and molecular lines (e.g. Stenflo 1980; Landi Degl'Innocenti 1998; Manso Sainz & Trujillo Bueno 2003; Berdyugina 2011; Alsina Ballester et al. 2021). The relevance of measuring those polarisation signals is also essential to understanding the dynamic radiative transfer processes that shape the emergent solar polarisation in its formation process along the line of sight. Simulations and theoretical models have shown that temperature and velocity gradients drive two independent but complementary physical mechanisms capable of modulating the amplitude, the sign, and the spectral morphology of the polarisation signals. These effects are called dynamic anisotropy (see, for example, Carlin et al. 2012, 2013; Carlin & Bianda 2017) and dynamic dichroism (Carlin 2019). Understanding those mechanisms complements the information we extract from analysing the magnetic Hanle and Zeeman effects. In this regard, the goal is to detect and resolve the dynamic quiet Sun polarisation at its natural spatio-temporal scales. A goal that would allow a better characterisation of the (sometimes anomalous) solar polarisation signals resulting from the combination of magnetic and dynamic effects.

In terms of observations, for many years, the second solar spectrum has represented an essential test bench for the theories of the generation and transfer of polarised radiation. Thanks to instruments like ZIMPOL, today it is possible to reach high, up to $10^{-5}$ of the continuum intensity, polarimetric sensitivity. However, the observations of scattering polarisation that are presently available still lack spatial and temporal resolution because the linear polarisation amplitude is generally low, and the measurements are photon starved. Detecting these signals and analysing their variations at high spatial and temporal resolutions are among the priorities in this research field. For instance, one would like to resolve sub-granular scales to study the polarisation emerging from the axial symmetry breaking of the radiation field introduced by the thermal inhomogeneity due to granulation. Thus, the best conditions for that are observations closer to the disk centre to avoid any projection effects that appear when observing at high heliocentric angles close to the solar limb (customary for scattering polarisation-driven campaigns). However, scattering polarisation signals are weaker when observing at disk centre conditions (e.g. del Pino Alemán et al. 2018; Dhara et al. 2019; Zeuner et al. 2020). Hence, this goal demands the use of large-aperture telescopes that can work at a moderate-high spatial resolution (e.g. $0.1''$) while allowing high signal-to-noise polarimetric measurements (around $5\times10^{-4}$ of $I_c$) with short integration times (between 10-20 s).

### 4.2. *A selection of representative science topics*

This section focuses on selected science sub-topics that serve as a natural connection to the properties of the telescope and the instrument suite from Sect. 4 onwards. Additional science topics are thoroughly described in the EST SRD, and we refer the reader to it for further information.

#### 4.2.1. Structure and evolution of quiet Sun magnetic fields

Magnetic fields are ubiquitous in the solar photosphere. They can be observed in sunspots, active regions, and the rest of the solar surface (i.e. the quiet Sun). The strongest fields occur in sunspots, where field strengths of the order of a few kilogauss are not uncommon. In the quiet Sun, magnetic fields create two distinct patterns: the photospheric network, outlining the edges of super-granular cells, and the inter-network, which corresponds to the interior of super-granular cells. The fields are strong (of the order of 1 kG) and vertical in the network, and weaker and more horizontal in the inter-network, with strengths of the order of 100 G (see, for instance, Bommier 2016). In all cases, the intrinsic spatial scales are 100 km or less. That inherent scale is the size of the smallest magnetic elements that can be detected with current solar facilities in the quiet Sun inter-network and the size of the internal structure displayed by larger magnetic features, such as network flux concentrations or sunspot penumbral filaments.

Quiet Sun fields have revealed themselves as important contributors to the flux and energy budget of the solar photosphere (e.g. Trujillo Bueno et al. 2004). The total quiet Sun magnetic flux has been estimated to be about $7.9 \times 10^{23}$ Mx using observations taken by the narrowband filter imager (NFI) installed on the Solar Optical Telescope (SOT; Suematsu et al. 2008) on board the Hinode mission (Kosugi et al. 2007). The results indicate that the network contributes to around 85% of the total flux while the inter-network to the remaining 15% (Gošić 2015). This flux content is comparable to the flux carried by active regions during the maximum of the solar cycle. For example, in Cycle 23, the active region flux was estimated to be $6 \times 10^{23}$ Mx (Jin et al. 2011). Moreover, inter-network magnetic features bring flux to the solar surface at rates from 120 Mx cm$^{-2}$ day$^{-1}$ (Gošić et al. 2016) to 1100 Mx cm$^{-2}$ day$^{-1}$ (Smitha et al. 2017), much faster than active regions at solar maximum (0.1 Mx cm$^{-2}$ day$^{-1}$; Schrijver & Harvey 1994). This large amount of flux would, in principle, be sufficient to maintain the photospheric network, as pointed out by Gošić et al. (2014) and Giannattasio et al. (2020). Thus, inter-network fields are essential ingredients of solar magnetism.

The quiet Sun is a dynamic place where magnetic fields continually emerge, evolve, interact with each other, and disappear from the surface (e.g. Bellot Rubio & Orozco Suárez 2019). The magnetic fields also interact with the surrounding granular convection, leading to braiding and twisting of the magnetic field lines (see, for example, Berger & Field 1984; Berger & Asgari-Targhi 2009). As a result, the magnetic topology changes at all heights in the atmosphere (see, for example, Aschwanden 2004). Such a reorganisation of the fields may lead to energy release through magnetic reconnection in the upper atmosphere (see, Priest & Forbes 2000, as a reference). Indeed, that energy release holds promise as one of the candidates for chromospheric or coronal heating mechanisms away from active regions. However, despite their importance for the energetics and dynamics of the solar atmosphere, the origin and evolution of the quiet Sun fields are still not well understood. The consensus is that they result from a dynamo process, but the details are scarce. Network fields would be supplied by decaying active regions, by ephemeral regions, or by inter-network fields. Inter-network fields, on the other hand, could be produced by a small-scale local dynamo (e.g. Petrovay & Szakaly 1993; Vögler & Schüssler 2007; Rempel 2014), by flux recycling from active regions (Ploner et al. 2001), by the emergence of horizontal fields into the solar surface (Steiner et al. 2008), or by dragging of canopy fields from the chromosphere (Pietarila et al. 2011). Distinguishing between the different scenarios requires a comparison of the observed field properties with those resulting from realistic MHD simulations, but also detailed analyses of the evolution of the fields from appearance to disappearance.

Unfortunately, a complete characterisation of the properties and temporal evolution of quiet Sun fields is not possible with current solar facilities. Indeed, observing quiet Sun fields is exceptionally challenging. They produce very weak polarisation signals, of the order of $10^{-3}$ of the continuum intensity ($I_{\rm QS}$) or even smaller. It is particularly challenging to detect linear polarisation signals (Lites et al. 2008), which are essential to determine the three components of the magnetic field vector with confidence (Borrero & Kobel 2012). With such small polarisation amplitudes, most signals have an amplitude below the noise of the observations. Therefore, temporal resolution is usually sacrificed in favour of long integration times to reduce the noise level and access those weak signals (see Figure 2). Therefore, an essential piece of information to distinguish between competing models is unavailable.

With integration times of 67 s, the Hinode measurements of Lites et al. (2008) achieved $3 \times 10^{-4}\, I_{\rm QS}$ and showed circular and linear polarisation signals in 70% and 27% of the field of view (FOV), respectively. Deeper integrations of 6.1 min reach $1.3 \times 10^{-4}\, I_{\rm QS}$ and displayed clear circular and linear polarisation signals in 88% and 53% of the FOV, respectively (Bellot Rubio & Orozco Suárez 2012). However, this result comes at the expense of no temporal resolution and degraded spatial resolution. In addition, different signals may contribute to the observed signal during the long integration, distorting the observed polarisation profiles. Similar values were obtained from near-infrared spectral lines at the 1.5m GREGOR telescope with shorter exposure times but essentially the same spatial resolution (Lagg et al. 2016; Martínez González et al. 2016). In addition, Campbell et al. (2021) used the integral field unit (IFU) mode (Dominguez-Tagle et al. 2018; Dominguez-Tagle et al. 2022) of the GREGOR Infrared Spectrograph at GREGOR to measure the temporal evolution of inter-network fields within a small FOV finding similar results. This work highlighted that two-dimensional spectrographs are a key instrument required to fully understand these small-scale fields as it allows a two-dimensional area to be measured strictly simultaneously while maintaining high spectral integrity, that is, all wavelengths in the observed spectral range of a given point in the FOV are recorded at the same time. Still, a higher polarimetric sensitivity is needed to detect both circular and linear polarisation signals everywhere within the FOV.

Even with relatively long integration times, the average field properties are not well characterised yet (see Bellot Rubio & Orozco Suárez 2019, for a review). Measurements with the Hinode spectropolarimeter (Lites et al. 2013) at 0$''$.3 revealed that inter-network fields are weak and highly inclined. The magnetic field strength distribution seems to peak at hG values while the inclination distribution shows a maximum at 90 degrees, corresponding to purely horizontal fields. However, the exact shape of the inclination distribution is still under debate, with some authors favouring a quasi-isotropic distribution (Asensio Ramos 2009; Asensio Ramos & Martínez González 2014) and others suggesting the presence of very inclined fields but not isotropic (Lites et al. 2008; Orozco Suárez & Bellot Rubio 2012; Danilovic et al. 2016; Lites et al. 2017). The disagreement is primarily due to the various ways the authors try to avoid the effects of noise in the polarisation measurements and the different models used to interpret the observations. Furthermore, the magnetic filling factor (the fractional area of the resolution element covered by magnetic fields) inferred from current observations amounts to 0.2-0.3, even at the resolution of Hinode/SOT (i.e. around 200 km). Thus, there is still room for sub-pixel structuring of the field on smaller scales as pointed out in Lagg et al. (2010).

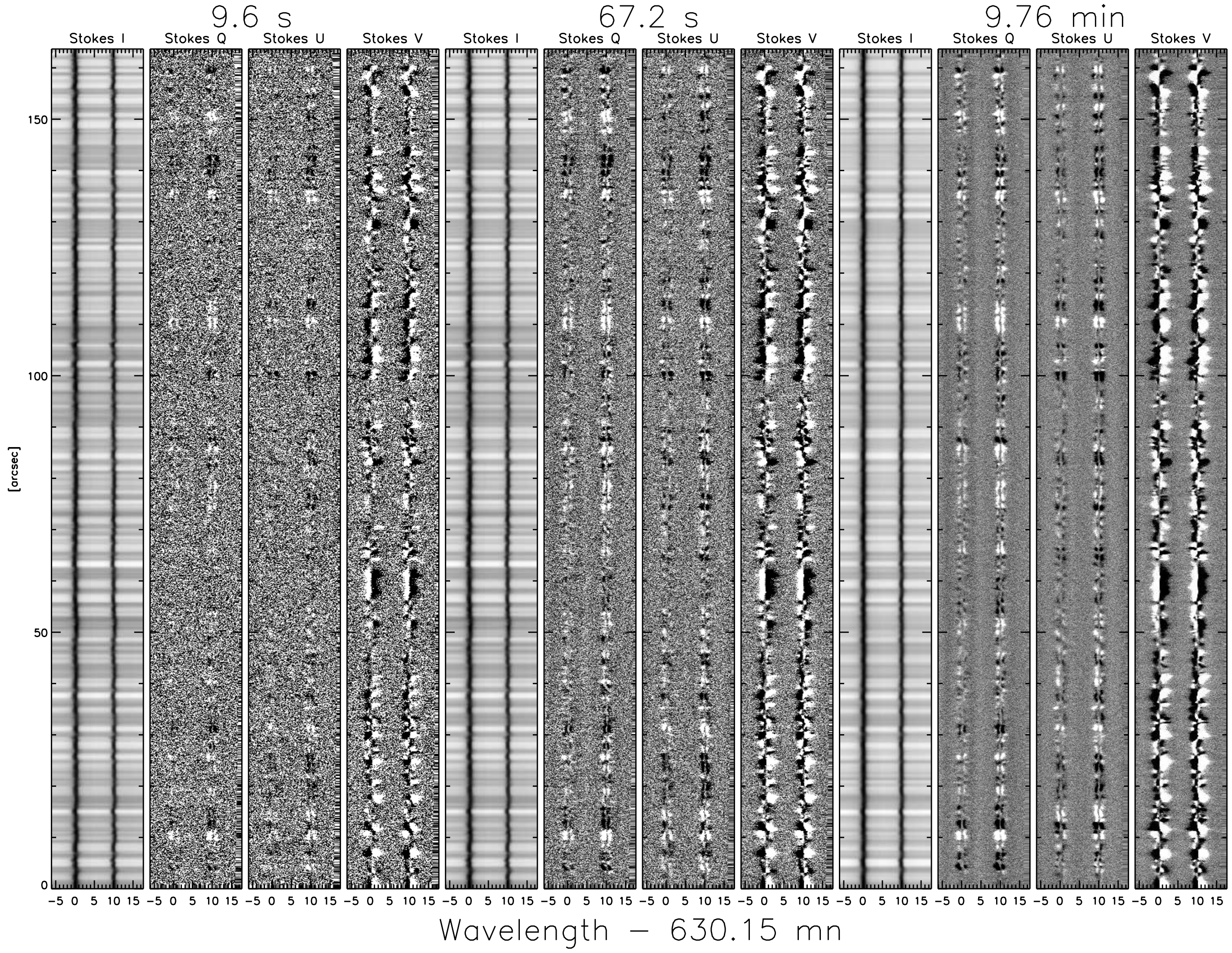

**Fig. 2.** Stokes $I$, $Q$, $U$, and $V$ spectra of the two Fe i lines at 630 nm recorded by the Hinode spectropolarimeter in a quiet Sun region with integration times of 9.6 s (left), 67.2 s (centre), and 9.8 min (right). The Stokes $Q$, $U$, and $V$ panels are saturated at $\pm 10^{-3}\, I_{\mathrm{QS}}$. The linear polarisation signals (Stokes $Q$ and $U$) stand out more prominently over the slit the longer the integration time, (i.e. as the noise decreases). From Bellot Rubio & Orozco Suárez (2012).

A further issue is the evolution of those ubiquitous small-scale features at high spatial and temporal resolution. If observations can be performed combining three elements – (1) observing a two-dimensional area (with a size in the range of 10-20×10-20 arcsec$^2$) strictly simultaneously, (2) with a high polarimetric sensitivity (of the order of 5-8×$10^{-4}$ of $I_c$) and spectral integrity, and (3) and a high spatial (50-100 km) and temporal resolution (10-20 s per two-dimensional scan) – a complete view of the magnetism of the quiet solar photosphere can be provided for the first time. These new observations will likely increase the estimates of the total flux and flux appearance rates in the quiet Sun.

Another fundamental process is the small-scale magnetic flux emergence in the quiet Sun. This is an essential process that occurs everywhere on the solar surface. It has been studied in detail in active regions (e.g. Guglielmino et al. 2010; Centeno 2012; Ortiz et al. 2014; Centeno et al. 2017), but not so much in the quiet Sun (with some notable exceptions, like Gömöry et al. 2010; Guglielmino et al. 2012; Palacios et al. 2012; Fischer et al. 2019, 2020; Guglielmino et al. 2020; Kontogiannis et al. 2020). In particular, flux emergence at the inter-network is still poorly understood; thus, our understanding of the process is limited. The lack of temporal resolution and sensitive-enough spectropolarimetry has hampered progress in this area. Two main aspects need to be clarified: the modes of appearance of the magnetic field and the rise of those magnetic fields into the atmosphere and their contribution to atmospheric heating. This topic adds the necessity to cover multiple atmospheric layers simultaneously. Hence, to adequately understand the emergence of small-scale magnetic features, access must be granted to high spatial resolution, high cadence, high signal-to-noise ratio spectropolarimetric observations of multiple spectral lines.

Inter-network fields are observed to appear on the surface as granular-sized magnetic loops (Centeno et al. 2007; Ishikawa et al. 2008; Martínez González & Bellot Rubio 2009) and clusters of opposite-polarity patches (Wang et al. 1995; Gošić et al. 2022), that is, as bipolar features, but also as isolated unipolar elements within intergranular lanes or above granules (Martin 1988; Lamb et al. 2008; Orozco Suárez et al. 2008; Lamb et al. 2010; Gošić et al. 2016; Smitha et al. 2017). Unipolar appearances are ubiquitous in the quiet Sun and may contribute significantly to the total inter-network flux (Gošić et al. 2022). However, their origin is a mystery because the opposite polarity associated with unipolar features has not been detected. This apparent violation of Maxwell equations might be due to insufficient sensitivity or to the specificities of the processes whereby those magnetic fields emerge on the surface. Coalescence of weak background flux hidden in the noise has been proposed as a possible explanation for the lack of detection of the opposite polarities (Lamb et al. 2008). However, it is difficult to verify the conjecture without better polarimetric sensitivity and temporal resolution. To solve this problem, high cadence multi-line spectropolarimetric measurements are required at the highest possible polarimetric sensitivity.

The observational characterisation of the magnetic properties (strengths, inclinations) and modes of appearance of both bipolar and unipolar quiet Sun features is necessary to investigate the origin of inter-network fields, in particular, the possibility of local dynamo action mediated by shallow recirculation in granular convection (Rempel 2014, 2018), a mechanism that has recently been detected using high resolution observations (Fischer et al. 2020). The details of the recirculation process, the fluxes involved, and the frequency of the events are open questions to be addressed to validate small-scale dynamo simulations, as well as flux emergence simulations (e.g. Moreno-Insertis et al. 2018). A detailed comparison of the distributions inferred from the observations and those from small-scale dynamo simulations should be carried out, along with a comparison of the evolution of the flux.

A very promising approach to obtain measures of magnetic field complexity, stochastic entropy production and evolution timescales is based on the stochastic thermodynamics, which was recently applied to solar magnetic fields (Gorobets et al. 2016, 2017; Gorobets & Berdyugina 2019). One fundamental conclusion is that small-scale solar magnetic fields represent a steady-state non-equilibrium system that evolves towards a maximum entropy limit on timescales that depend on the complexity of the underlying structures. Also, there is a non-negligible probability of occasional local violations of the second law of thermodynamics strictly according to the fluctuation theorem, which was proven before in laboratory experiments and now was found to exist also on the Sun.

Another aspect that needs to be clarified is the role of emerging inter-network loops and unipolar patches in heating the solar atmosphere. About 25% of the magnetic loops that are observed to emerge at the solar surface reach the upper photosphere and the chromosphere (Martínez González & Bellot Rubio 2009), producing measurable polarisation signals and brightenings on their way up. These brightenings are very likely the result of magnetic reconnection with pre-existing fields. Some of the loops may even reach the upper chromosphere and transition region, similarly to the granular-sized magnetic bubbles described by Ortiz et al. (2016) in active regions. For the first time, the recent study by Gošić et al. (2021) enables glimpses on the chromospheric signals produced by inter-network fields as they rise from the photosphere into the chromosphere, but only in circular polarisation and at very low signal-to-noise ratios. Still, this work demonstrates that inter-network fields can reach the chromosphere and interact with the magnetic fields there. It is imperative to carry out a similar analysis on a much larger sample of events, having access to multi-line observations with high polarimetric accuracy, to assess the role of inter-network fields in heating the quiet Sun chromosphere and corona.

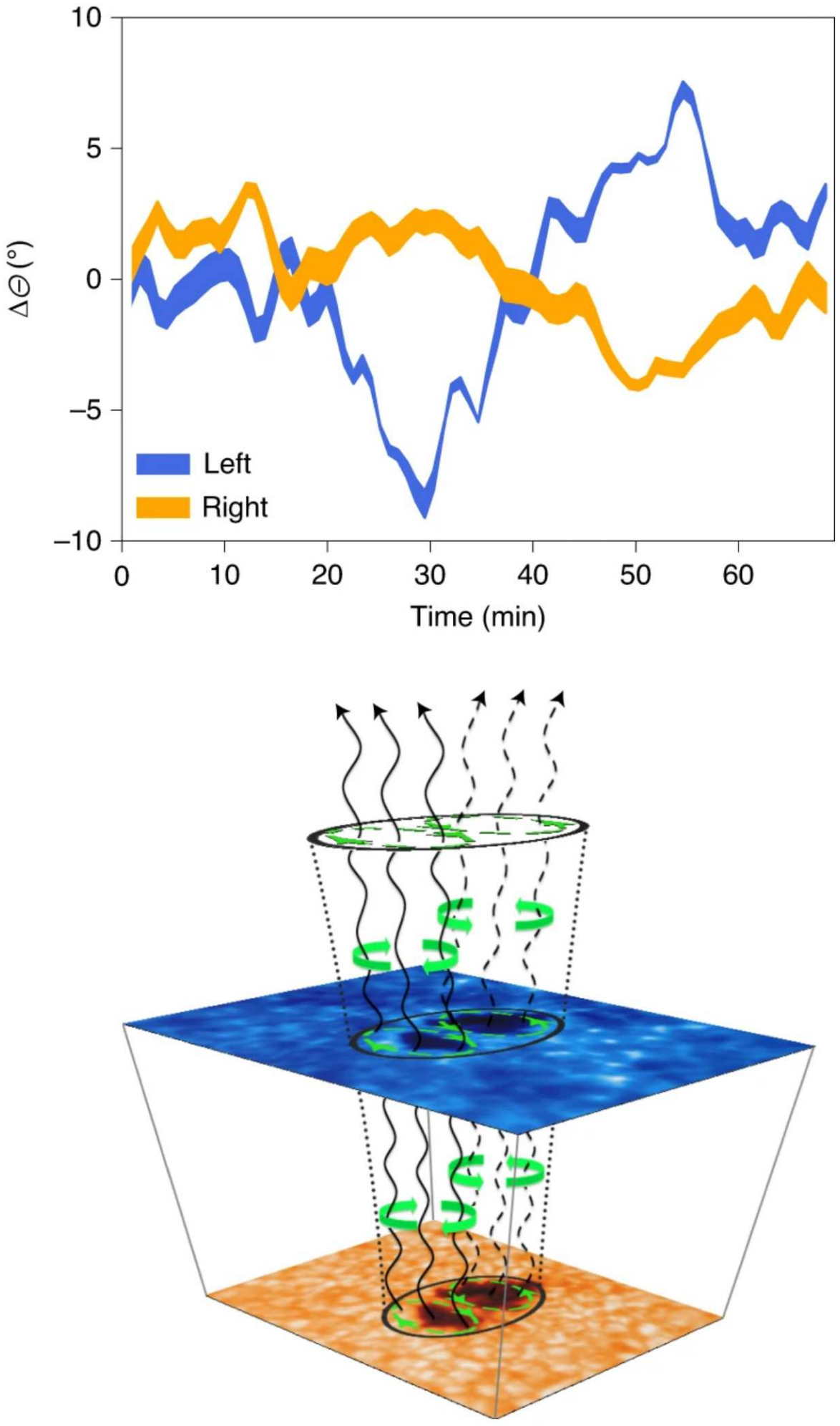

**Fig. 3.** Torsional oscillations of a pore. Upper panel: Angular rotation oscillations of the two lobes of the magnetic pore measured by following the edges of the structure in maps of circular polarisation in the Fe I 617.3 nm line using IBIS data. Bottom panel: Schematic representation of the conclusions of the above measurements on the propagation of the $m = 1$ torsional Alfvén mode in the observed pore. Adapted from Stangalini et al. (2021b).

#### 4.2.2. Wave coupling throughout the solar atmosphere

Studying observationally processes related to wave coupling requires the detection of the magnetic wave guides, namely, resolving and measuring the magnetic field strength and topology at the smallest scales. Unfortunately, these measurements are at the very limit of the current instrumentation, especially when dealing with the chromosphere and higher up. Wave diagnostics based on polarimetry or filter imaging have advantages and shortcomings, and the information is often masked by radiative transfer effects and limited spatial and temporal resolution. Below we discuss the recent progress at some selected research fronts concerning solar waves, emphasising the current observational limitations.

Alfvén waves are expected to play a crucial role in energy dissipation and heating in the solar atmosphere (e.g. De Pontieu et al. 2007c; Grant et al. 2018). Still, they represent one of the most elusive MHD waves, and direct detection is challenging. In the corona, observations by the Coronal Multi-channel

Polarimeter (CoMP) reported by Tomczyk et al. (2007) were among the few claiming direct detection of Alfvén waves. In addition, this first direct detection by the CoMP instrument was later criticised by Van Doorsselaere et al. (2008), who argued that small intensity perturbations and the collective behaviour of the line-of-sight velocity of the structures were rather suggestive of the interpretation in terms of the fast kink mode. The detection of those waves seems also to be difficult at lower atmospheric layers, although some works have pointed out their presence on observation through the analysis of the swaying and torsional motions on the disk counterpart of Type II spicules (Sekse et al. 2013), which is compatible with the presence and propagation of Alfvén waves (see also, Pereira et al. 2014; Freij et al. 2014; Verth & Jess 2016, as reference).

Alfvén waves can be generated by various mechanisms, such as flares (Fletcher & Hudson 2008), by the magnetic tension that is amplified in the formation of spicules and released to drive flows and heats the plasma through ambipolar diffusion (Martínez-Sykora et al. 2017), or by convective buffeting of the magnetic structures (Narain & Ulmschneider 1990) or swirling convective downdrafts in the sub-surface layers. For instance, the latter case is very interesting from the wave coupling point of view. The Alfvén waves generated below may propagate with very little dissipation through the lower atmosphere since they are not compressible and do not dissipate through viscosity or radiation, which efficiently eliminates magneto-acoustic waves. Therefore, the Alfvén waves may carry their energy to the chromosphere, corona, and higher up to the coronal wind (for example, De Pontieu et al. 2007c). To detect Alfvén waves, we usually rely on predictions from theoretical works and numerical simulations showing particular Doppler and magnetic behaviour, characteristic of these waves.

Several complementary procedures for detecting Alfvén waves in the photosphere and the chromosphere have been proposed in the literature. Torsional and kink Alfvén waves can be measured by following the time evolution of individual solar magnetic structures, typically small-scale flux tubes, pores or fibrils (Fujimura & Tsuneta 2009; Jess et al. 2009; Morton et al. 2012). This method requires simultaneous observations at several heights with extremely high spatial resolution and a sufficiently large FOV, allowing the detection of well-defined isolated flux tubes. Recent examples of such observations are those by Stangalini et al. (2021b), where the authors detected anti-phase incompressible torsional oscillations in a magnetic pore in the photosphere by using IBIS data at the DST In their observations, an isolated solar pore was detected composed of two lobes, each of them undergoing a rotational motion in the horizontal plane with opposite phase, see Figure 3. The conclusion that Alfvén waves were responsible for these motions was reached by following the movement of the edges of the structure in circular polarisation images of the photospheric Fe I 617.3 nm line. These intriguing observations were compared to numerical simulations to get indirect estimates for the energy content of the torsional oscillations. However, observational limitations due to the noise level and spatial resolution did not allow inference of the complete magnetic field vector. Besides, only photospheric information was available to the authors. Thus, future telescopes must simultaneously provide the evolution of the velocity and magnetic field vector at several heights to understand the propagation properties of such a torsional mode and any associated magnetic energy flux to the chromosphere.

Following a similar strategy of feature tracking, transverse motions of spicules and fibrils have been repeatedly detected and interpreted as Alfvén waves (De Pontieu et al. 2007c; McIntosh et al. 2011; Morton et al. 2012; Mooroogen et al. 2017). In the first work, the authors analysed Ca II H 396.8 nm broadband images taken with the Hinode/SOT. They found that many chromospheric spicules undergo significant transverse displacements with an amplitude of order 500-1000 km during their lifetimes of 10-300 s. These transverse motions were observed in the direction perpendicular to the long axis of the spicule and hence interpreted as Alfvén waves. The amount of energy contained in these motions could not be evaluated just from the observational data, and simulations again were used to show the importance of these waves for the energy balance of the solar chromosphere. McIntosh et al. (2011) reported on the propagation of Alfvénic motions into the transition region and corona by using data taken with the Atmospheric Imaging Assembly (Lemen et al. 2012) on board the Solar Dynamics Observatory mission (Pesnell et al. 2012). There is also an extensive list of publications that have measured rotational and transverse motions in spicules since the early work of Beckers (1968), to recent ones based on higher spatial and temporal resolution observations (among others, Sekse et al. 2013; De Pontieu et al. 2012, 2014a) that seem to validate further the interpretation of those motions with the presence of Alfvén waves in spicules. However, more work is needed to further understand the properties of those transverse pure-magnetic waves and better quantify their impact on the solar atmosphere, analysing the properties of the magnetic field topology and its fluctuations in spicules. However, magnetic field inference from spectropolarimetric observations of faint features at the limb, such as spicules, is at the very edge of the current instrumental capabilities (Trujillo Bueno et al. 2005; López Ariste & Casini 2005; Orozco Suárez et al. 2015; Kuridze et al. 2021). Thus, these challenging observations require a clear improvement of our current capabilities.

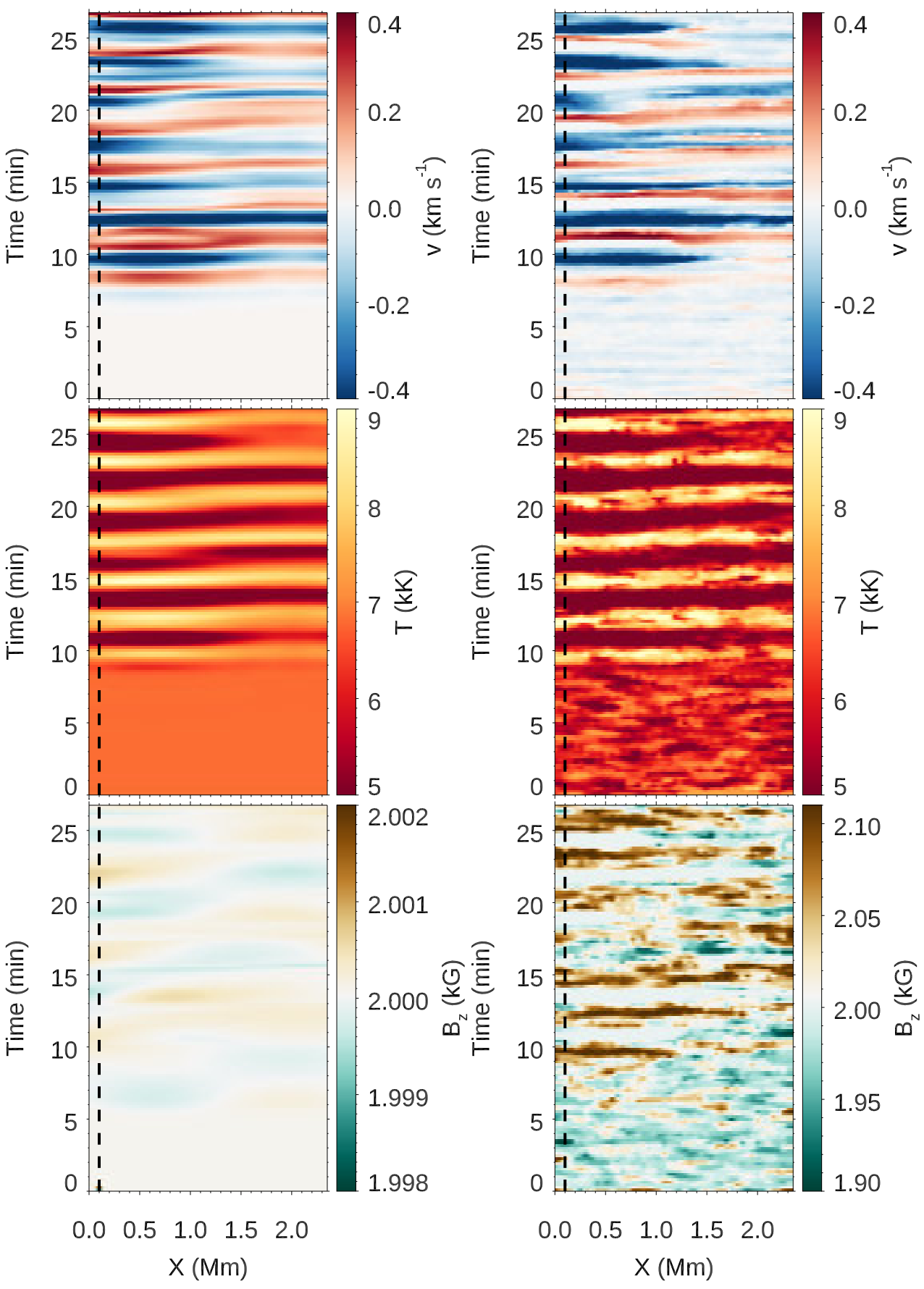

**Fig. 4.** Simulations of wave propagation in a sunspot umbra by Felipe et al. (2021a). Left panels: Time-distance maps of variations in the vertical velocity (top), temperature (middle), and the vertical magnetic field component (bottom) in the chromosphere. Right panels: Same quantities obtained after NICOLE inversions of the Ca II 854.2 nm line synthesised from the simulations. From Felipe et al. (2021b).

Spectral and phase difference analysis between simultaneous observations at different heights of the solar atmosphere are other complementary techniques for the detection of Alfvénic wave signatures in solar magnetic structures. The existence of torsional Alfvén waves in magnetic bright points has been suggested by Jess et al. (2009) based on the analysis of H$\alpha$ full width half maximum (FWHM) oscillations. Morton et al. (2013) concluded, using ROSA observations and complementary simulations, that vortex motions in strong photospheric magnetic flux concentrations can excite torsional Alfvén and kink waves. Wave analyses of chromospheric rotating structures in IBIS Ca II 854.2 nm spectropolarimetric observations (Murabito et al. 2020) suggested that the observed rotational vortex pattern results from the complementary action of a slow actual rotation and a faster azimuthal phase speed pattern of a magneto-acoustic mode. Using H$\alpha$ and Ca II 854.2 nm CRISP observations, Tziotziou et al. (2019), and Tziotziou et al. (2020) reported the existence of fast kink waves within a chromospheric vortex flow with significant substructure appearing as intermittent chromospheric swirls that were attributed to localised Alfvénic torsional waves. Vortex flows, as simulations indicate (Fedun et al. 2011a,b; Shelyag et al. 2013; Battaglia et al. 2021), are natural drivers of several types of MHD waves, such as torsional Alfvén, kink, or sausage waves, that can transport significant energy to higher solar layers (Wedemeyer-Böhm et al. 2012; Liu et al. 2019; Yadav et al. 2020). High-precision spectropolarimetry combined with multi-line observations seems to be the critical element to enhance our capabilities for the observational identification of the mentioned variety of wave types.

Detecting oscillations of the magnetic field has been elusive (Bellot Rubio et al. 2000; Fujimura & Tsuneta 2009). Apparent oscillations in the magnetic field are not always due to 'real' oscillations in the magnetic field vector. Opacity effects can shift the formation of spectral lines and mask actual oscillations of the magnetic field (Rüedi & Cally 2003; Khomenko et al. 2003). These effects are not so severe in photospheric lines (Felipe et al. 2014) and can be overcome by applying inversion codes on multiple spectral lines with different sensitivity to the atmospheric parameters and different heights of formation. In chromospheric lines, such as Ca II 854.2 nm, the effects of shifts of line formation heights on the passage of shocks can produce the detection of 'false' magnetic field oscillations with amplitudes as large as $50-100$ G, while intrinsic oscillations do not exceed a few Gauss in amplitude (Felipe et al. 2014, 2021b). This effect is illustrated in Figure 4. In Felipe et al. (2021b), the authors used numerical simulations of umbral flashes to investigate magnetic field oscillations in sunspot umbrae. The authors assessed the error in the inferred magnetic field oscillations by varying the parameters of the NICOLE (Socas-Navarro et al. 2015) inversions, concluding that inferred oscillations have to be treated with caution because of the effects explained above. To separate real oscillations from these effects, one must have information about magnetic field gradients and compare data from different spectral lines, providing complementary information. Another way to assess the opacity effects is to look at the phase shifts between oscillations of different quantities. In a recent work, Stangalini et al. (2021a) claimed to detect Alfvénic fluctuations by measuring the phase shift between the circular polarisation and the intensity signals in the Ca II 854.2 nm line in a sunspot observed with IBIS. Oscillations of both quantities were correlated in a region at the umbra-penumbra boundary, with a very well-defined phase shift between both quantities. These works show that magnetic field oscillations, oscillations in other quantities, and the phase information of propagating waves at several layers are required to identify the wave modes observed in sunspots, depending on the region (umbra, penumbra) and height. Therefore, several spectral lines with different sensitivity to temperature need to be observed simultaneously. While the spatial resolution can be moderate, the main requirement future facilities need to fulfil to overcome the difficulties mentioned above is to perform high signal-to-noise polarimetric observations.

Another way of producing Alfvén waves in the solar atmosphere is through mode conversion. Theoretical wave mode conversion models suggest that it is a two-step process (Cally & Goossens 2008; Khomenko & Cally 2012). In the first place, acoustic $p$-modes get converted to fast and slow magneto-acoustic waves at heights where the plasma-$\beta$ is equal to unity. The slow mode (acoustic in nature) would continue along the magnetic field lines to the upper chromosphere. The fast (mainly magnetic) mode would refract and reflect due to the gradients in the Alfvén speed. Around the heights where this reflection occurs, a second mode transformation would generate the Alfvén mode. These processes depend on the wave parameters (e.g. frequency), magnetic field inclination, and azimuth. This theoretical process could provide a way of transferring the energy to the chromosphere efficiently via the generation of Alfvén waves at heights close to the transition region, amplifying their possibility to escape this barrier without reflection. Observational confirmation of this process is mostly missing. Related to this topic, Chae et al. (2021), in a recent publication, used spectra of the H$\alpha$ 656 nm and Ca II 854.2 nm lines taken by the Fast Imaging Solar Spectrograph (Chae et al. 2013) installed on the 1.6 m Goode Solar Telescope (Goode et al. 2010). These authors identified transverse MHD waves propagating in the direction parallel to the super-penumbral fibrils with periods of 2.5-4.5 min and supersonic propagation speeds of 45-145 km s$^{-1}$. Due to the close association of these waves with the umbral oscillations and running penumbral waves in the observed sunspot, Chae et al. (2021) concluded that they are the signature of Alfvénic waves excited by the mode conversion of the upward-propagating slow magneto-acoustic waves.

Similarly, Grant et al. (2018) employed IBIS Ca II 854.2 nm observations together with photospheric vector magnetograms taken by the Helioseismic and Magnetic Imager (Schou et al. 2012) to report on the signatures of Alfvén waves. Those authors stated that those waves were heating the sunspot umbral chromosphere through the formation of shock fronts. The shocks observed in this work had a tangential velocity component making them different from umbral flashes. Grant et al. (2018) claimed that the observed heating events are consistent with the dissipation of mode-converted Alfvén waves, generated from upwardly propagating magneto-acoustic waves. Again, the works mentioned above underline the need for high resolution, high cadence, multi-line observations for identifying the process of mode conversion and production of Alfvén waves.

Additionally, interactions of waves with magnetic structures and the efficiency of the wave energy transfer through the solar atmosphere are frequency-dependent processes. However, it is not yet established how the high-frequency and small-wavelength waves contribute to the energetic connectivity between the various layers of the solar atmosphere (Fossum &

Carlsson 2006; Wedemeyer-Böhm et al. 2007; Bello González et al. 2009; Fleck et al. 2010), especially when it comes to magnetic waves. Simulations and observations have been controversial in this respect, mainly because observational detections are challenging to make (Srivastava et al. 2021). Theoretical predictions suggest that dissipation mechanisms, such as resonant absorption, are very efficient at high frequencies (Verth et al. 2010). Other dissipation mechanisms of magnetic waves involving Ohmic or ambipolar diffusion also operate at high frequencies, yielding substantial highly localised energy deposits (Arber et al. 2016; Shelyag et al. 2016). Dissipation of Alfvén waves is extremely difficult to test observationally with the existing instrumentation. Very few observational works explore the high-frequency end of the MHD wave spectrum. Intensity observations obtained with the ROSA instrument (Jess et al. 2010) provide hints of the frequency-dependent velocity power of transverse waves (Morton et al. 2014). These data show that the velocity power of transverse waves in the chromosphere increases at higher frequencies; that is to say, in specific small-scale solar structures, higher frequency waves transport more energy through the chromosphere.

Finally, Srivastava et al. (2017) reported the detection of high-frequency torsional motions of chromospheric fibrils in H$\alpha$ data obtained at the SST, with inferred power enough to account for coronal heating and solar wind acceleration. These works demonstrated that the knowledge of the energy budget in the high-frequency domain is crucial for atmospheric heating. Hence, by performing simultaneous observations of multiple photospheric and chromospheric spectral lines with high cadence polarimetry, access will be given to the pieces that can solve the most critical puzzles, for example, how energy is transferred between atmospheric layers, where and how wave mode conversion occurs, and what the energy budget of magnetic oscillations truly is.

#### 4.2.3. Chromospheric dynamics, magnetism, and heating

The solar chromosphere remains a layer of the solar atmosphere that is challenging to understand. In this layer, the dominant force changes from gas pressure to the Lorentz force, the ionisation degree ranges from close to neutral to fully ionised, radiation transport occurs under non-equilibrium conditions, and the MHD approximation is not sufficient to fully describe the behaviour of the solar plasma.

The critical parameter that governs the behaviour of the chromosphere is the magnetic field. Typical field strengths inside active regions are of the order of hectogauss, sufficient to measure at least the vertical field strength at the diffraction limit of current 1m class telescopes (e.g. Morosin et al. 2020). However, in the quiet Sun, the chromospheric magnetic field has a typical strength of only a few tens of gauss, demanding long integration times or severe spatial degradation of the observations to yield a measurable polarisation signal. In addition, the horizontal component of the chromospheric field leaves an even weaker polarisation imprint in the observed spectrum, which means that it is only measured, in general, around sunspots and active regions.

Thus, it is evident that, to further extend our knowledge of the chromosphere, an improvement of how it is observed and probed is needed. Particularly, present capabilities for collecting photons in a given observing time has to be enhanced and the accuracy with which the polarisation degree of the incoming light can be measured has to be improved. A large aperture telescope (larger than the current 1m class) is needed to address both points. On the one hand, a large aperture allows us to collect more photons if we aim to achieve a similar spatial resolution than that of current facilities. On the other hand, the fact that we collect more photons per unit area means that the signal-to-noise ratio can be increased with reasonable integration times. Combining both elements promises to vastly improve the quantification of the magnetic field in structures of active regions and quiet Sun. These improvements, in turn, lead to a variety of other questions, some of which are summarised below.

The continuous flux emergence in the quiet Sun (Gošić et al. 2014) leads to magnetic loops permeating the chromosphere ranging from granular to super-granular scales, called the magnetic carpet (Title & Schrijver 1998). This magnetic field should play an essential role in the heating of the quiet Sun chromosphere and replenishing the average radiative losses of 4 kW m$^{-2}$ (e.g. Withbroe & Noyes 1977). Also, the determination of the magnetic field will constrain theories of quiet Sun chromospheric heating. In resistive MHD, electric currents lead to heating. The currents themselves cannot be observed but can be inferred by taking the curl of the magnetic field. On the observational side, this is hampered by a lack of sensitivity, leading to noise in the field determination. That uncertainty on the magnetic field determination is amplified by the curl operator and a lack of (mainly vertical) resolution in the inferred field, leading to the underestimation of the resulting currents. Currently, the determination of electric currents in the chromosphere is very limited due to the difficulty in inferring the chromospheric magnetic field, with some notable exceptions (Louis et al. 2021). One key requirement for improving the capabilities of inferring electric currents is to observe multiple spectral lines that are sensitive to the atmospheric parameters at different heights in the solar atmosphere.

A promising heating theory beyond standard resistive MHD is the efficient dissipation of cross-field currents through the friction between ions and neutrals (for instance, Zweibel 1989; Khomenko & Collados 2012; González-Morales et al. 2020; Martínez-Sykora et al. 2020). In this scenario, the magnetic field structure appears subtly different from standard MHD models. However, to deepen on this theory, simultaneous measurements of the detailed spectral line shapes of neutral and ionised species needs to be done strictly simultaneously.

Another area where substantial progress needs to be made is understanding how magnetic flux rises through the atmosphere and how magnetic reconnection in the chromosphere leads to plasma heating and flows. Reconnection of vertical field yields Ellerman bombs if it happens in the low chromosphere (Vissers et al. 2019) and UV-bursts when occurring higher up (Guglielmino et al. 2018; Ortiz et al. 2020). Reconnection of more horizontal fields in active regions leads to gentle, spatially extended, and persistent heating, but the magnetic field configurations and heating mechanisms themselves are poorly understood (Leenaarts et al. 2018).

Another reference solar feature still not entirely understood are plasmoid instabilities. Current observations have provided information on a plasmoid instability in a UV-burst (among others, Innes et al. 2015; Rouppe van der Voort et al. 2017; Díaz Baso et al. 2021, see also Figure 5). However, to infer the detailed magnetic field configuration in such events, a higher signal-to-noise ratio than that of current telescopes over short integration times is required. Besides a larger aperture, new instrumentation in the form of integral field spectrographs is needed. These instruments record a given surface on the Sun plus the solar spectrum with high spectral resolution and signal-to-noise ratio. This would allow the probing of the internal structure and time evolution of plasmoid instabilities currently beyond our reach.

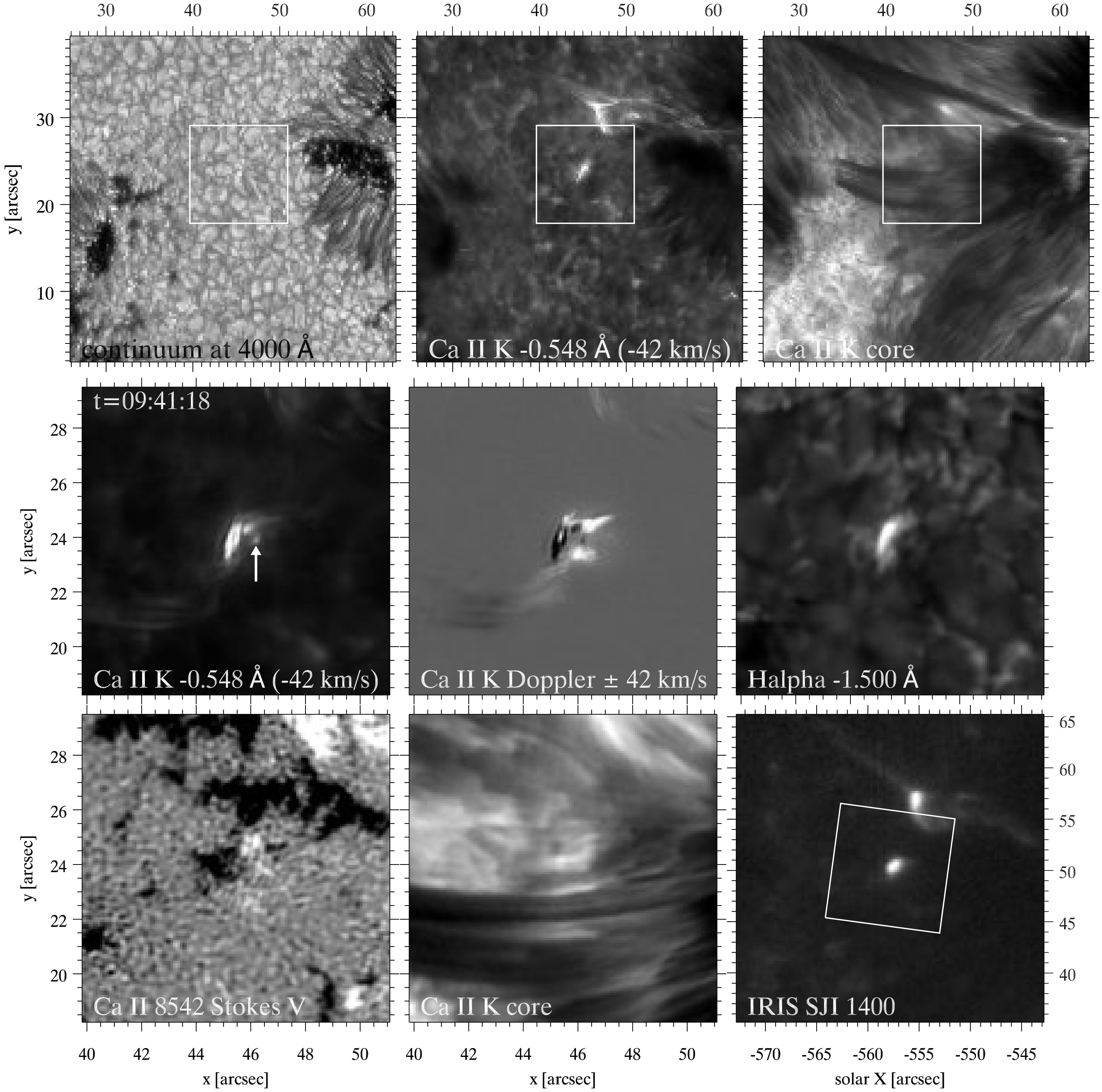

**Fig. 5.** Example of a plasmoid instability. Panels correspond to SST observations, except the bottom right panel, which displays the IRIS slit-jaw image (SJI) sampled using the 140 nm filter. The white box in the top row and the IRIS SJI 140 nm image marks the area centred on the UV burst, which is shown in more detail in the SST images in the two bottom rows. The white arrow marks an isolated plasmoid-like blob. From Rouppe van der Voort et al. (2017).

The chromosphere produces a plethora of jet phenomena that protrude into the corona, ranging in scale from large surges (e.g. Robustini et al. 2016), to smaller Type I and Type II spicules (De Pontieu et al. 2007b). The acceleration mechanism of surges is related to magnetic reconnection, but the exact magnetic field configurations and acceleration mechanism are still unknown. To progress more on this topic, it is necessary to better infer the temperature and velocity structure in the acceleration region and the magnetic field stratification. For that purpose, high spatial resolution (to solve the surge), fast cadence (to be able to track its evolution on the surface of the Sun), multi-line (to be able to track its impact on different layers) observations are needed.

Type I spicules are driven by magneto-acoustic shocks, but the acceleration mechanisms of Type II spicules are under debate. One theory proposes the release of magnetic tension without the need for reconnection to occur (for example, Martínez-Sykora et al. 2017). On the other hand, other works proposed that magnetic reconnection plays a role in the formation of spicules (e.g. Shibata et al. 2007; De Pontieu et al. 2007b, 2011). Thus, it is crucial to infer the magnetic field structure inside spicules to constrain or settle what accelerates Type II spicules. For that purpose, high signal-to-noise spectropolarimetric observations are needed with high cadence to track the beginning and the end of these high-speed plasma flows. Integral field units seem to be the right choice for this kind of observations because they can cover a FOV several times larger than the size of a spicule in a single exposure.

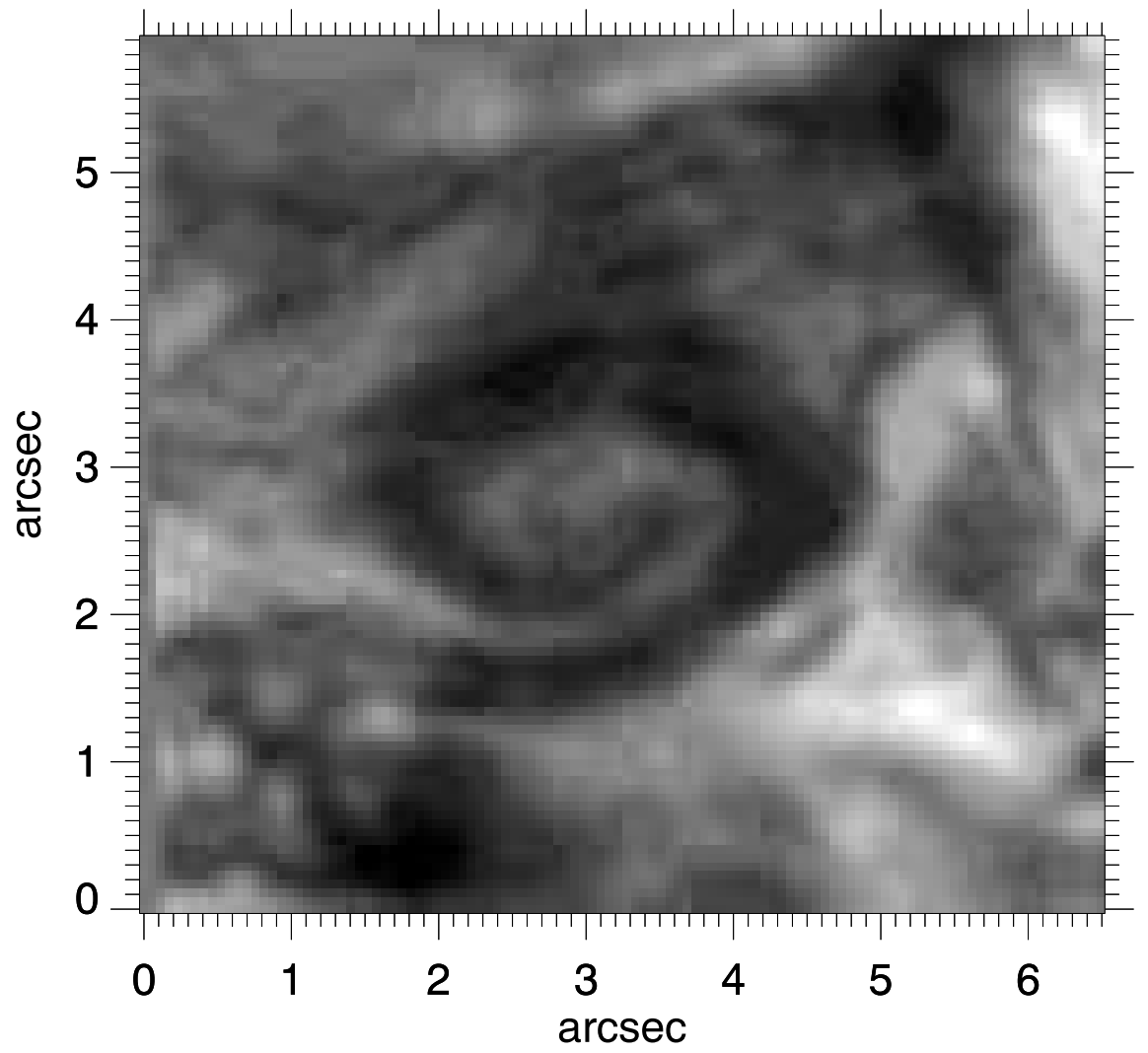

**Fig. 6.** Chromospheric swirl of spiral shape, observed in the blue wing of H$\alpha$ at -0.02 nm. The observations were taken with CRISP/SST on August 13, 2019.

Integral field units also seem to be ideal for the study of chromospheric swirls, observed in the Ca II 854.2 nm line (Wedemeyer-Böhm & Rouppe van der Voort 2009; Wedemeyer-Böhm et al. 2012) and recently in H$\alpha$ (Park et al. 2016; Tziotziou et al. 2018; Shetye et al. 2019). Figure 6 illustrates an observation of a chromospheric swirl in H$\alpha$. These ubiquitous structures are the observational chromospheric signatures of magnetic tornadoes (Wedemeyer-Böhm et al. 2012) that are created at intergranular downdrafts where the plasma returns to the solar interior due to an interaction between convective flows and photospheric magnetic field concentrations (Wedemeyer & Steiner 2014). Magnetic tornadoes can foster a wide variety of waves and channel mass, momentum and energy from the photosphere to the low corona. It is, therefore, crucial to infer their vertical structure through multi-wavelength observations and understand to what degree and in what way the magnetic field governs the dynamics of such vortical structures, including the relation to the observed plasma motions and wave dynamics within them.

Finally, another exciting topic is mass cycling in the solar atmosphere. Observations of the Doppler shift of transition region spectral lines indicate an overall downflow at temperatures below 250 kK. The mass flux in these downflows must be compensated by heating of chromospheric material to transition region and coronal temperatures. Theoretical works (e.g. Guerreiro et al. 2013) indicate that transition region material cycles between chromospheric and transition region temperatures in low-lying loops, heated by dissipation of impulsive electric currents, and cooling down fast through radiative losses. Polito et al. (2020) found correlations between the Doppler shift of chromospheric lines and spectral lines that form in the corona in active regions, indicating that the chromosphere plays a role in setting the circumstances that drive mass flow into the corona. Whether this is related to spicules or reconnection between closed chromospheric loops and open field lines remains unclear. Observations at the spatial resolution of the IRIS satellite, around 0.4″, show a finely structured velocity pattern. To better understand such phenomena, simultaneous spectropolarimetric observations of the upper chromosphere are needed in spectral lines such as Ca II K 393 nm or He I 1083.0 nm with a high signal-to-noise ratio coordinated with space observatories, such as Solar Orbiter and the upcoming Solar-C EUVST and MUSE satellites, that will have access to the transition region and the corona observing the EUV part of the solar spectrum.

The chromospheric heating terms associated with these processes cannot be directly measured from observational data. But since the radiative losses are sustained by heating mechanisms, they represent a lower-limit estimate of the heating terms. A model of the atmosphere is required to calculate the radiative losses. Multi-line non-local thermodynamic equilibrium (NLTE) inversion codes can be used to reconstruct the physical parameters of a depth-stratified model atmosphere of the photosphere and chromosphere (e.g. Socas-Navarro et al. 2015; Milić & van Noort 2018; de la Cruz Rodríguez et al. 2019; Ruiz Cobo et al. 2022; Li et al. 2022). The radiative losses must be estimated in the main chromospheric coolers originating from hydrogen, magnesium and calcium atoms (e.g. Athay 1976; Vernazza et al. 1981).

This approach was utilised with the VALC model by analysing spatially averaged spectra (Vernazza et al. 1981). Recently, Abbasvand et al. (2020), Díaz Baso et al. (2021) and Morosin et al. (2022) have used models that were reconstructed from spatially resolved observations, in order to estimate the net radiative losses in the chromosphere (e.g. see Fig. 7). Their results have shown finely structured maps down to the diffraction limit of the telescope. The benefits EST can provide to estimate the radiative losses are twofold. First, the multi-wavelength capabilities of EST will allow for the simultaneous co-observation of many spectral lines. Adding more lines in the inversion can improve the fidelity and depth-resolution of the resulting model. Second, the very high spatial resolution of EST observations will minimise the mixture of information between different chromospheric features, which is necessary to discriminate between other heating mechanisms.

The combined utilisation of EST datasets with data from space missions and the overall quality of the inferred models can be optimised by spatially coupled and multi-resolution inversion methods (van Noort 2012; de la Cruz Rodríguez 2019) that take into account the instrumental spatial degradation and resolution effects of the different spectral windows.

#### 4.2.4. Large-scale magnetic structures

Sunspots are one of the largest known and most prominent manifestations of solar activity. Hale (1908) discovered that sunspots have a magnetic nature. They comprise a dark central area, known as the umbra, that harbours the strongest and most vertical magnetic field on the solar surface. The umbra is surrounded by the filamentary penumbra, where the magnetic field becomes weaker and more horizontal towards the outer edge of the sunspot. The presence of a penumbra distinguishes sunspots from pores, another type of magnetic flux concentration that appears dark on the solar surface and whose properties are similar to those of umbrae.

Sunspots and pores appear dark on the solar surface because the strong vertical magnetic field within these structures inhibits overturning convection. However, recent analyses of high spatial resolution observations and numerical simulations reveal that fine-scale features in sunspots, such as umbral dots, light bridges, and penumbral filaments, result from magneto-convective motions that are heavily affected by the strong magnetic field. Therefore, studying the fine structure of sunspots helps us understand magneto-convection mechanisms in different field configurations. Furthermore, the fine-scale interaction of plasma and magnetic field also results in small-scale events observed in the chromosphere above sunspots, such as umbral and penumbral microjets and various jet-like events in sunspots light bridges (e.g. Louis et al. 2014). However, in general, we lack detailed information related to the magnetic field's height dependence, as discussed in the review of Balthasar (2018). For example, gradient determinations based on the formation height of spectral lines yield a decrease with height of 2 – 3 G km$^{-1}$, while determinations based on div $B = 0$ lead only to 0.3 – 1 G km$^{-1}$ for photospheric layers. Thus, it is important to understand better how the magnetic field rooted in the interior of the Sun extends to higher atmospheric layers. This is particularly important to any studies of coronal activity that are based on magnetic field extrapolations (e.g. Altschuler & Newkirk 1969). The previous technique infers how the magnetic field can evolve through different atmospheric layers. Hence, if we can input the magnetic field information not only on the photosphere but also at the chromosphere, one can expect higher accuracy when modelling the coronal magnetic field.

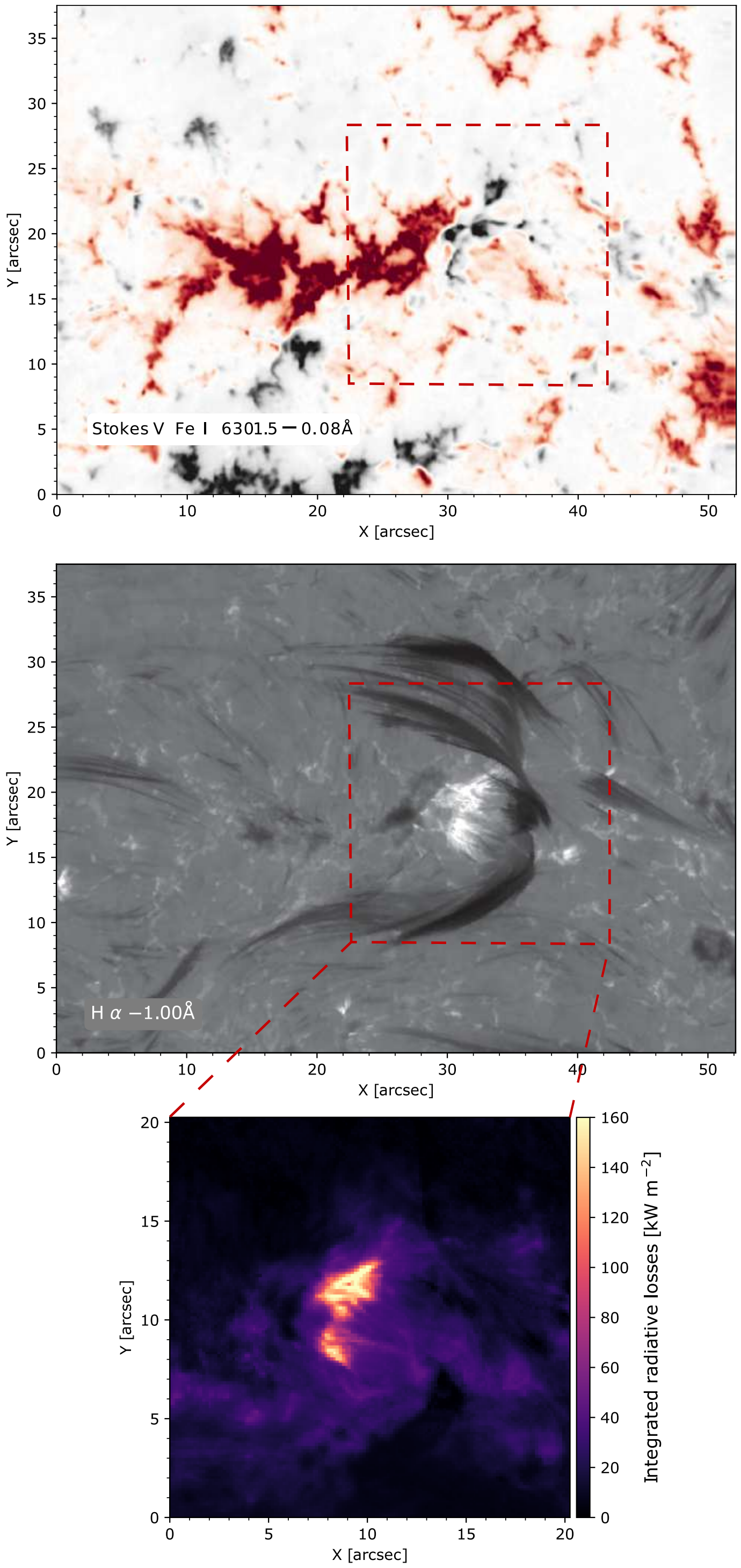

**Fig. 7.** Spatially resolved net radiative losses in the chromosphere of an emerging-flux region (bottom), a context H$\alpha$ image (middle), and the line-of-sight component of the photospheric magnetic field (top). The red box indicates the FOV used to calculate the radiative losses. The observations were acquired at the SST in the Ca II K line, Ca II 854 nm, Fe I 630.1, and 630.2 nm and in the H$\alpha$ line with the CRISP and CHROMIS instruments. Adapted from Díaz Baso et al. (2021).

Other aspects are also crucial for our understanding of sunspots and their evolution. The magnetic field that creates sunspots on the solar surface extends from the interior of the Sun to higher atmospheric layers. Understanding the formation, dynamics and decay of sunspots helps us understand the global solar dynamo and the flux emergence throughout the convection zone of the Sun (Berdyugina & Usoskin 2003). Relative motions of sunspots with respect to each other causes shearing or twisting of magnetic field lines in the higher layers of the solar atmosphere and thus a build-up of energy, stored in a non-potential magnetic field. This energy can be suddenly released in the form of eruptive events due to reconnection of the magnetic field lines. In this case, we need not only a higher spatial resolution to have a better view of the location where the magnetic reconnection events take place, but also access to different atmospheric layers through multi-line observations to better understand the impact of those events on the global magnetic field even beyond the solar corona (for instance, Owens & Forsyth 2013).

For an extensive review of sunspot structure, we refer to Solanki (2003) and more recent reviews by Borrero & Ichimoto (2011) and Rempel & Schlichenmaier (2011). Despite observing sunspots for centuries, our understanding of the sunspot fine-scale structure continuously advances as new telescopes with higher angular resolution become available. Even if sunspots are large structures of more than 100-300 Mm in some cases, they are composed of small substructures, as small as those in the quietest Sun. Interestingly, some of those small-scale features produce dynamic phenomena such as jets and flashes that can significantly impact the energy balance of the surrounding atmosphere, and whose origin requires better and more complete observational data. Therefore, it is essential to move beyond the 1m class telescopes currently available to telescopes with a larger aperture. Realistic MHD simulations made that upgrade in the early 2010s (e.g. Rempel 2012), and upcoming telescopes such as DKIST are expected to follow that path.

In active regions, granules are on average smaller and have longer lifetimes than those in quiet Sun areas (Hirzberger et al. 2002; Lagg et al. 2014; Falco et al. 2017). However, we still observe large convective cells that visually resemble quiet Sun granulation in regions where the magnetic field is weak and not highly inclined. Such conditions are found close to the boundaries of pores that are surrounded by magnetised granulation, in plage regions, and broad light bridges. If the horizontal compo-nent of the magnetic field is sufficiently strong, the magneto-convective cells become aligned with the magnetic field and elongated in its direction. Such a configuration is typical for flux emergence regions, where the magnetic field has to be stronger than 700 G to cause the elongation of the granule (see the case study of Centeno et al. 2017). A similar magnetic field configuration was also found in orphan penumbrae, where the field is horizontal and has strengths typically around 1000 G and as strong as 1500 G (Kuckein et al. 2012; Jurčák et al. 2014; Zuccarello et al. 2014). However, it is not clear whether the fundamental differences in the appearance of the elongated granules and very narrow penumbral-like filaments (e.g. Guglielmino et al. 2014) are caused just by the difference in magnetic field strength in the photosphere or by a different morphology of the magnetic field in the chromosphere. The main limitation delaying progress on this front is the lack of spatial resolution and high signal-to-noise simultaneous polarimetric observations of photospheric and chromospheric lines. Having access to the latter observations, the evolution of light bridges can be traced from their first appearance until they disappear, while probing the vertical stratification of the magnetic field, from the underlying sunspot up to above the light bridge itself. To that aim, it is essential to simultaneously observe spectral lines with different heights of formation over extended periods, for example, a few hours.

Filaments in orphan penumbrae are in all aspects comparable to more frequently observed filaments in sunspot penumbrae (Jurčák et al. 2014). Penumbrae have been thoroughly analysed in the past, and there are indications that penumbral filaments are highly elongated convective cells harbouring a horizontal field (Tiwari et al. 2013), a magnetic field that is embedded in the surrounding stronger and more vertical sunspot magnetic field. Even analyses of observations with low spatial resolution implied this uncombed magnetic field configuration (Solanki & Montavon 1993). In addition, the analyses of observations that sample the fine structure of penumbral filaments are mostly in agreement with numerical simulations of penumbral fine structure (see e.g. Rempel 2012). See Figure 8 for a comparison of fine-scale line-of-sight velocities determined from observations with vertical velocities resulting from numerical simulations. However, even the best observations nowadays do not fully resolve the penumbral fine structure, and sophisticated techniques are needed to infer the information of the small-scale features in the penumbra (van Noort 2012; Ruiz Cobo & Asensio Ramos 2013). Having access to a higher spatial resolution, the magnetic field can be determined along penumbral filaments and between them. Previous works are generally based on the knowledge derived from observations of spectral lines mainly sensitive to the photosphere. Thus, to better understand how the penumbra (a highly inclined magnetic field structure) forms as an extension of an extremely vertical concentration of magnetic field (i.e. the umbra), access to the atmospheric parameters with height resolution (through multi-line observations) from the bottom of the photosphere to the chromosphere is required.

Convective cells become smaller and fainter as one moves into the umbra, where the magnetic field is stronger and more vertical than the penumbra. However, the transition between properties of convective cells seem to be smooth when moving from penumbrae to the cores of umbrae (Sobotka & Jurčák 2009; Löptien et al. 2021). Analyses of umbral dots are the most challenging as the structures are spatially small, and their physical properties barely influence the line-forming regions that are observable (Ortiz et al. 2010; Riethmüller et al. 2013). Nevertheless, the conclusions based on observations are in good agreement with simulations of magnetoconvection in strong and vertical magnetic fields (Schüssler & Vögler 2006). We note, however, that, although the simulations of fine-scale structures in sunspots match the observed properties, the global configuration of simulated sunspots magnetic fields does not correspond to that of observed sunspots (Jurčák et al. 2020). Thus, it is important to observe sunspots at a similar spatial resolution to that of the mentioned simulations to understand what ingredients could be missing from the theory behind the sunspot formation, particularly the formation of the penumbra.

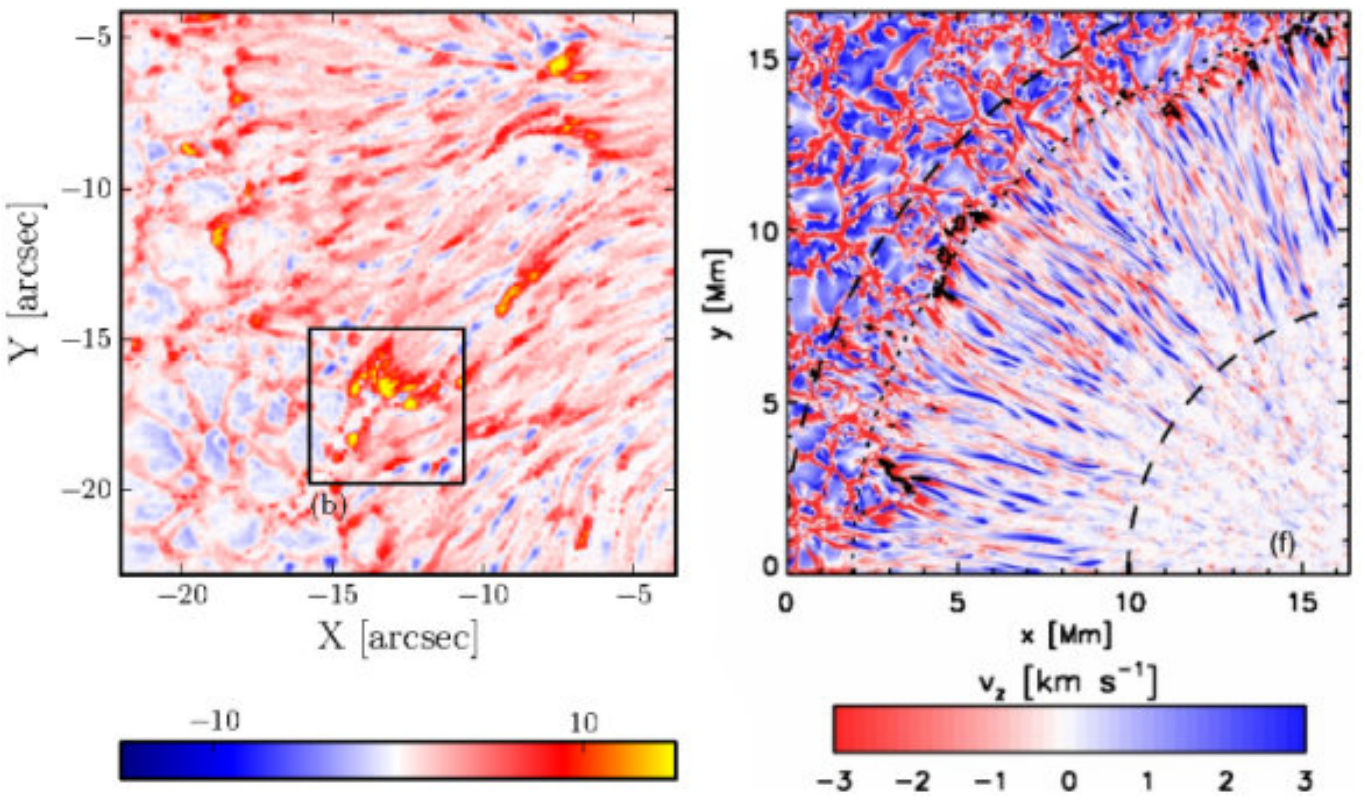

**Fig. 8.** Observations in comparison with simulations. The left panel shows the line-of-sight velocity determined with inversions of Hinode spectropolarimetric observations (adapted from van Noort et al. 2013), and the right panel displays the line-of-sight velocity resulting from an MHD simulation of a sunspot with a grid size of 16 km (adapted from Rempel 2012). Both panels display the spatial distribution of the velocity at the optical depth unity at continuum wavelengths.

From a theoretical point of view, it is expected that the vertical component of the magnetic field ($B_{\rm ver}$) is the crucial parameter for inhibiting the overturning convection (Chandrasekhar 1961; Gough & Tayler 1966). Such behaviour was observationally confirmed by the statistical analysis of umbral-penumbral boundaries where the intensity of convective cells drops significantly (Jurčák et al. 2018). García-Rivas et al. (2021) confirmed the importance of $B_{\rm ver}$ on the pore boundary, and now we understand the critical $B_{\rm ver}$ value as the photospheric counterpart of the critical vertical field stabilising the sub-photospheric layers against more vigorous modes of magneto-convection. This condition was recently confirmed by the analysis of MHD simulations of sunspots by Schmassmann et al. (2021) who found that the stabilising role of $B_{\rm ver}$ can be identified for depths below 7 Mm under the solar surface. It remains to be observationally clarified if the elongation of convective cells is proportional only to the horizontal component of the magnetic field or if the ratio between the horizontal and vertical components is of importance. Moreover, their relationship with flux emergence areas (e.g. Schlichenmaier et al. 2010; Murabito et al. 2017) and with the onset of the Evershed flow (Murabito et al. 2016) needs to be clarified. The main limitation is, again, the lack of spatial resolution. However, if it can be improved and combined with spectropolarimetric observations of various spectral lines with different heights of formation, it will be feasible to assess the influence of the magnetic field on the properties of convective cells. As a result, the evolution of the fine-scale structures of sunspots will be better understood and thus its role in the global evolution of active regions, their formation, and decay.

The most frequent chromospheric activity around sunspots is observed above light bridges. They harbour weak and highly inclined fields surrounded by the strong and vertical fields of the adjacent umbra (Jurčák et al. 2006; Lagg et al. 2014; Toriumi et al. 2015). Tian et al. (2018) differentiate between two types of jets above light bridges. First, a short type that is virtually continuously present and is caused by upward leakage of magneto-acoustic waves from the photosphere, similar as for dynamic fibrils (De Pontieu et al. 2007a). Second, less frequent long and fast surges caused by intermittent magnetic reconnection. The comprehensive study of the chromospheric and transition-region properties of light bridges by Rezaei (2018) confirms that they are complex multi-temperature structures associated with enhanced energy deposition. However, to better understand the roots of that proposed magnetic reconnection and their impact on the energy balance of the solar atmosphere, we need to get access to the magnetic field vector on these structures. This target requires observing multiple spectral lines with sensitivity to various atmospheric layers. Moreover, if magnetic reconnection occurs, it will probably happen at smaller scales than the observed jets, so these observations also require a higher spatial resolution than that achieved with current telescopes. It is crucial to observe these phenomena with two-dimensional instruments to be able to track the evolution of the jets accurately. With these observations, it may be possible to identify the roots of active region heating mechanisms and the height at which the energy is deposited.

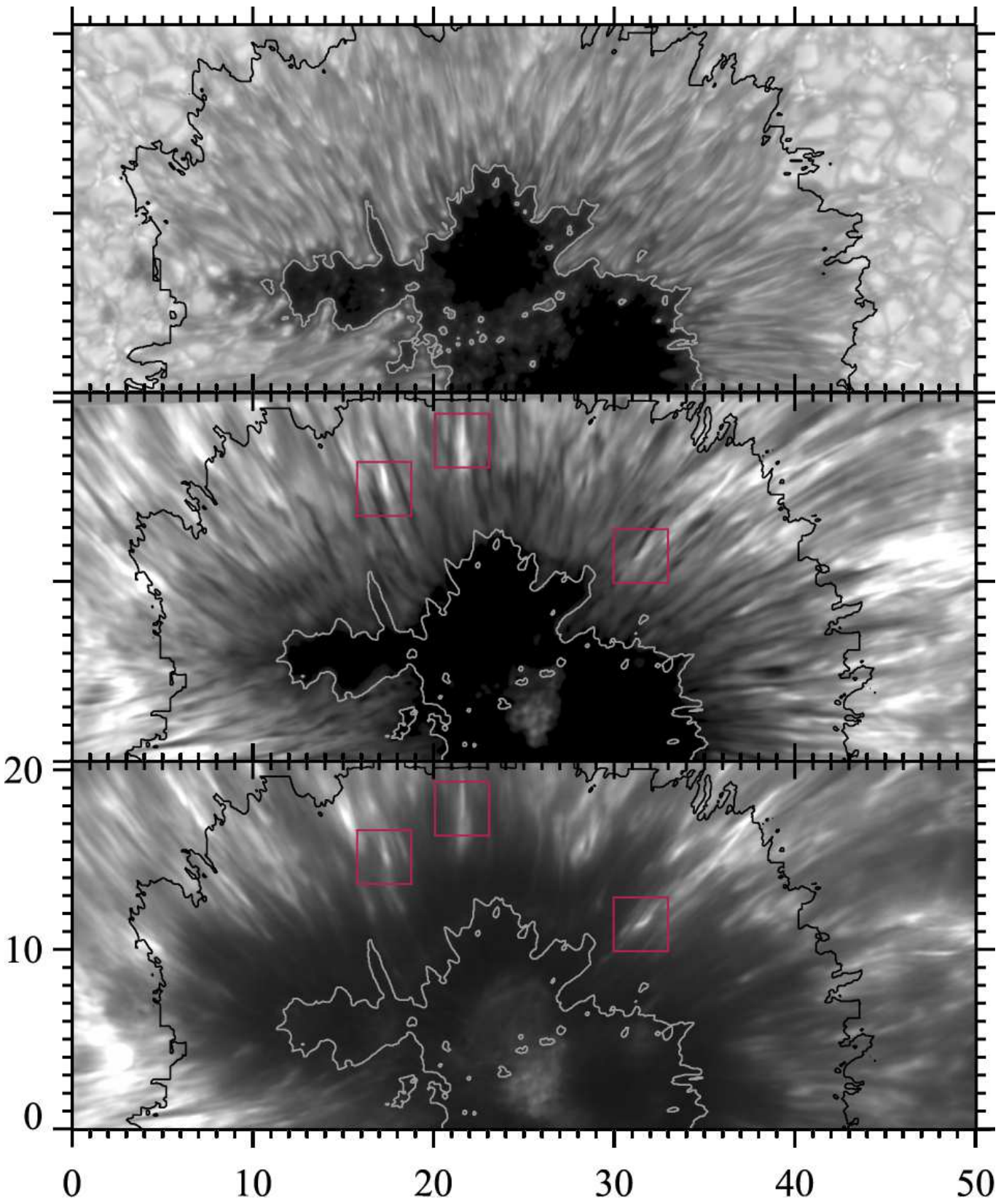

**Fig. 9.** Filtergram images of the continuum around the Fe I 630.1 nm line (top), the blue wing of the Ca II 854.2 nm line at -0.021 nm (middle), and the blue wing of the Ca II K line at -0.024 nm (bottom). Pink squares show the position of penumbral microjets in the middle and bottom panels. Axes are in arcseconds. Contours outline the inner and outer boundaries of the penumbra. Adapted from Esteban Pozuelo et al. (2019).

Fine-scale jets are also observed above sunspot penumbrae, where there is a complex magnetic topology caused by the uncombed magnetic configuration between horizontal filaments that carry the Evershed flow and the surrounding more vertical magnetic field (Solanki & Montavon 1993; Tiwari et al. 2013). These jets were detected for the first time by Katsukawa et al. (2007) and the authors speculated that these penumbral microjets are a consequence of magnetic reconnection. The observed fast apparent motions can either be actual mass flows from the reconnection site or the propagation of a thermal front caused by the reconnection. Since the follow-up studies have not identified any significant Doppler velocities within these structures (see e.g. Reardon et al. 2013; Drews & Rouppe van der Voort 2017; Esteban Pozuelo et al. 2019; Drews & Rouppe van der Voort 2020), the second scenario is now favoured. However, the propagation of these events is still not evident due to the lack of knowledge of the magnetic field at atmospheric layers above the photosphere. Hence, as mentioned above, another critical missing point for better understanding microjet phenomena is having access to multiple atmospheric layers strictly simultaneously. That condition can only be achieved by performing spectropolarimetric observations of various spectral lines at the same time. Jets are fast-moving phenomena and spectropolarimetric observations of multiple spectral lines need to be performed with high cadence and over a two-dimensional area. The onset phase appears to happen very rapidly, with brightenings occurring over hundreds of kilometres in only a few seconds (Rouppe van der Voort & Drews 2019). A recent study of the magnetic field evolution in penumbral microjets by Siu-Tapia et al. (2020) concluded that the 17 s cadence of their observations was insufficient to resolve the temporal evolution correctly. In Figure 9, the appearance of penumbral microjets in different wavelength bands is shown. If reconnection is happening at such a rapid rate, it is critical to estimate the energy contribution it can have on the active region's atmosphere.

Umbral flashes, a large-scale oscillatory pattern with periodicity around 3 min (Beckers & Tallant 1969), dominate observations of the chromosphere above sunspot umbrae. These oscillations are caused by $p$-modes that generate acoustic waves that are expected to steepen into shocks in the chromosphere (see the reviews by Jess et al. 2015; Khomenko & Collados 2015). Fine structure is observed within umbral flashes. The first hints of fine structures seen in absorption were reported by Centeno et al. (2005) and confirmed later by, for example, Henriques & Kiselman (2013), Rouppe van der Voort & de la Cruz Rodríguez (2013), and Yurchyshyn et al. (2014). These structures are nowadays called short dynamic fibrils and exhibit parabolic profiles in time-distance plots consistent with the magneto-acoustic nature of these events. Bright fine-scale, short-lived events above umbrae were reported by Bharti et al. (2013) who interpreted them as reconnection events occurring above umbral dots. However, follow-up studies by, for example, Nelson et al. (2017) and Henriques et al. (2020) showed that the properties of these small-scale umbral brightenings are not consistent with jets but rather with localised compression shocks related to the propagation of the large-scale umbral flashes.

Besides the jet-like events in sunspots, another topic of interest related to active regions is the role of the chromospheric magnetic field configuration on the formation and decay of the sunspot penumbrae. There are numerous studies showing that the chromospheric magnetic field is related to penumbra formation (see e.g. Shimizu et al. 2012; Romano et al. 2013, 2014). However, in MHD simulations of sunspots, the magnetic field has to be set more horizontal on the upper boundary of the simulation domain to create a penumbra comparable to the observed one (Rempel 2012). Additionally, such simulations still do not have a magnetic configuration as rich and complex as that detected in current state-of-the-art observations (Jurčák et al. 2020). Therefore, to progress in understanding sunspots, observations must be improved hand-in-hand with simulations. In the case of observations, more detailed analyses of the chromospheric magnetic field configuration, its temporal evolution, fine structure, and connectivity with the photospheric layers are necessary to understand its role in the penumbra formation and decay (Romano et al. 2020; Murabito et al. 2021). In the case of simulations, the input from observations can help set better boundaries for the formation of the sunspot's features, which can eventually help better understand the physics that produces the active regions.

#### 4.2.5. Filaments and prominences

Filaments and prominences are names for the same large-scale solar phenomena composed of dense and cool clouds of plasma embedded in the hotter and less dense chromosphere and corona. The elevation of the plasma is possible thanks to the magnetic field, which supports the dense plasma against gravity and insulates it from the hot surroundings. This phenomenon is known as filaments when observed on the solar disk and as prominences when observed beyond the solar limb. Filaments appear as dark structures in contrast to the bright solar disk, while prominences are visible as bright structures against the dark background of the sky. Often both terms are used interchangeably in the literature. Depending on the location on the Sun, filaments are classified into four groups: quiescent filaments (in the quiet Sun, see an example of a quiescent prominence in Figure 10), active region filaments (inside active regions, see Figure 11), intermediate filaments (next to active regions), and polar crown filaments (close to the poles). Comprehensive information on the structure of filaments and prominences, their plasma and magnetic field properties, and the modelling of prominences can be found in, among others, Mackay et al. (2010), Labrosse et al. (2010), Parenti (2014), Schmieder et al. (2014a), Vial & Engvold (2015), and Gibson (2018).High spatial resolution observations of filaments and prominences show that they are not homogeneous clouds of plasma but harbour a fine structure that is often highly dynamic. The smallest observed dimensions of filament and prominence fine structure elements are as fine as $0.16''$(e.g. Lin et al. 2005; Okamoto et al. 2007; Lin et al. 2008; Vourlidas et al. 2010). That is near the diffraction limit of current 1m class telescopes. This means, therefore, that the true dimensions of these fine structures might be even smaller, and current telescopes cannot resolve them. In filaments, the smallest dimensions are typically the widths of the filament threads. The length of the threads depends on the size of the filament itself, and they can show a twisted pattern. Active region filaments are typically smaller than quiet Sun filaments. The latter usually have a length in the range of 60 – 600 Mm (Tandberg-Hanssen 1995), but even larger filaments were reported crossing the solar disk (e.g. Kuckein et al. 2016).

While the overall structure of quiescent and polar-crown filaments remains stable, their fine structures exhibit a wide range of motions visible both in the filament view and the prominence view. The situation is more dynamic in the case of active region filaments. The nature of these motions, which often appear to counteract what we know about the filaments and their magnetic field configurations, remains largely an open question. One example of such dynamic behaviour is the counter-streaming flows observed in filaments. Such flows have been documented, for example, by Schmieder et al. (1991); Zirker et al. (1998); Lin et al. (2003, 2005, 2008); Chae et al. (2007); Schmieder et al. (2010); Alexander et al. (2013); Diercke et al. (2018); Ruan et al. (2018) and can be seen even in adjacent fine-structure threads of otherwise quiescent filaments. Another example is the seemingly vertical movements of prominence fine structures, sometimes forming rising plumes (see e.g. Berger et al. 2008) that closely resemble the manifestations of the Rayleigh-Taylor instabilities (see e.g. Hillier 2018). Moreover, the apparent helical fine structure of prominence or tornado may strongly depend on projection effects (e.g. Schmieder et al. 2017; Levens et al. 2017). Thus, the true nature of filament and prominence fine structure dynamics may have several causes including, for example: local (e.g. Okamoto et al. 2007; Zapiór et al. 2015) or bulk (see e.g. the review by Luna et al. 2018) oscillations; movements caused by the evolution of the underlying photospheric flux distribution (Feynman & Martin 1995; Gunár & Mackay 2015; Roudier et al. 2018; Joshi et al. 2020); external disturbances (Zhou et al. 2020; Luna & Moreno-Insertis 2021); or various projections of the field-aligned plasma flows into the observed plane-of-the-sky (Luna et al. 2012; Gunár et al. 2018). Understanding these small-scale dynamics will require coordinated multi-wavelength observations with an unprecedented spatial and temporal resolution, both from the ground and space.

Detailed understanding of the thermodynamic properties of filament and prominence plasma requires analyses of multi-wavelength spectral observations. Such analyses are possible thanks to generations of radiative transfer models that provide us with synthetic spectra to compare with the observations. These radiative transfer models are increasingly sophisticated and multi-dimensional (Heinzel & Anzer 2001; Gouttebroze 2006, 2007, 2008) and can be used for complex statistical analyses of spectral observations of filaments (Schwartz et al. 2019) and prominences (Gunár et al. 2010, 2014; Schwartz et al. 2015; Peat et al. 2021; Barczynski et al. 2021). Reviews of the radiative transfer modelling of filaments and prominences can be found in Labrosse et al. (2010) and Gunár (2014). The high spatial and temporal resolutions that will be provided by the next generation of 4m class solar telescopes, combined with spectral and imaging capabilities of instruments like the multichannel subtractive double pass (Mein et al. 2021), will greatly benefit the radiative transfer modelling of not only filaments and prominences, but also the chromosphere more generally.

The plasma of filaments is embedded in the magnetic field, which plays a crucial role in its stability. The modelling of the magnetic field in filaments and prominences was comprehensively reviewed by Mackay et al. (2010) and Gibson (2018). To infer the properties of the filament magnetic field from spectropolarimetric observations, one must analyse the imprint the Hanle and Zeeman effects leave on specific spectral lines. Some of the spectral lines most suitable for the inference of prominence magnetic fields are the He I $D_3$ line and the He I triplet around 1083 nm (see e.g. Paletou et al. 2001; Trujillo Bueno et al. 2002; Kuckein et al. 2009; Casini et al. 2009; Léger & Paletou 2009; Orozco Suárez et al. 2014; Schmieder et al. 2014b; Levens et al. 2016; Díaz Baso et al. 2019a; Wang et al. 2020; Di Campli et al. 2020; Kuckein et al. 2020). A review of the prominence magnetometry can be found, for example, in López Ariste (2015). To further increase our knowledge of the magnetic field configuration of filaments and prominences, not only should multiple spectral lines be observed simultaneously, but this should also be done with high spatial resolution (better than that provided by the 1m class telescopes). Moreover, high cadence is necessary to completely scan the area of interest in timescales of the order of minutes to understand how the fila-ments come to the solar surface, evolve, and eventually erupt. In addition, high signal-to-noise polarimetric observations are needed to access the weak chromospheric Zeeman and even weaker Hanle modulated polarimetric signals.

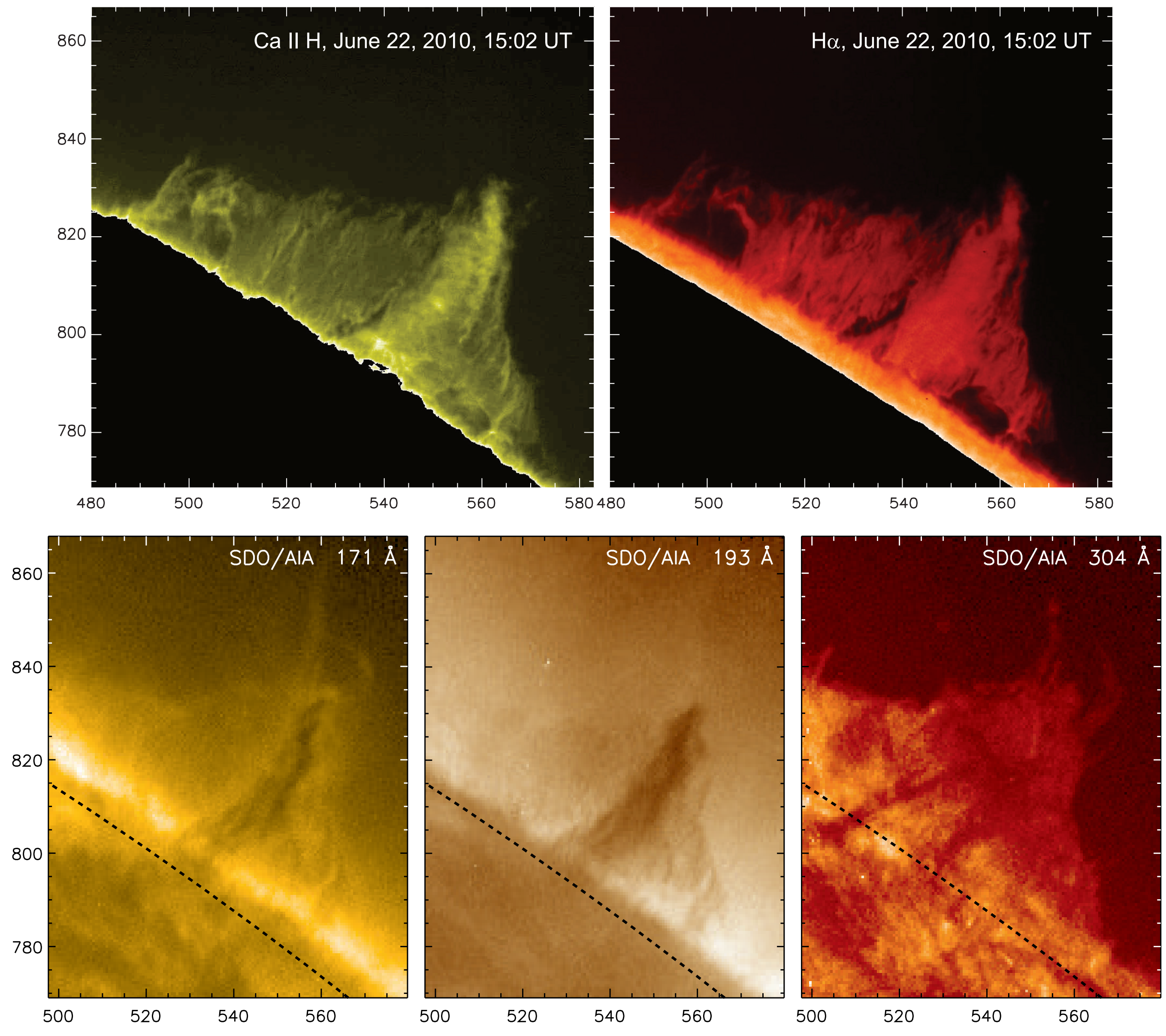

**Fig. 10.** Example of a quiescent prominence, observed on June 22, 2010. Top row: Hinode/SOT observations in the Ca II H line (left) and the H$\alpha$ line 0.0208 nm from the line centre (right). Images are aligned to show the same FOV of 103$''$ × 99 $''$. Bottom row: SDO/AIA 17.1 nm, 19.3 nm, and 30.4 nm channel observations. Labels indicate arcseconds. Adapted from Gunár et al. (2018).

Additionally, prominences are excellent targets to study the two-fluid scenario. There are several works related to the interaction between the plasma and neutral elements in the Sun (see, for instance, Ballester et al. 2018; Wiehr et al. 2019). Both species are strongly coupled collisionally in the photosphere. Thus, the plasma can be described as a single fluid. The most significant non-ideal effect of having neutrals is ambipolar diffusion (Martínez-Sykora et al. 2012; Khomenko & Collados Vera 2012). Under photospheric conditions, the velocities and temperatures of different plasma components are likely to be the same. However, in the chromosphere the collisional coupling is much weaker. Therefore, it is not possible to assume a single fluid plasma for describing fast processes because of the interaction between ions and neutrals, that is to say, their velocities and temperatures are expected to be slightly different (decoupled) at small spatial and temporal scales (Khomenko et al. 2014).

The ion-neutral decoupling and its timescales can be determined theoretically by performing realistic MHD numerical simulations. Popescu Braileanu et al. (2019a,b) have produced chromospheric two-fluid simulations related to the propagation of shocks. The results demonstrate that the decoupling between ions and neutrals can appear at the wavefronts. The magnitude found in the chromosphere may be a non-negligible fraction of the wave amplitude. Popescu Braileanu et al. (2021) have also carried-out two-fluid simulations of the Rayleigh-Taylor instability in prominences. The results show analogous ion-neutral velocity drifts. In the case of observations, there is a similar scenario, although not as clear as in the numerical experiments. The detection of the decoupling was described by Khomenko et al. (2016) using the Ca ii 854.2 nm and He i 1083 nm spectral lines while observing prominences (see also Díaz Baso et al. 2019b). The authors found large spatial and temporal velocity gradients. Wiehr et al. (2019) also reported higher velocities of ions compared to neutrals, from observations of the Sr ii 407.8 nm and Na i $D_2$ 589 nm lines. Wiehr et al. (2021) also found velocity drifts between ions and neutral observing the He i 501.5 nm and Fe ii 501.8 nm transitions. In addition, González Manrique et al. (2022) found high velocity drifts at the edges of a prominence while Anan et al. (2017) interpreted the difference of observed Doppler velocities as being a result of the motions of different components in the prominence along the line of sight, rather than the decoupling of neutral atoms from the plasma.

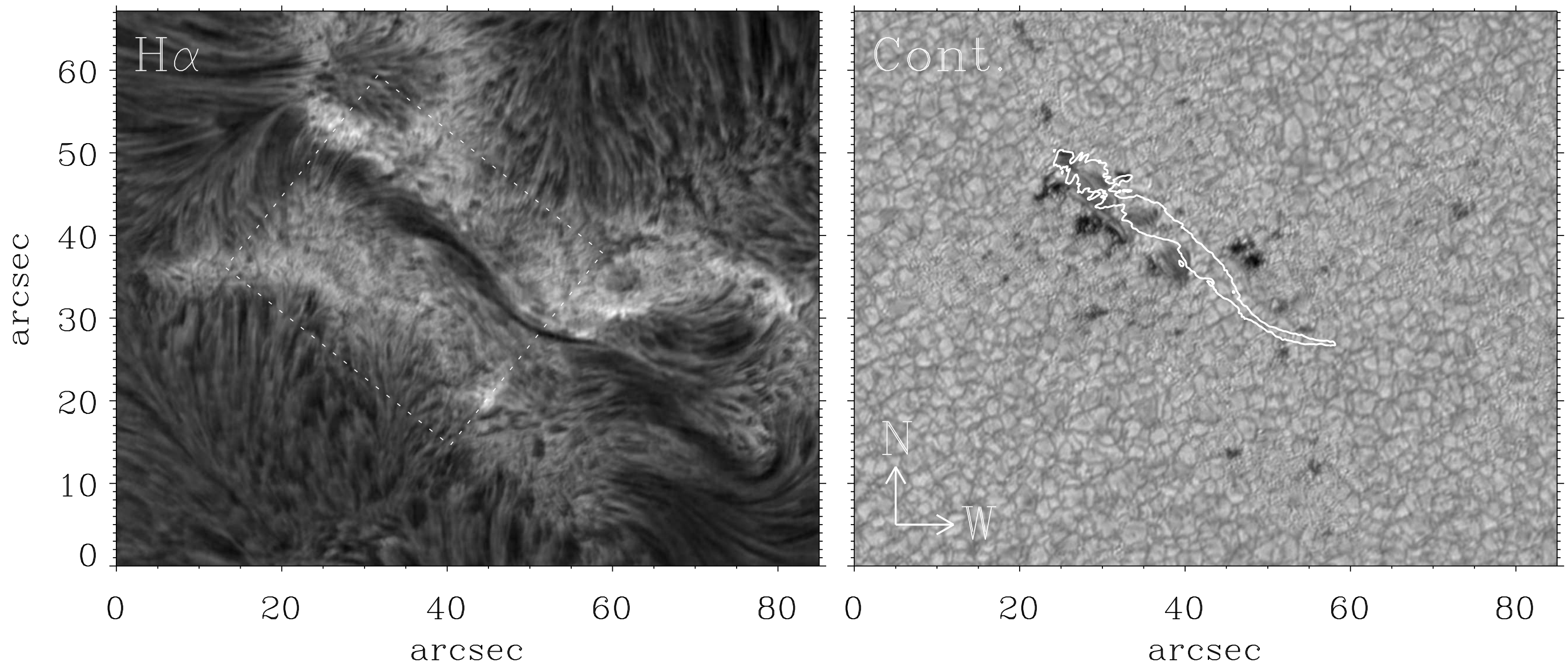

**Fig. 11.** Active region filament seen in Hα (left) and in the continuum together with the contours of the filament on top (right). The data were recorded with the DOT telescope on July 5, 2005, at 8:44 UT. Adapted from Kuckein et al. (2012).

In any case, the main requirement to improve our current knowledge of the possible decoupling between neutrals and ions in the solar atmosphere comes from multi-wavelength observations at a higher spatial resolution than that achieved by 1m class telescopes and closer to that used in the numerical experiments described above. Additionally, upcoming facilities need to provide observations with a high spectral resolution and sampling, so the spectral imprint due to differences in the temperature and plasma velocity between ions and neutrals is accurately detected.

It is true that when analysing the spectral properties of transitions with different heights of formation and sensitivity, the inferred, for instance, velocity drifts can be ambiguous since it is not a priori the case that the emission originates from the same plasma parcel. One possibility is to use inversion codes like those presented in de la Cruz Rodríguez et al. (2019), andRuiz Cobo et al. (2022) to solve the radiative transfer equation in NLTE for multiple atomic species to infer the atmospheric parameters. They might allow the observed profiles to be reproduced via the decoupling of neutrals and ions.

## 5. Technical requirements

The previous section (and the SRD) highlights current observational limitations that inhibit a complete understanding of a range of solar phenomena with existing facilities. These limitations have been used as the primary source for the definition of the technical requirements for EST (summarised in Table 1), which act as the cornerstone of the instrumental developments explained in the following section. We describe some of these requirements in more detail as follows.

The telescope FOV shall have a diameter of 125″ (equivalent to a square of 90″ × 90″ on a given instrument sensor). That squared area fulfils the FOV requirements of the observing programmes presented in the SRD. An external auxiliary telescope may provide larger FOV context images.

Atmospheric perturbation shall be corrected using an MCAO system. The corrected area shall reach the diffraction limit at 500 nm over a circular FOV with a diameter of 60″ (around 40″ × 40″ on a squared sensor). The MCAO system shall be able to work on-disk and also off-disk (see, for instance, the recent work of Schmidt et al. 2018). Future upgrades on the MCAO system will expand the corrected area closer to the maximum FOV the telescope provides.

The telescope shall be optimised for high photon flux, minimising the number of optical surfaces in the light path. The latter can be achieved using novel technologies such as an adaptive secondary mirror (ASM).

For the Coudé platform, the SRD states that image rotation is acceptable as the required instruments are imagers and two-dimensional spectrographs (see Section 6.11.4). Hence, EST shall not use a rotating platform and shall instead be designed to minimise image rotation during the morning that corresponds to the time period when the seeing is usually best.

As mentioned in previous sections, one of the critical drivers for EST is the ability to observe weak polarisation signals above the noise level. Defining the polarimetric sensitivity as the ability to detect a signal above the noise (e.g. the ratio between the root-mean-square noise value of a given Stokes parameter $Q$, $U$, or $V$ and the average intensity $I$) and without crosstalk between Stokes parameters, EST shall achieve a sensitivity of $3{\times}10^{-5}$ normalised to the continuum intensity. Polarimetric accuracy can be defined as the residual errors in establishing the zero polari-sation level. In that case, EST shall achieve a value of $5 \times 10^{-4}$ normalised to the continuum intensity.

| | |
|---|---|
| **Telescope** | On axis Gregorian telescope |
| **Aperture** | 4.2 m with a central obscuration of 1.1 m |
| **Secondary mirror** | ASM with 5 degrees of freedom (piston, $\delta$x, $\delta$y, and tip-tilt) |
| **Mount** | Altitude-azimuth mount |
| **FOV** | 125″ diameter |
| **AO** | MCAO |
| **Spatial resolution** | Diffraction limited at 0.025″ at 500 nm |
| **Polarimetric accuracy** | $5 \times 10^{-4}$ of $I_c$ |
| **Polarimetric sensitivity** | $3 \times 10^{-5}$ of $I_c$ |
| **Spectral range** | 380-2300 nm |
| **Observations** | Multi-wavelength simultaneous observations |
| **Coudé lab** | Non-rotating platform |
| **Instruments** | 1. Integral Field Spectropolarimeters, |
| | 2. Tunable Imaging Spectropolarimeters, |
| | 3. Fixed Band Imagers |
| **Polarimeters** | Polarimeter(s) in the blue, visible, red and near-infrared |
| **Lifetime** | At least two Hale solar cycles, i.e. 44 years |

**Table 1.** Summary of EST future capabilities based on the technical requirements presented in Section 5.

The EST wavelength range shall span from 380 nm to 2300 nm, and the telescope transmission shall be optimised for scanning the Ca II 854.2 nm spectral line (i.e. the most demanded transition in the SRD observing programmes). This transition is sensitive to a wide range of heights from the photosphere to the lower-middle chromosphere (e.g. Uitenbroek 2006; Pietarila et al. 2007; Cauzzi et al. 2008; de la Cruz Rodríguez et al. 2012; de la Cruz Rodríguez et al. 2015; Quintero Noda et al. 2016), and has a relatively high sensitivity to the Zeeman effect (at least compared to other chromospheric lines). It can be modelled with a simple and fast atom (for example, Shine & Linsky 1974) because calcium is almost entirely singly ionised under typical chromospheric conditions. Additionally, non-equilibrium and partial redistribution effects are negligible for that spectral line (Uitenbroek 1989; Wedemeyer-Böhm & Carlsson 2011), which simplifies and speeds up the radiative transfer process for modelling it. Up to three photospheric and chromospheric transitions shall be observable strictly simultaneously.

Instruments shall include large FOV NB imagers and two-dimensional spectrographs that offer a wider spectral range (albeit with a small FOV). In some cases, the spatial resolution of those instruments will be close to EST 4m diffraction limit. However, various science cases ask for a high signal-to-noise ratio that will require increasing the collection area (reducing the spatial resolution). The latter condition may be achieved by a variable image scale (for instance, see DL-NIRSP observing modes in Elmore et al. 2014) or through binning of the recorded pixels.

A simple calculation can show that a signal acquisition time of 15 s is required to achieve a polarimetric sensitivity of $10^{-4}$ of $I_c$ for the diffraction limited area $(0.025'')^2$ at a wavelength of 500 nm and a spectral resolution 100000 for a telescope with a 4.2 m aperture. This value is based on the assumption of a 10% total (telescope+instrument) throughput and that noise in Stokes parameters $Q$, $U$ and $V$ is limited by shot noise of the parent signal ($I$). If a higher temporal resolution is needed to study, for example, fast phenomena such as the propagation of shock events, requirements for polarimetric sensitivity, spatial resolution, or spectral resolution have to be adjusted accordingly at the instrument level.

## 6. Telescope and instruments

### 6.1. Geographical location and proposed site

The proposed site for EST is at the Roque de Los Muchachos Observatory (ORM) in La Palma, Spain. This site was suggested after extensive studies of the characteristics of the best observatories in the world. It started with the information gathered during the Large Earth-based Solar Telescope (LEST) project (Engvold 1991) that compared the conditions at three candidate observatories (Hawaii, Teide, and Roque de Los Muchachos). The LEST team ran a campaign covering around five years of seeing characterisation, concluding that ORM was one of the best sites for solar observations. This conclusion has been verified by the excellent performance of the Swedish Vacuum Solar Telescope (Scharmer et al. 1985) and by its successor, the 1m SST (Scharmer et al. 2003a), located close to the proposed site for the LEST telescope, during their more than 30 years of operation.

Moreover, updated in-depth studies of the characteristics of the two Canarian observatories, the ORM and the Teide Observatory (OT), were performed in the last two decades. The main features of both observatories were analysed, from sky conditions to existing infrastructures. In addition, climate and meteorological conditions have also been specifically studied by the Sky Team of the Institute of Astrophysics of the Canary Islands. These studies concluded that both observatories possess excellent qualities for performing solar observations. Thus, the particular location at each of the observatories and the height of the building above the ground are the most critical parameters.

After considering all the available information, two locations were pre-selected at each observatory. Finally, the EST Board proposed a location near the SST at ORM as the preferential site for EST (see Figure 12) on October 4, 2019. This proposal was evaluated by the International Scientific Committee of the Canarian Observatories and approved on May 21, 2021. The resolution explains that the site in the area of the DOT (see Figure 12) is the best site for solar observations and with the least impact on the surrounding infrastructures.

### 6.2. Enclosure, pier, and building

Solar telescopes are often placed on the top of a tower to improve the local seeing conditions. EST will follow this philosophy set-ting the telescope structure holding its primary mirror M1 (see yellow in Figure 13) on a tower or pier (green in the same figure) at about 38 m from the ground. The pier will provide enough stiffness to achieve high pointing accuracy and stability. In addition, the shape of the pier will be optimised to reduce wind loads and avoid degrading the local seeing while allowing the M1 maintenance manoeuvres.

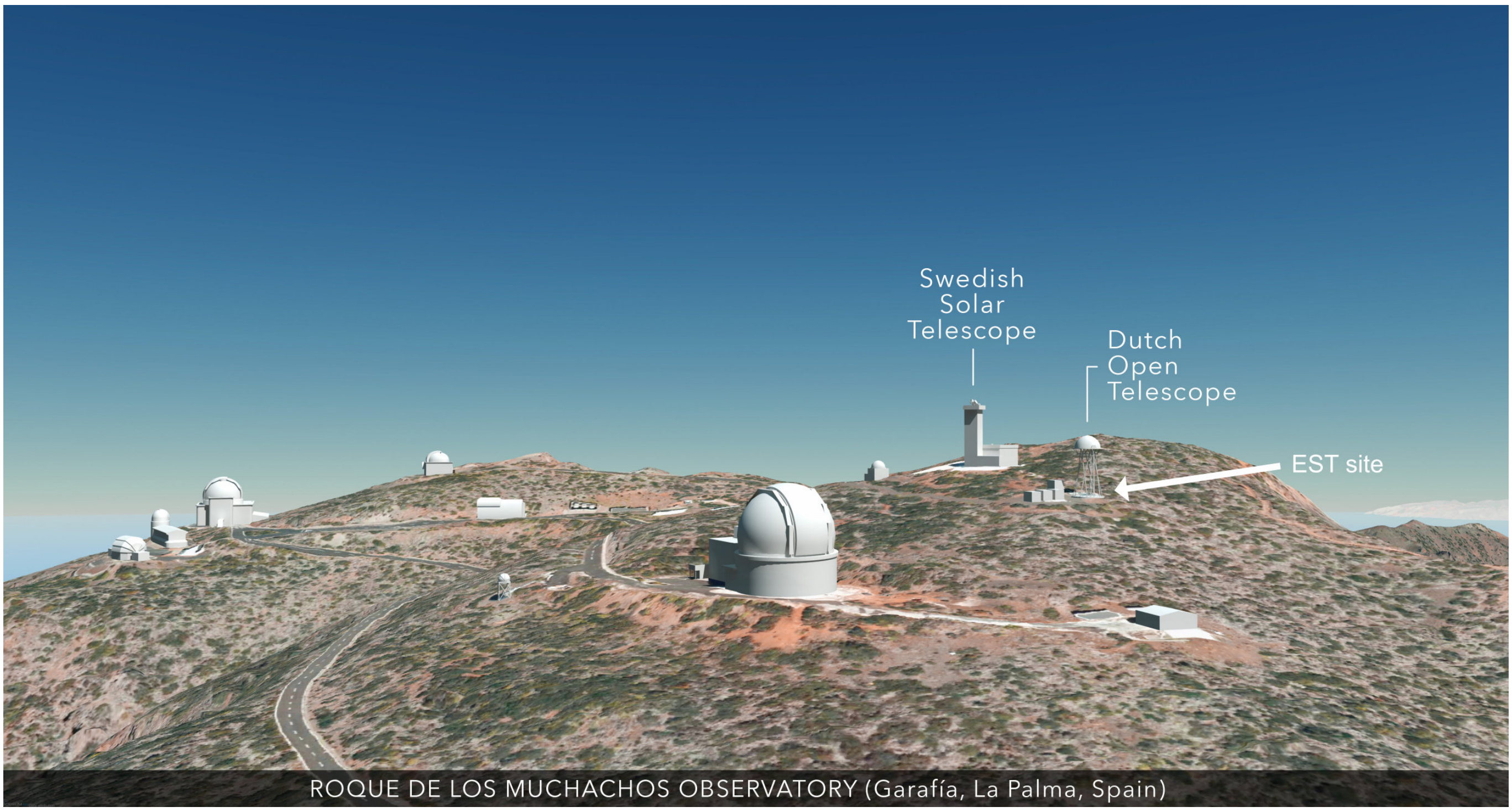

**Fig. 12.** EST site at Roque de Los Muchachos Observatory (La Palma, Spain). It was approved by the International Scientific Committee of the Canarian Observatories on May 21, 2021. The area is close to the DOT (see white arrow), next to the SST. Image credit: Gabriel Pérez (IAC) and originally published on the EST website on June 8, 2021.

At the top of the pier, there is the telescope enclosure (see red in Figure 13) that will protect the telescope structure and the optics subsystems when the telescope is parked and when the environmental conditions are out of limits for operation. The enclosure will also provide a secure area for maintenance tasks and keep the telescope temperature under control during the night. The enclosure will be fully retracted when operating, allowing for natural air flushing. Also, accurate thermal control around the telescope structure will be in place.

The pier also hosts the pier optical path (POP; see Sec. 6.10), which transfers the telescope focus to the Coudé instrument laboratory, placed at the bottom of the structure (see the arrow in Figure 13). The laboratory comprises three independent floors, and each one is configured as a clean room with all the facilities required to run the operations of the instruments. The baseline for the telescope pier is a concrete tower that will enclose the Coudé room and the transfer optics while providing the necessary stiffness to the telescope azimuth base.

The main building that includes the facilities and services required for the telescope operation, support, and maintenance is attached to the telescope pier. There, the control room, workshops, laboratories (mechanical, electrical, optics and instrumentation, mirrors coating and cleaning facilities, etc.), storage, staff, and visitor areas will be located.

In addition to the main building, there will be a separate auxiliary building located at a certain distance. This supporting facility will host those services that produce heat, smoke or vibrations, minimising the disturbances and the local seeing degradation they can induce during the observations. For instance, the auxiliary building will host the power and water supply rooms, cooling and air conditioning equipment rooms, fire prevention equipment rooms, hydraulic pumps and air compressor rooms.

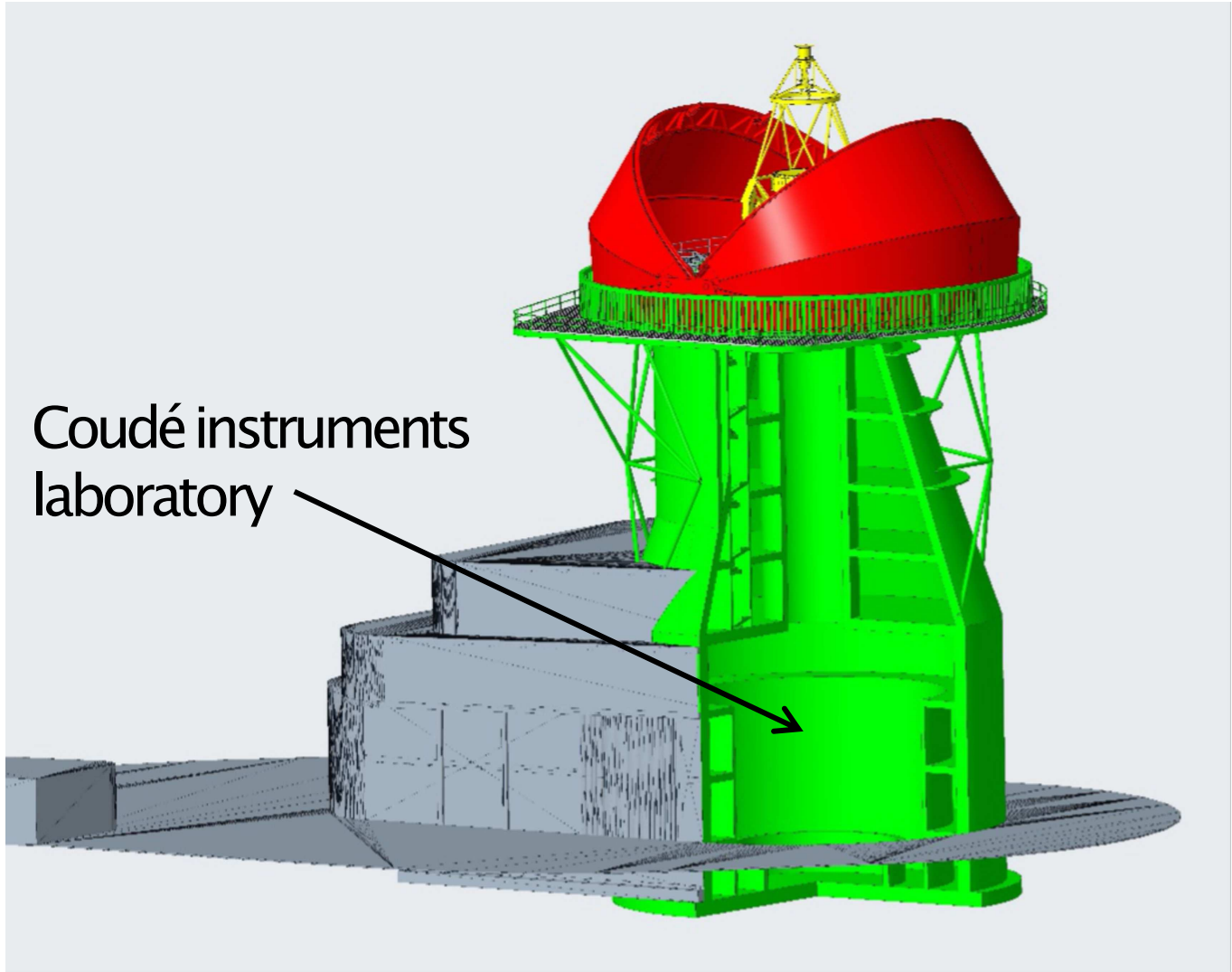

**Fig. 13.** Telescope structure (yellow), the pier (green) and the enclosure (red) of the EST.

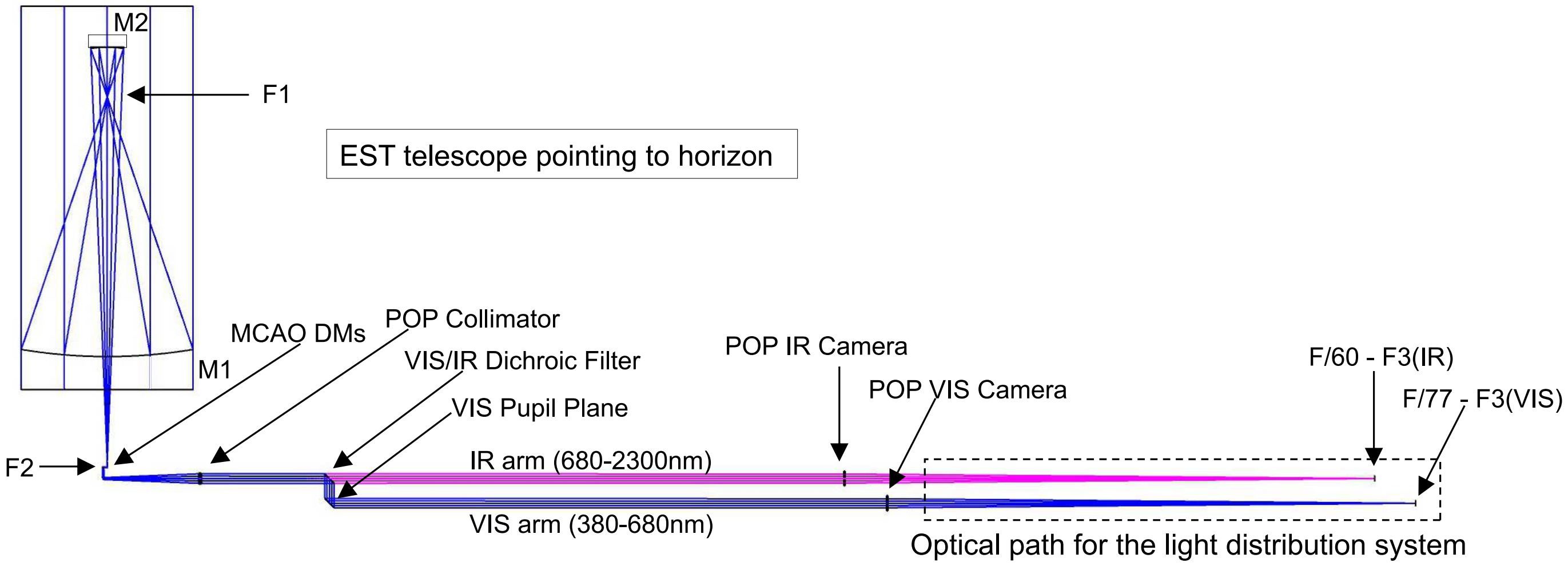

**Fig. 14.** Complete optical layout of EST. The telescope delivers the focal plane F3 to the instruments after passing the MCAO set of deformable mirrors (DMs) and the Pier Optical Path (POP) system. The VIS and IR labels define whether an arm samples the visible or infrared parts of the spectrum, with the former applying to the wavelength range from 380-680 nm and the latter applying to the wavelength range from 680-2300 nm.

| | Diameter [m] | Focal length [m] | Conic constant | Coating |
|---|---|---|---|---|
| M1 | 4.2 | 6.3 | -0.9954 | Ag/Al |
| M2 | 0.8 | 1/DM | -0.6344 | Ag |
| M3 | 0.065 | DM | Flat | Ag |
| M4 | 0.077 | DM | Flat | Ag |
| M5 | 0.082 | DM | Flat | Ag |
| M6 | 0.116 | DM | Flat | Ag |

**Table 2.** Specifications of the mirrors included in the optical path of EST. DM stands for deformable mirror. Mirrors M3 to M6 are flat.

### 6.3. Optical design

The optical design is presented in Figure 14. It is based on an aplanatic Gregorian telescope with only six reflections (see Table 2), arranged in mirror pairs with incidence-reflection planes perpendicular to one another (see also Table 2) to compensate for the instrumental polarisation induced by each individual mirror. Thanks to that configuration, the telescope has a polarimetrically compensated layout, with a telescope Mueller matrix that is diagonal (except for known rotation matrices), independent of wavelength and telescope pointing. Consequently, the Mueller matrix will not change during a given observation if the coatings of each mirror pair have identical optical properties.

### 6.4. Telescope structure

The telescope mechanical configuration is alt-azimuthal and is shown in Figure 15. The elevation axis has been placed below the M1 vertex to facilitate a natural air flushing on the mirror. This configuration also provides enough space below M1 for adequate placement of the transfer optics (mirrors from M3 to M6, which provide the telescope motion around the elevation and azimuth axes) and the calibration assembly (see Section 6.8). The telescope structure provides a pointing accuracy of $\pm 2.7''$ and a solar tracking error of $\pm 0.9''$. The telescope structure surfaces will be thermally controlled and maintained at around $\pm$ 1 K with respect to the surrounding air temperature. This structure supports the primary and secondary mirrors, the heat rejecter (HR), the transfer optics, and the calibration assembly.

| Movement | Piston [$\mu$m] | Tip-tilt [$\mu$rad] | Surface [nm] |
|---|---|---|---|
| Range | $\pm$2000 | $\pm$1000 | – |
| Accuracy | 5 | 5 | 60 |

**Table 3.** Degrees of freedom for the M1 system. Range indicates the maximum amplitude each actuator can move, while accuracy represents the maximum acceptable positioning error. Those corrections are performed by the, approximately, 80 actuators shown in Figure 16.

### 6.5. Primary mirror

The EST M1 Assembly is composed of a 4.2 m diameter mirror with near-zero thermal expansion coefficient, which will be supported by the M1 Cell, which includes the M1 Mirror Support and the M1 Thermal Control (see Figure 16). The M1 Cell also provides the interface that fixes the M1 Assembly to the telescope structure. The mirror comprises a polished blank made from glass-ceramic or optical glass and its coating. The mirror is a thin solid meniscus blank, with a coating based either on aluminium or silver, aiming to maximise the throughput at 854 nm, the spectral region given the highest priority in the SRD.

The M1 Mirror Support system holds the mirror into the cell. It includes about 80 active and passive actuators that will allow for correcting M1 surface figure errors due to slow changing effects (gravity, temperature and partly wind), and will also enable piston and tip-tilt corrections to be performed (see Table 3 for more information). The goal is to compensate for various optical aberrations using look-up tables and real-time wavefront sensor measurements. The requirements for the range of corrections of typical optical aberrations are included in Table 4. Those corrections shall be achieved within around 1 s after receiving the command. The M1 Thermal Control System maintains the mirror and the cell close to ambient temperature to minimise image degradation due to local seeing. The mirror will be at around +0.5/-2 K with respect to the surrounding air and the cell at approximately +1.5/-2 K with respect to the surrounding air.

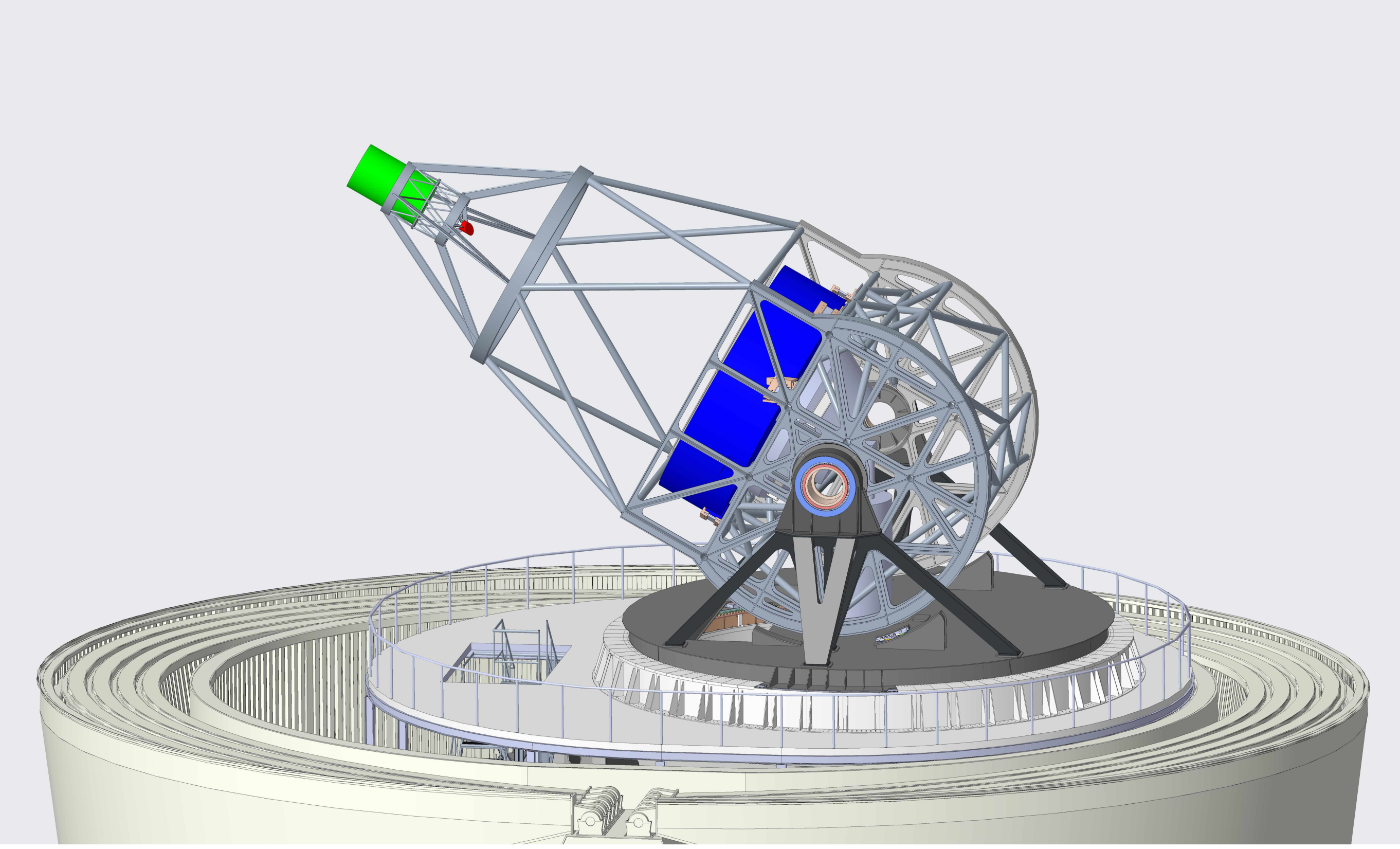

**Fig. 15.** Telescope structure of EST. Blue, red, and green highlight the primary mirror, the HR, and the adaptive secondary mirror, respectively.

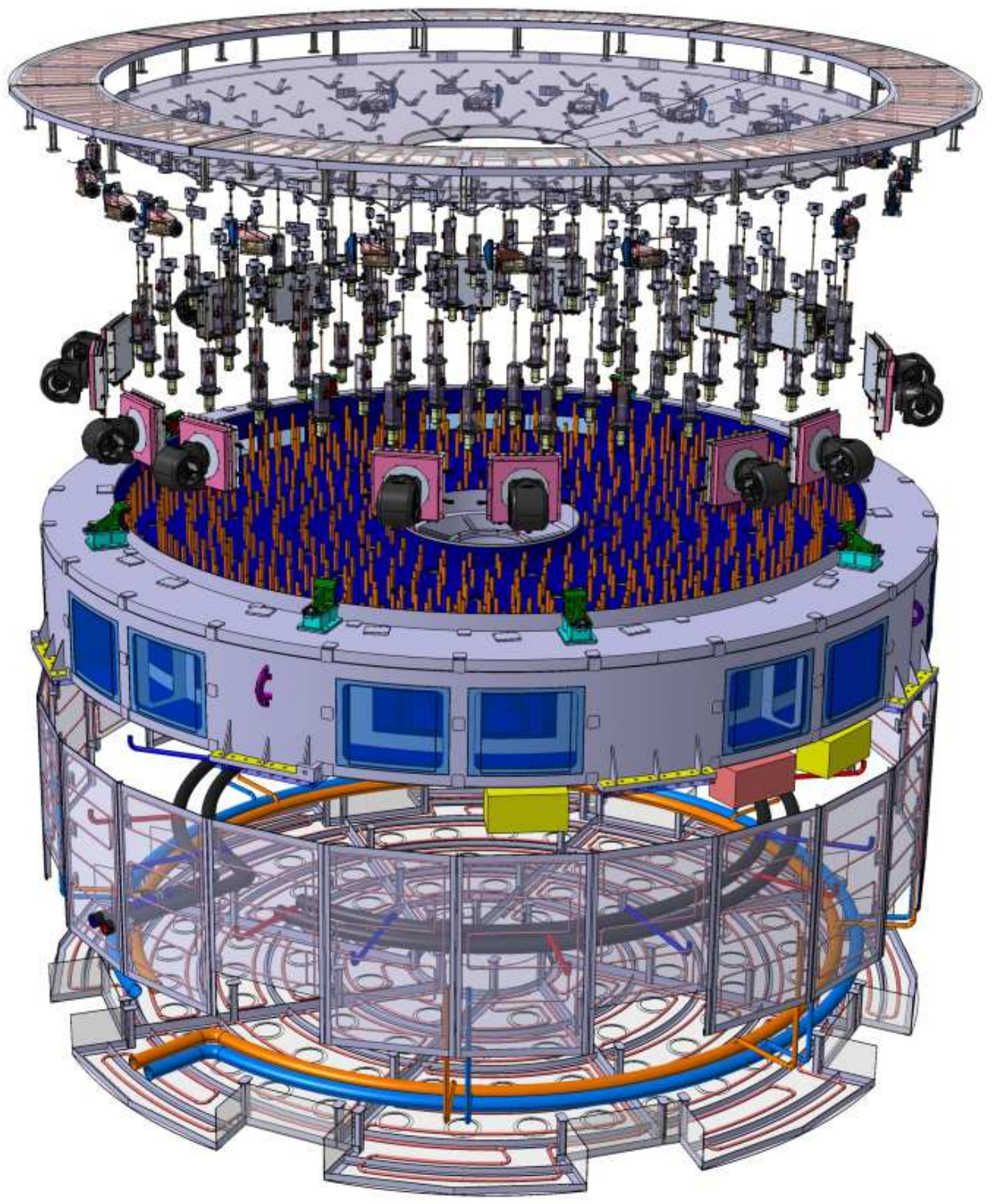

**Fig. 16.** Preliminary design of the primary mirror assembly courtesy of SENER-aerospace. Labels designate the four main elements of the system: (a) the M1 mirror, (b) the M1 support system, (c) the M1 cooling system, and (d) and the M1 cell structure.

| Term | Astigmatism | Coma | Spherical | Trefoil | Quadrafoil |
|---|---|---|---|---|---|
| Req. [$\mu$m] | 5 | 0.5 | 0.5 | 0.5 | 0.5 |

**Table 4.** Range for required corrections of optical aberrations external to the M1 Assembly. Those corrections shall be achieved within 1 s.

### *6.6. Heat rejecter*

The HR assembly lies at the telescope prime focus. It operates as a field stop, selecting the FOV of the telescope and reflecting more than 99% of the solar disk radiation (see the concept presented in Figure 17) to the outside. The HR aims to stop the solar radiation from falling outside the transmitted FOV, avoiding the development of thermal plumes that can cause internal seeing. A considerable heat load (approximately 13 kW) is concentrated by the primary mirror (M1) in the solar disk image at the telescope focal plane. This heat load implies a very high power density, equal to about 4 MW m$^{-2}$. The energy that cannot be rejected is absorbed by the HR mirror and ranges between 700-2000 W depending on the HR reflectivity. This absorbed energy is dissipated by the cooling system, which requires high heat transfer coefficients (HTCs) to maintain the temperature difference between the ambient temperature and the HR front face within a few degrees. Elevated HTCs can be achieved through the use of suitable heat removal techniques such as multiple impinging jets (e.g. Berrilli et al. 2010). It seems adequate to demand the lowest temperature differences between the HR mirror and the ambient air. To establish that requirement for the HR mirror, the maximum temperature difference was defined based on previous works for other solar telescopes (see, for instance, Dalrymple et al. (2004) for DKIST or Volkmer et al. (2003a,b) for GREGOR). In the latter case, which uses a type of HR similar to that expected on EST (e.g. Berrilli et al. 2010), HR prototype temperature gradients have been measured as low as ±1 K with respect to the ambient temperature (Volkmer et al. 2010) although larger values, up to ±10 K, were measured on the telescope during commissioning (Soltau et al. 2012). So, we have defined a baseline requirement for the HR mirror to work with a deviation from ambient air within ±3 K during observations. That value seems a reasonable middle ground between a feasible optomechanical design and high-performance image quality. The requirement will be updated when additional studies are performed with the EST HR prototype that is currently under development. Finally, the HR spider (see Figure 17) has four structural links that connect the HR assembly to the telescope elevation structure, and it serves as a support for the ducts for coolant and cabling.

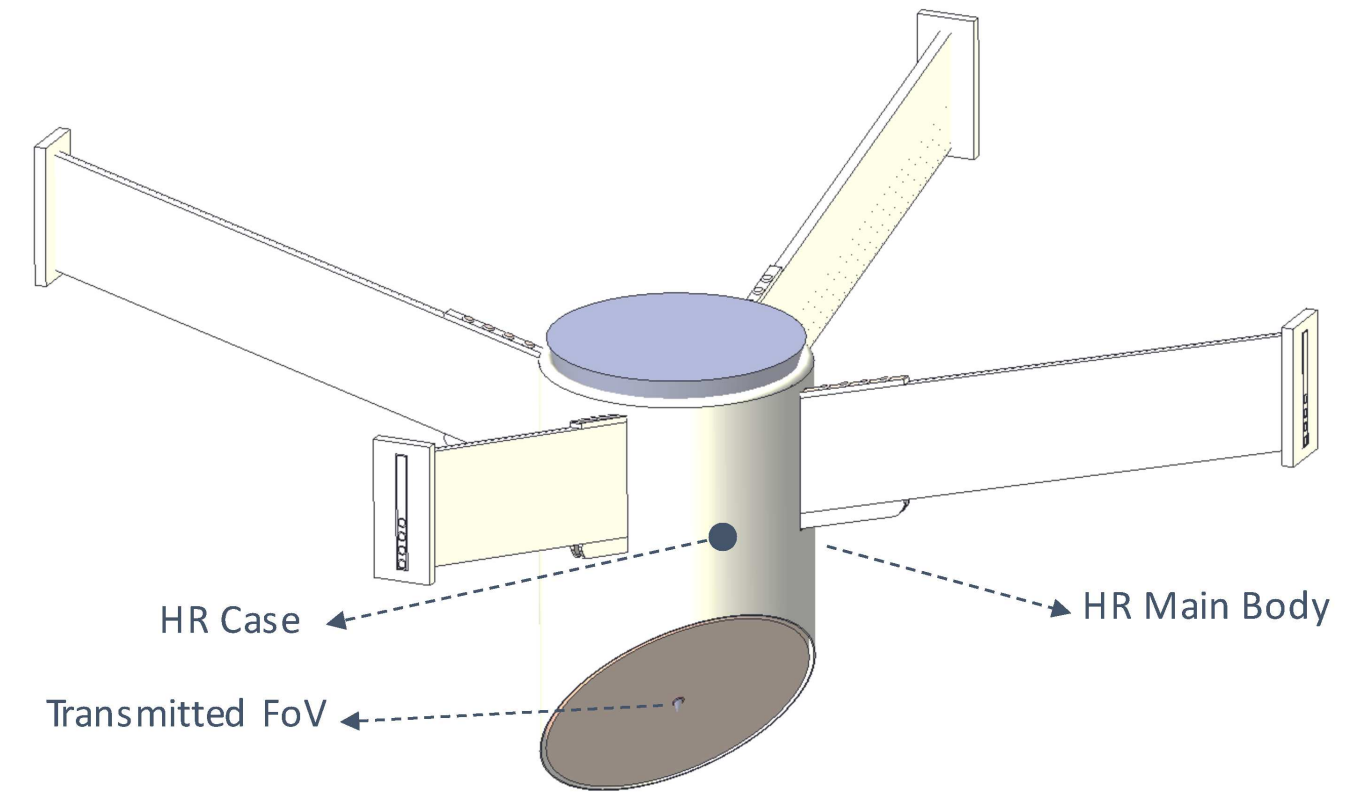

**Fig. 17.** Heat rejecter concept. It has a flat reflective cooled surface inclined 45° with respect to the optical axis that rejects all the incoming light that falls outside the observed FOV.

### 6.7. Adaptive secondary mirror

The adaptive secondary mirror (ASM) is a 0.8 m on-axis ellipsoid and is defined as the whole system aperture stop. The latter property allows the secondary mirror to be assembled as an adaptive mirror to provide AO capabilities for correcting atmospheric turbulence. The baseline for the ASM specifications is to have around 2000 actuators (approximately 50 actuators along the pupil diameter) with a stroke of the order of 16 microns to correct high-order aberrations. The ASM also needs to correct the image jitter caused by, for example, atmospheric turbulence, wind disturbance, or vibrations. This image jitter correction is about ±6.5″ at a bandwidth correction of 20 Hz and ±1.2″ at a bandwidth correction of 350 Hz. The ASM will be assembled on a hexapod mounting with 5 degrees of freedom (piston, $\delta$x, $\delta$y, and tip-tilt) to perform AO tasks as well (see the concept of the system in Figure 18). The M2 coating is protected silver for maximising the throughput at 854 nm, the spectral region with the highest science potential on the SRD.

### 6.8. Transfer optics and calibration assembly

The transfer optics and calibration assembly (TOCA) is shown in Figure 19. It is composed of 4 mirrors (M3 through M6) located below M1 to guide the light from F2 to the Coudé lab. The mirrors follow the movement of the telescope elevation and azimuth axes. The elevation axis is defined by the line that connects M4 and M5, while the azimuth axis is vertical after the beam is reflected on M6 (see Figure 19). The four mirrors are deformable and, together with the ASM, form the MCAO system (see more in Sec. 6.9). The mirrors are oriented perpendicular to each other to compensate for the polarisation induced by each individual mirror. In particular, M4 compensates for the polarisation caused by M3, while M6 represents the same role with respect to M5. This compensation is significant in this part of the optical path because the orientation of the mirrors changes with time, and so does the instrumental polarisation they induce individually. The orthogonal configuration removes the time-dependent instrumental polarisation induced during the observations, with a diagonal Mueller matrix down to M6 (except for known rotation matrices). The TOCA includes the ASM calibration unit located at the telescope focal plane F2 to calibrate the ASM, a polarisation calibration unit to characterise any (constant) instrumental polarisation of the fixed optical elements of the rest of the optical path, alignment elements, and room for a wavefront sensor unit, if needed.

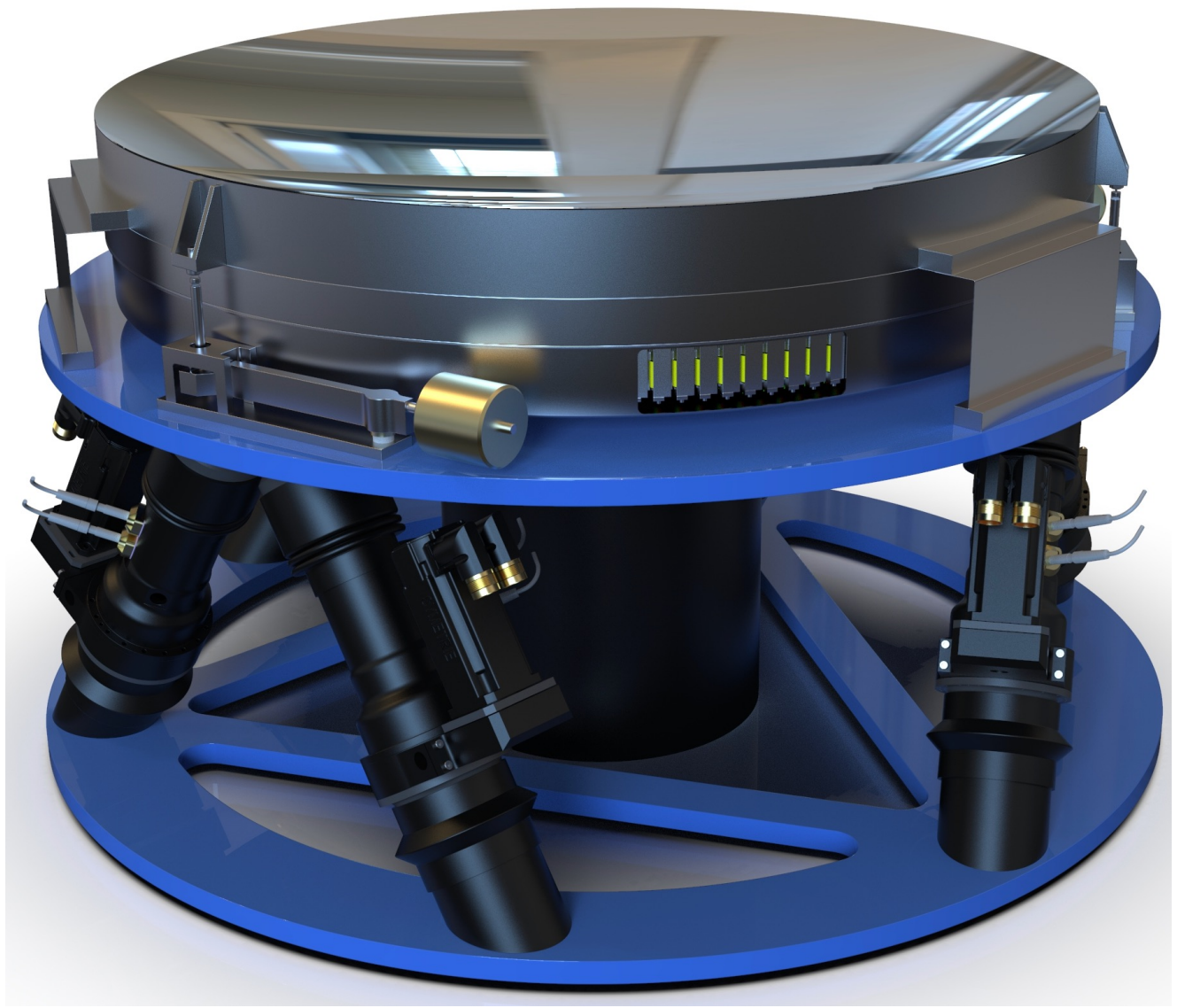

**Fig. 18.** Concept of the ASM. Courtesy of TNO (Netherlands Organisation for Applied Scientific Research) and B. Dekker.

### 6.9. MCAO

The MCAO system of EST will correct the perturbations produced by the Earth's atmosphere. The final effect of this correction is that the solar image has a real-time improvement over a larger FOV than traditional single conjugated AO systems. The MCAO system features five high altitude deformable mirrors (DMs) tentatively conjugated to 0 (ASM), 5 (M6), 9 (M5), 12 (M4), and 20 km (M3) (Montoya et al. 2015), and the current design proposes two wavefront sensors (WFSs). The first is a narrow field high-order WFS provisionally located inside the visible optical arm (see Section 6.11). It receives a small amount of the incoming light (around 10 %) in a narrow bandpass in the wavelength range from 500-680 nm. This WFS allows the traditional single conjugated AO operation to correct the atmospheric ground layer when working under MCAO or ground layer AO conditions. A second WFS is devised for complete

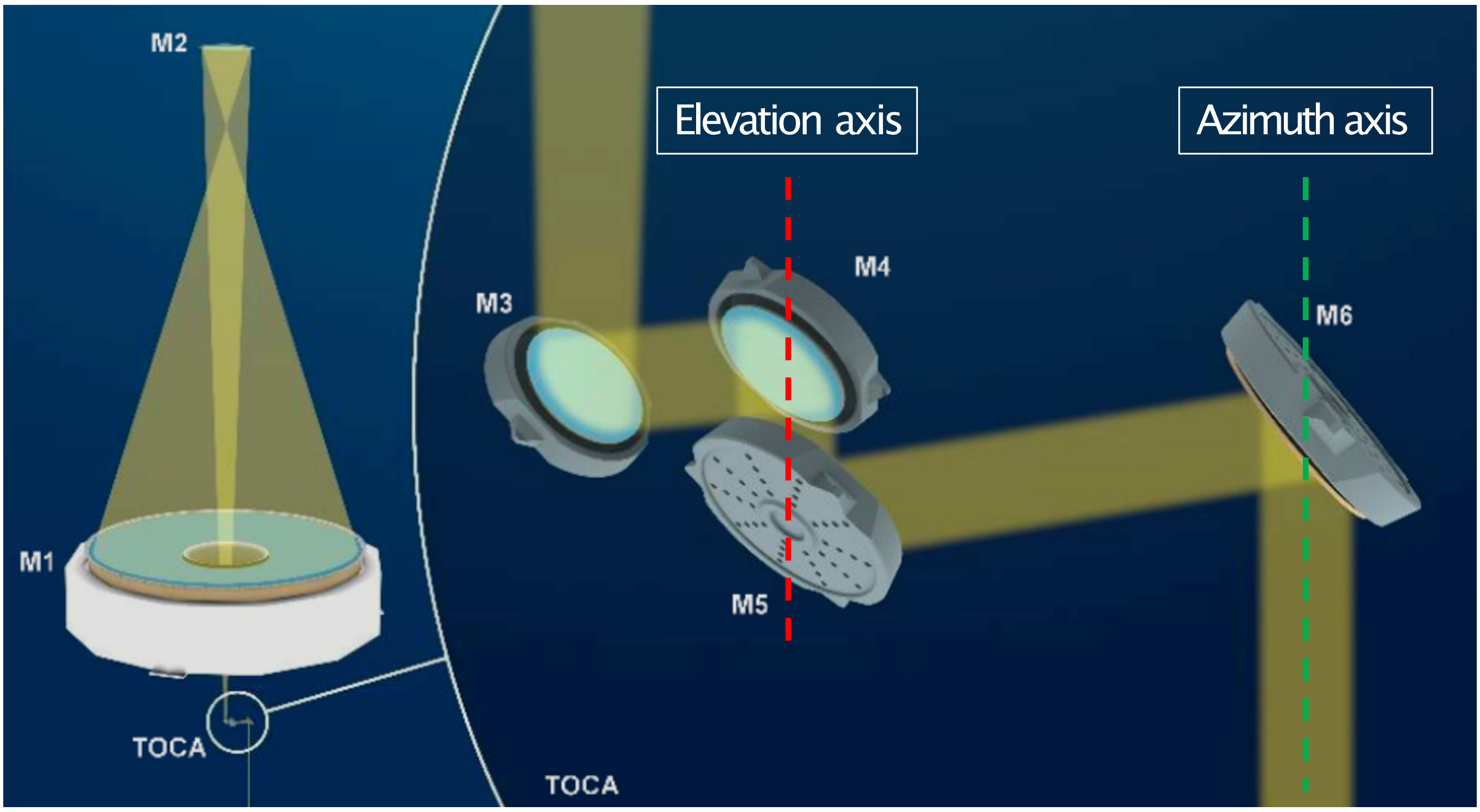

**Fig. 19.** TOCA set-up (M3-M6) after the Adaptive Secondary Mirror (M2) in the light path.

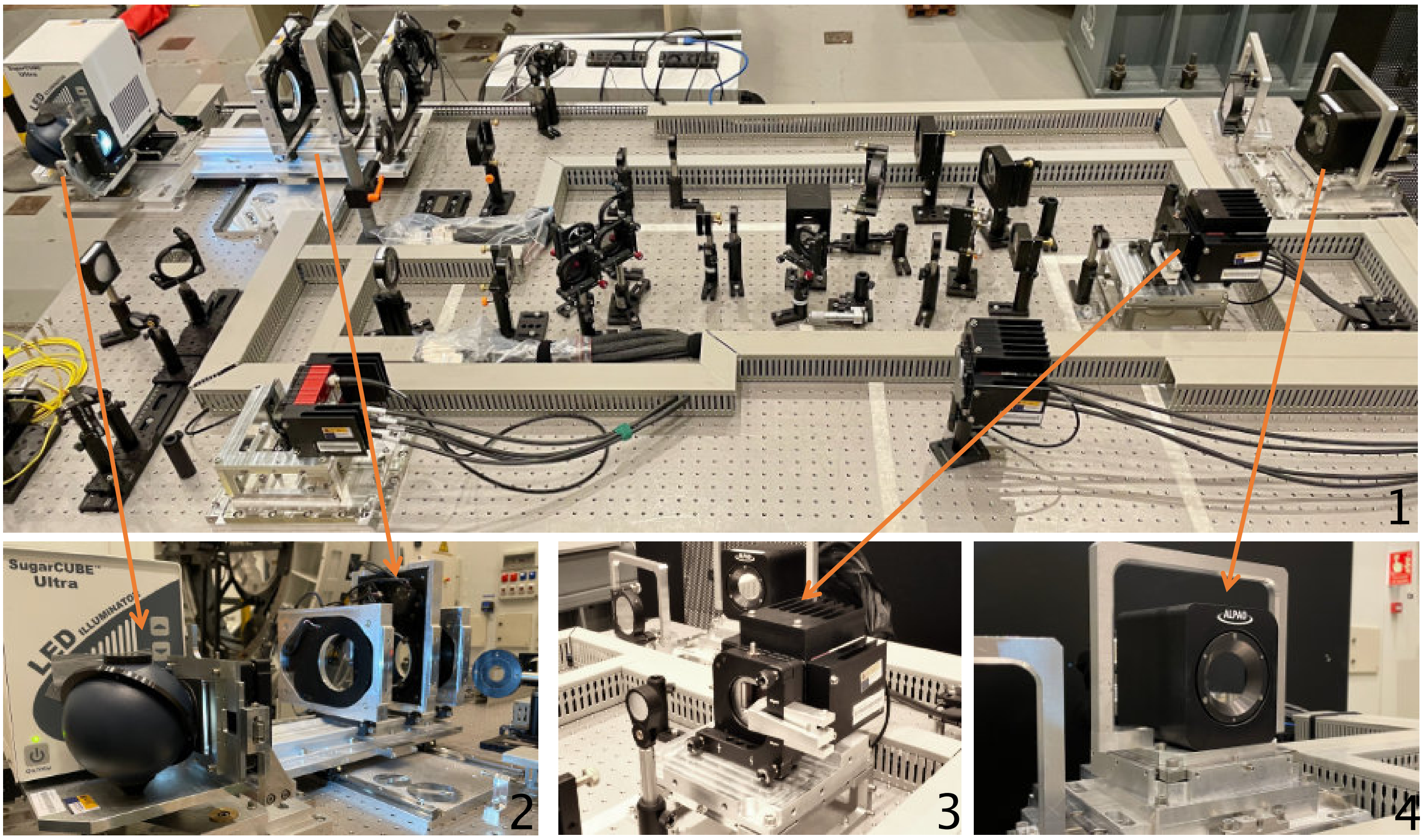

**Fig. 20.** MCAO test bench set-up at the IAC lab. The upper panel (label 1) shows the current, complete set-up, while the bottom panels display a zoom-in of the main elements highlighted with arrows in the upper panel. From left to right of the bottom row, there is the extended light source and phase screens (label 2), the wavefront sensor (label 3), and the pupil deformable mirror (label 4). MCAO DMs will be added at a later stage.

| Subsystem | Lens | Optical glass | Clear Diameter (mm) | Clear Thickness (mm) |
|---|---|---|---|---|
| Collimator | 1 | S-FPL55 | 239.265 | 30 |
| Collimator | 2 | PBL35Y | 238.781 | 30 |
| Collimator | 3 | N-LAF35 | 239.675 | 35 |
| Blue-visible Camera | 1 | N-FK58 | 390.212 | 25 |
| Blue-visible Camera | 2 | N-BK7HT | 389.810 | 27 |
| Red-infrared Camera | 1 | N-PK52A | 345.839 | 25 |
| Red-infrared Camera | 2 | N-KZFS11 | 345.593 | 25 |

**Table 5.** Preliminary properties of the different lenses belonging to the POP system.

MCAO corrections to drive the high altitude deformable mirrors (M3-M6). The current baseline considers a multi-directional wavefront sensor.

The requirement of the MCAO system is to reach the 4.2 m diffraction limit at 500 nm over a FOV of 40″ × 40″, when the seeing conditions are adequate (Fried parameter $r_0 > 7$ cm). Although a lot of progress has been made with the CLEAR system (for instance, Schmidt et al. 2017, 2018, 2021), pathfinder for DKIST's MCAO, and the MCAO experiments at the New Vacuum Solar Telescope (Rao et al. 2018). Presently there is no solar telescope with an MCAO system that is offered to the observer as a baseline AO system. In the following years, more telescopes are expected to provide this option for the community, particularly the DKIST team, who have made much progress with the mentioned CLEAR system. In the case of the EST team, as the MCAO system is aimed for the first light of the telescope, the optomechanical design, control software and additional components are developed in the lab with an MCAO demonstrator (see Figure 20). Among different goals, different strategies will be validated for wavefront measurement and corrections, new sensor technologies tested, computational resources defined to establish a consistent control between the various deformable mirrors and the wavefront sensor or even study the possibility of having multiple wavefront sensors working on different orders.

### *6.10. POP system*

The scientific instruments will be installed at the base of the pier in the Coudé room. The vertical distance between the telescope mount and the Coudé room is around 30 metres. This long-distance requires a complementary optical system, the POP, to transfer the telescope F2 focal plane from the top of the building to the scientific laboratory at the bottom of the structure (F3). The POP starts after the MCAO DMs highlighted in Figure 14 and extends up to the two F3 focus points located inside the Coudé room.

The optical system works in the spectral range from 380 to 2300 nm, and the image quality is limited by diffraction over the whole telescope FOV. The current version of the POP uses a colour corrected lens-based relay consisting of a collimator-camera set-up (see Figure 14). The properties of each lens are described in Table 5. The collimator subsystem is a lens triplet (top panel of Figure 22), complemented with a dual camera subsystem based on lens doublets for the visible and infrared optical arms (see middle and bottom panels in Figure 22). The beginning of the POP (i.e. the location of the collimator lens) is about 2.5 m below F2. In the case of the visible and IR camera lenses, they are situated at a distance of around 13 and 10 m below the collimator lens, respectively. The F/# of each optical arm is slightly different (see Figure 14).

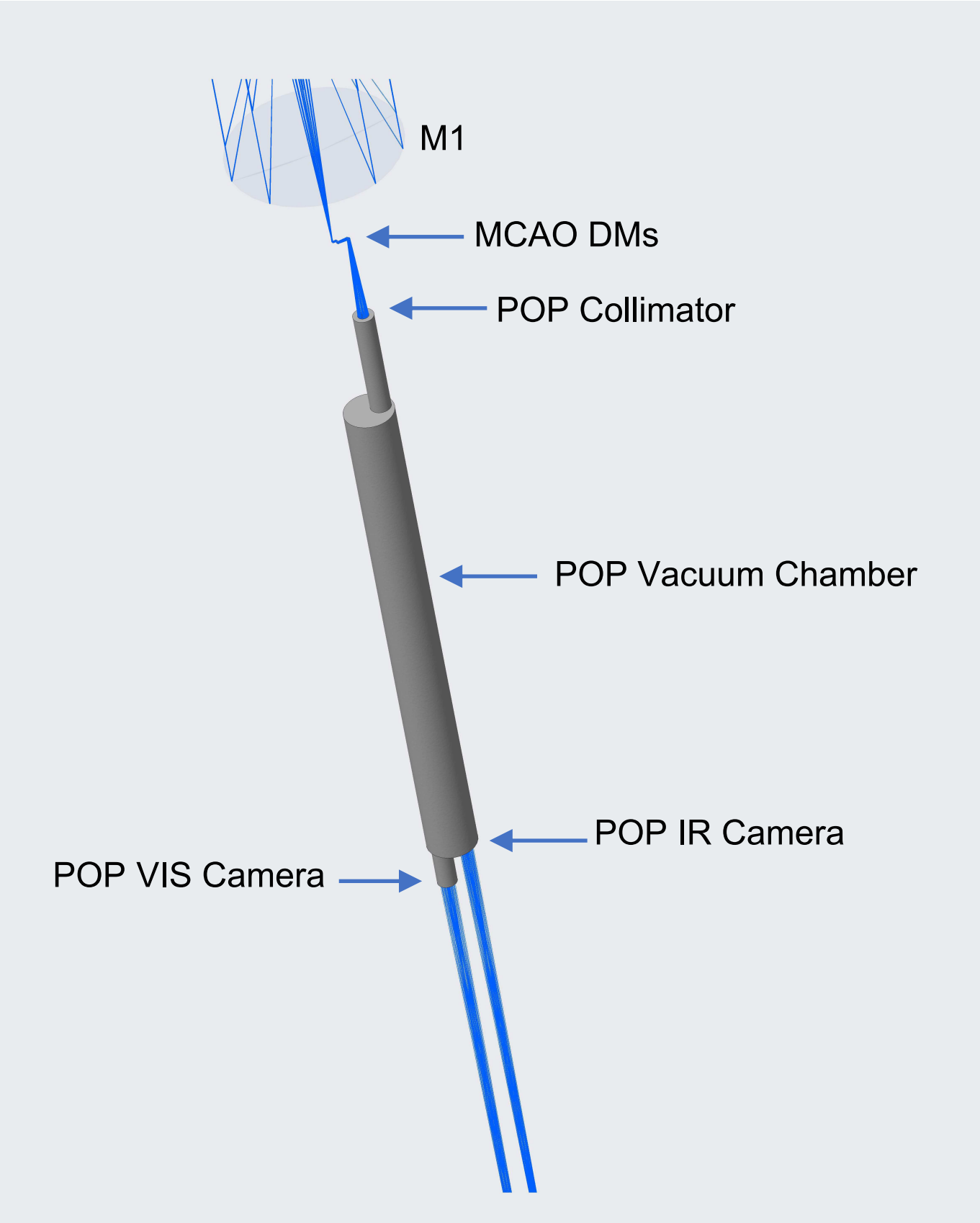

**Fig. 21.** Concept of the POP system inside the vacuum chamber. The incoming beam is divided into two (blue and visible, and red and infrared) wavelengths regions inside the vacuum tube. The collimator lens and the dual camera lens system act, respectively, as entrance and exit windows.

Lens-based systems are affected by chromatic aberration, which arise from the relation between the optical properties of lenses (such as focal length, magnification and spherical aberration) and the refractive index characterising each optical glass. Since the refractive index varies with wavelength, the chromatic variation of all these properties of lens systems results in aberrated images. Combining glasses with different optical properties constitutes the main strategy to overcome this aberration effect. The approach involves splitting the beam after the collimation stage with a dichroic beamsplitter. The division is currently made at 680 nm so that the blue-visible beam and the red-infrared beams will go through two independent camera doublets (see Figure 14). As a result, each camera doublet deals with a reduced spectral range, facilitating the diffraction-limited performance over the whole FOV. The entire POP system will be located inside a vacuum vessel isolated from the rest of the pier (see Figure 21), to provide stable pressure and temperature conditions for the optical train. The entrance and exit windows of the vacuum vessel will correspond to the collimator and camera lenses (see Figures 14 and 21).

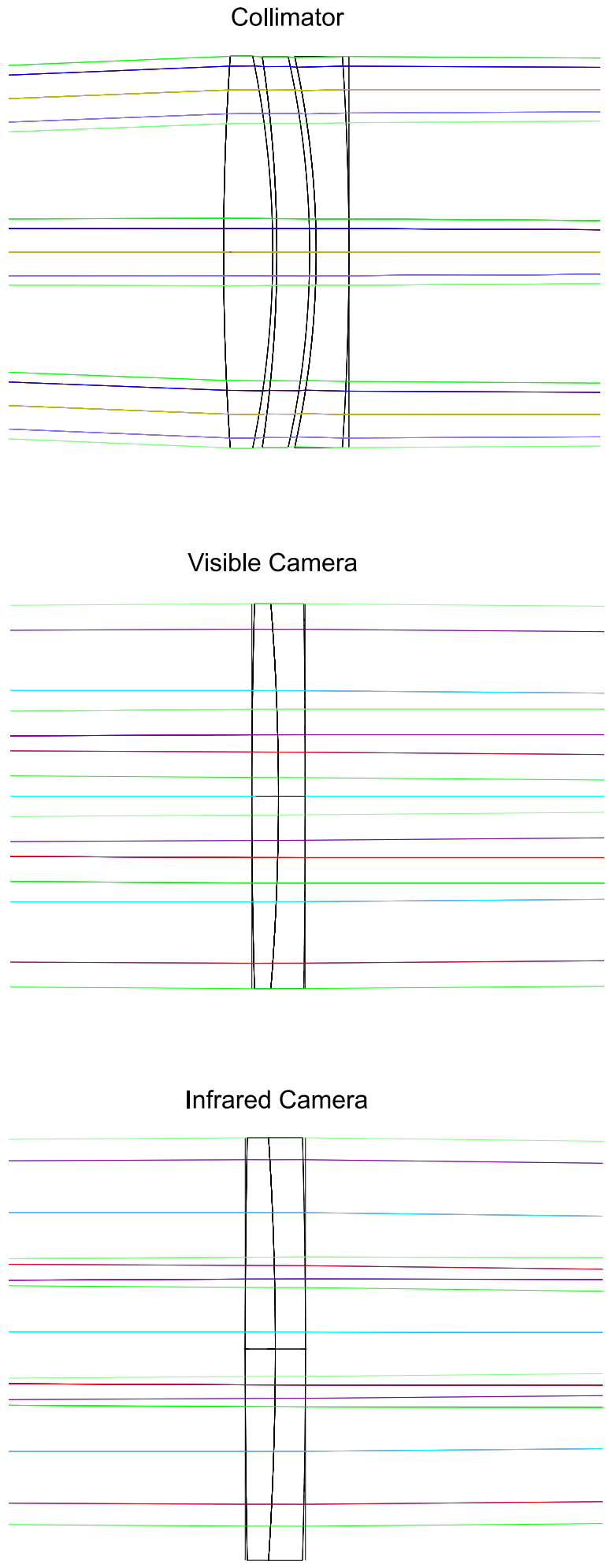

**Fig. 22.** Concept of the lenses proposed for the POP. From top to bottom are the collimator, the visible, and the infrared cameras. The light coming from the Sun travels from the left to the right side of each lens.

### *6.11. Coudé room*

#### 6.11.1. Light distribution system

The most up-to-date version of the light distribution system is shown in Figure 23. Based on the observing programmes included in the SRD, the light is divided into four optical arms, two in blue-visible beam and two in the red-infrared beam. This wavelength division by arm facilitates the realisation of simultaneous observations at various wavelengths. Each individual arm hosts the required spectroscopic and imaging instruments to cover the scientific targets described in previous sections and in the SRD. The core instruments are integral field spectropolarimeters (IFSs) and tunable imaging spectropolarimeters (TISs). A brief description of the requirements for both types of instruments is given below.

#### 6.11.2. Integral field spectropolarimeters

Integral field spectropolarimeter systems represent the future for observations in solar physics (see e.g. Calcines et al. 2013a,b; Iglesias & Feller 2019). They are three-dimensional spectrographs that combine the two-dimensional capabilities of traditional imaging instruments (broad or narrow bands) with the spectral capabilities that only long-slit spectrographs can provide (e.g. spectral resolution and coverage). The components of a traditional spectrograph are a slit, a collimator, a diffraction grating, and a camera optical system that focuses the wavelength region of interest on the instrument sensor. The integral field spectropolarimeter replace the standard long slit with an IFU to reformat an input two-dimensional FOV and leave room at the detector to expand each imaging pixel with its spectrum. This way, the inclusion of the IFU enables the spectra of all points to be observed in a given two-dimensional surface on the Sun, around $10'' \times 10''$ for EST, strictly simultaneously. The specifications for the IFS are presented in Table 6.

So far, four types of IFU devices have been successfully tested in solar physics: a subtractive double pass (e.g. Beck et al. 2018; Mein et al. 2021); image slicers; microlens arrays; and optical fibres. The latter will be used in the Diffraction Limited Near Infrared Spectropolarimeter (DL-NIRSP) (see the general information in Elmore et al. 2014) on DKIST. Image slicers and microlens arrays are the two preferred candidates for EST in the present preliminary design phase. The SOLARNET FP7, GREST H2020, and SOLARNET H2020 projects have proven the viability of both techniques using prototypes in current solar telescopes. The micro-lens prototype (MiHi[2]) has been tested at the SST (van Noort et al., in preparation). A slicer-based unit is installed at GREGOR as an upgrade of the GRIS spectrograph (Collados et al. 2012) and is the first solar IFU system that is offered openly to the community (Dominguez-Tagle et al. 2022). A consortium of institutions was formed in 2021 including the Max Planck Institute for Solar System Research, the Institute of Astrophysics of the Canary Islands, Stockholm University, the University of Coimbra, the Istituto Ricerche Solari Locarno (IRSOL), Scuola Universitaria Professionale della Svizzera Italiana, Haute Ecole d'Ingénierie et de Gestion du Canton de Vaud, the Palacky University Olomouc, and the Astronomical Institute of the Czech Academy of Sciences with the goal of making progress for the definition of these instruments. Following the light distribution presented in Figure 23, the aim is to have one IFS instrument per optical arm from the visible to the infrared, enabling strictly simultaneous multi-wavelength IFS observations within the spectral range from 380 to 2300 nm.

#### 6.11.3. Tunable imaging spectropolarimeters

Tunable imaging spectropolarimeter instruments are based on traditional Fabry-Pérot systems, where tunable etalons are used to observe a given spectral region over a two-dimensional FOV. Fabry-Pérot systems have been extensively used in recent years by solar observatories. At ground-based telescopes, examples of this type of instruments developed by EST partners are the Telecentric Etalon SOlar Spectrometer (TESOS; Kentischer et al. 1998), IBIS (Cavallini 2006), the GREGOR Fabry Perot Interferometer (GFPI; Puschmann et al. 2012), CRISP (Scharmer et al. 2008), and CHROMIS (Scharmer 2017). In balloon missions like Sunrise (Solanki et al. 2010), there is the Imaging Magnetograph eXperiment (IMaX; Martínez Pillet et al. 2011). Finally, in space, there is the Polarimetric and Helioseismic Imager (PHI, Solanki et al. 2020) on board Solar Orbiter. Fabry-Perot systems remain crucial instruments for the next generation 4m class of telescopes, with the visible tunable filter (VTF; Schmidt et al. 2014) being one of the first-light instruments of DKIST.

[2] https://www.mps.mpg.de/solar-physics/mls

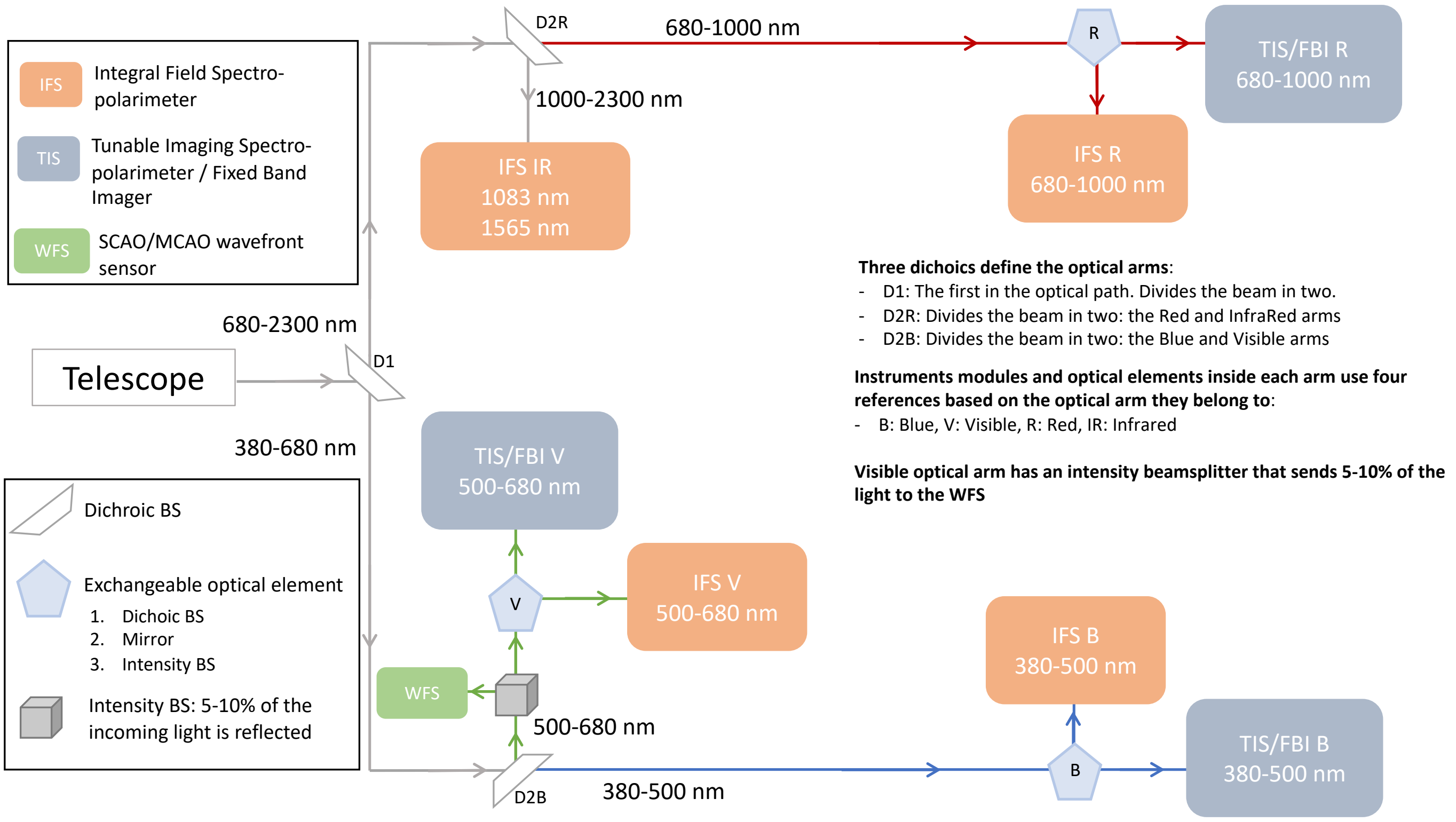

**Fig. 23.** Light distribution and instrument suite inside the Coudé room of EST.

| | |
|---|---|
| **Spatial resolution** | Diffraction limit in each optical arm |
| **FOV** | $10'' \times 10''$ |
| **Cadence** | 1. Up to around 60 FPS in spectroscopy mode |
| | 2. 15 s at diffraction limit, polarimetry with $5 \times 10^{-4}$ of $I_{cont}$ |
| **Integral Field Unit** | 1. Microlens arrays in the blue, visible, and red arms. |
| | 2. Image slicer in the infrared arms |
| **Spectral resolution** | Minimum of 150000 |
| **Spectral range** | Minimum of 1 nm |
| **Reference spectral lines** | 1. Blue: Ca II 396 nm, Ba II 455 nm, Sr I 461 nm, $H_\beta$ 486 nm |
| | 2. Visible: Mg I 517 nm, Na I 589 nm, Fe I 630 nm, $H_\alpha$ 656 nm |
| | 3. Red: K I 770 nm, Ca II 854 nm |
| | 4. Infrared: He I 1083 nm, Fe I 1565 nm |
| **Polarimetry** | Dual-beam to reduce the seeing-induced crosstalk |

**Table 6.** Summary of the general requirements of the integral field spectropolarimeters .

TIS instruments will be responsible for providing large FOV observations of the thermal and magnetic properties of solar features. They will operate in three optical arms strictly simultaneously (see Figure 23), with the general specifications described in Table 7. TIS instruments will have two main operation modes. In the first one, they will work as traditional NB spectropolarimeters covering a given FOV for each tuned spectral wavelength. In a parallel operation mode, the instrument will use broadband filters to work as a broadband context imager, corresponding to what is termed as fixed band imagers in Table 1. In this mode, broad filters will be used with bandwidths ranging from 0.15 nm for observing the line core of chromospheric lines to wider ones up to 1.5 nm to serve as a context for the NB optical arm (see, for instance, Figure 4 and Table 2 in Löfdahl et al. 2021, as reference). In both cases, observations may have a set of reference cameras (one in focus and the second one de-focused) to facilitate compensation for residual optical aberrations in post-processing. The baseline is that both operation modes will use the multi-object multi-frame blind-deconvolution (MOMFBD; Löfdahl 2002; van Noort et al. 2005) image reconstruction technique. This technique has been successfully used for more than 15 years on instruments such as CRISP and CHROMIS installed at the SST. Löfdahl et al. (2021) describe the data- and metadata-processing pipeline for CHROMIS and CRISP (SSTRED), which is publicly available through git repositories.

| | |
|---|---|
| **Spatial resolution** | Diffraction limit in each optical arm |
| **FOV** | $60'' \times 60''$ ($60''$ diameter for polarimetry) |
| **Cycle time** | 1. Up to around 60 FPS in spectroscopy mode |
| | 2. 10~20 s per spectral line and $1 \times 10^{-3}$ of $I_{cont}$ in polarimetry |
| **Spectral resolution** | Minimum of 50000 |
| **Wavelength samples** | 10 per line including a nearby continuum point |
| **Number of filters per module** | At least 5 |
| **Reference spectral lines** | 1. Blue: Ca II 396 nm, Ba II 455 nm, Sr I 461 nm, $H_\beta$ 486 nm |
| | 2. Visible: Mg I 517 nm, Na I 589 nm, Fe I 630 nm, $H_\alpha$ 656 nm |
| | 3. Red: Fe I 709 nm, K I 770 nm, Ca II 854 nm |
| **Broadband reference camera** | Each module has 2 reference broadband cameras to perform |
| | image reconstruction techniques |
| **Operation modes** | 1. Narrowband spectropolarimeter |
| | 2. Broadband context imager |
| **Polarimetry** | Dual-beam to reduce the seeing-induced crosstalk |

**Table 7.** Summary of the TIS general requirements.

The TIS instruments are key to understanding the properties of solar features that occur over larger spatial scales than those covered by the relatively limited FOV of IFS instruments. The task of designing TIS instruments has been assigned to a consortium of institutions led by the Spanish Space Solar Physics Consortium (S3PC), with partners in the Leibniz-Institut für Sonnenphysik (KIS), the University of Rome Tor Vergata, Stockholm University, Queen's University Belfast, Mullard Space Science Laboratory, IRSOL, the University of Catania, and the National Institute of Astrophysics of Italy (INAF).

#### 6.11.4. Image rotation

The Coudé lab, where the instruments are located, is a non-rotating platform. As a consequence, the solar image will be continuously rotating at F3 due to the alt-az mount of the telescope structure. The performance of classical long-slit spectrographs, for which the slit must be kept along the same direction on the sky to allow extended time-series observations, are strongly affected by a lack of rotating Coudé platform. However, two-dimensional instruments, like classical Fabry-Pérot based systems and spectrographs with IFUs, deal much better with image rotation. As the entire set of EST first-generation instruments (see Figure 23) is based on two-dimensional instruments, the requirement of having a rotating Coudé platform can be relaxed. The advantages are multiple, including reducing development and construction costs, stability, and increasing the available space at the instruments room.

Image rotation can be easily calculated to ascertain whether any instrument would require an internal image de-rotator. The predicted image rotation, $\alpha$, at the Coudé room is shown in Figure 24 for the equinoxes and solstices of 2020. These represent the extrema in slow and fast image rotation and thus the best- and worst-case scenarios. Image rotation speed values in the worst case do not exceed $0.^\circ003/\mathrm{s}$ in the first 5 hours of observing time (starting at 8:00 UT) and have a maximum of

$$(d\alpha/dt)_{\max} = 0.^\circ006/\mathrm{s} \tag{1}$$

in the afternoons around the summer solstice.

Based on those results, the TIS team evaluated the impact of image rotation on the TIS systems. A typical scan through a spectral line is assumed to take ~10 s, while the time spent collecting data at any spectral position is $\lesssim 1$ s. For MOMFBD image restoration, the 1 s timescale is the most relevant because the processing combines exposures collected during that length of time. Image restoration is usually done on subfields of approximately the size of an isoplanatic patch (say, $4''$) and then mosaicked to form a restored version of the whole FOV. In 1 s, the maximum rotation $(d\alpha/dt)_{\max}$ will move the outer parts of such a subfield by

$$h_{1\,\mathrm{s}} = 2'' \cdot \sin(0.^\circ006) \approx 0.''0002. \tag{2}$$

This displacement is negligible compared to the size of the resolution element in the blue at, for example, 400 nm:

$$h_{\mathrm{res}} = \frac{\lambda}{D} \approx 0.''02. \tag{3}$$

We note that for the purpose of the discussion in this section, angles denoted in degrees refer to rotation. At the same time, arcminutes and arcseconds are related to coordinates and spatial dimensions within the FOV.

The second time reference (i.e. the 10 s scale) is potentially relevant for image restoration because the wide-band (WB) reference images are collected in synchronisation with the NB images during the entire scan. The WB images provide a reference for the alignment of the NB data and most of the information to identify momentaneous aberrations. A 10 s integration will yield an image displacement of $10h_{1\,\mathrm{s}} = 0.''002$, which is closer to $h_{\mathrm{res}}$, but still negligible compared to the evolution of structures in the photosphere.

When examining the impact of the 10 s timescale over the entire FOV, the maximum rotation can be as large as

$$\alpha_{10\,\mathrm{s}} = (d\alpha/dt)_{\max} \cdot 10\ \mathrm{s} = 0.^\circ06, \tag{4}$$

which means outer areas in the FOV will move

$$H_{10\,\mathrm{s}} = \frac{FOV}{2} \cdot \sin(\alpha_{10\,\mathrm{s}}) = 30'' \cdot \sin(0.^\circ06) \approx 0.''03 \tag{5}$$

during a scan. This is similar to $h_{\mathrm{res}}$ and needs to be corrected for, to properly align images acquired at different spectral tuning positions. On the timescale of time-sequences (movies) of scans, potentially hours of data, images also have to be de-rotated to a common orientation. In this sense, the actual angles do not matter very much for the impact on image quality, as any re-sampling of the images is a blurring operation that affects contrast and power spectrum. All necessary re-samplings should be combined into as few such operations as possible.

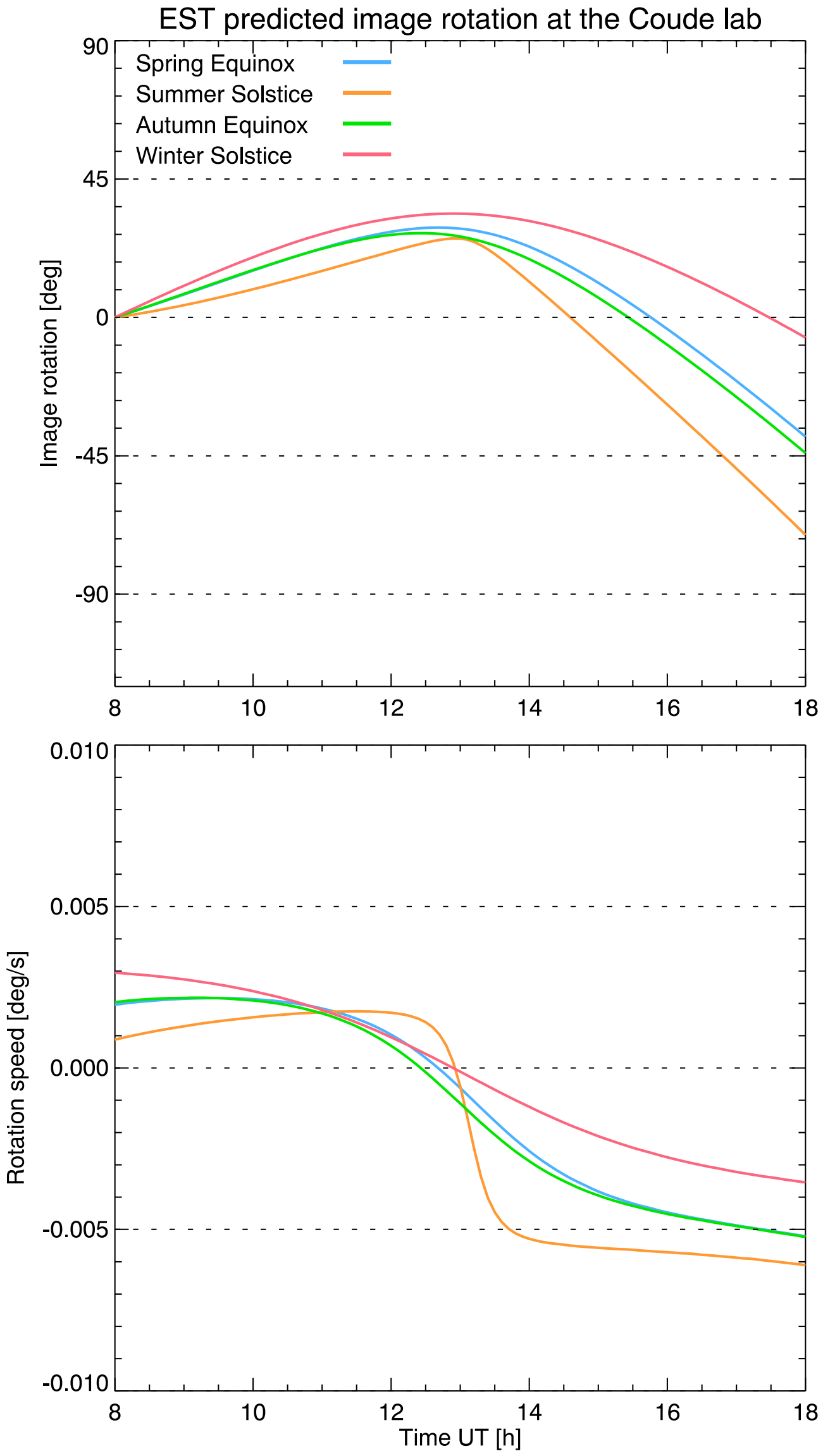

**Fig. 24.** Image rotation predicted for EST. Colours designate the dates picked as reference.

Additional work is needed to decide on the best approach regarding the data-processing pipelines for TIS and IFS instruments. There is a trade-off between rotating image data before or after image restoration. Pre-rotation is more computationally expensive because the number of raw data frames is much larger than the number of restored images. On the other hand, pre-rotation takes care of all image rotation issues, both for image restoration and the orientation of the science-ready data. In any case, the current conclusion is that multiple paths seem possible with no obvious show stoppers.

## 7. Conclusions

EST is an ambitious project that aspires to facilitate a great leap forward with respect to current European telescopes in multiple areas. It will have a primary mirror with a diameter three to four times larger than those of current state-of-the-art European solar observatories. Combined with a state-of-the-art MCAO system, it will deliver high spatial resolution observations over a wide FOV. The telescope has an optical design that minimises the time-dependent instrumental polarisation and facilitates the measurement of weak polarisation signals associated with many solar phenomena. A complete suite of TISs and IFSs will operate simultaneously to facilitate the inference of the magnetic, thermal, and dynamic properties of the plasma at different heights and their spatial and temporal fluctuations.

Among other scientific targets, EST has been especially designed for the study of the quiet Sun magnetism and its impact on the chromospheric energy balance, magneto-acoustic and Alfvén wave propagation and mode conversion, spicules, swirls and tornadoes, chromospheric heating, localised reconnection events, flares, filament eruptions, prominence-corona instabilities, and non-ideal MHD effects.

## Acknowledgements

C. Quintero Noda was supported by the EST Project Office, funded by the Canary Islands Government (file SD 17/01) under a direct grant awarded to the IAC on ground of public interest. This project has received funding from the European Union's Horizon 2020 Research and Innovation programme under Grant Agreements No 739500 (PRE-EST) and 653982 (GREST). This project was supported by the European Commission's FP7 Capacities Programme under Grant Agreements No 212482 (EST Design Study) and No. 312495 (SOLARNET). It was also supported by the European Union's Horizon 2020 Research and Innovation programme under Grant Agreement No. 824135 (SOLARNET). This work has been partially funded by the Spanish Ministry of Science and Innovation through project RTI2018-096886-B-C51, including a percentage from FEDER funds, and through the Centro de Excelencia Severo Ochoa grant SEV-2017-0709 awarded to the Instituto de Astrofísica de Andalucía in the period 2018–2022. The EST preparatory phase was supported by a grant for research infrastructures of national importance from the Swedish Research Council (registration number 2017-00625). Jan Jurčák was supported by the Ministry of Education, Youth and Sports of the Czech Republic through the EST-CZ project (LM2018095). C. Kuckein acknowledges funding received from the European Union's Horizon 2020 research and innovation programme under the Marie Skłodowska-Curie grant agreement No. 895955. Queen's University Belfast acknowledges support from the Science and Technology Facilities Council (STFC) under grant No. ST/V003739/1. This work was supported by Fundação para a Ciência e a Tecnologia (FCT) through the research grants [UID/FIS/04434/2019,] UIDB/04434/2020, UIDP/04434/2020, UIDB/00611/2020 and UIDP/00611/2020. This work was partly funded by a grant of the Austrian Science Fund (FWF): P 32958-N. J. de la Cruz Rodríguez, C. J. Díaz Baso and A. Pastor Yabar gratefully acknowledge financial support from the European Research Council (ERC) under the European Union's Horizon 2020 research and innovation programme (SUNMAG, grant agreement 759548). J. Aboudarham was supported by Centre National d'Etudes Spatiales (CNES). G. Aulanier, M. Carlsson, L. Fletcher, V. Hansteen, A. Ortiz and L. Rouppe van der Voort acknowledge support by the Research Council of Norway through its Centres of Excellence scheme, project number 262622. L. Belluzzi, M. Bianda, R. Ramelli acknowledge State Secretariat for Education, Research, and Innovation (SERI), Canton Ticino and Swiss National Science Foundation (grants 200020_184952, 200021_175997, CRSII5_180238) for the financial support. R. Brajša and D. Sudar acknowledge the support by the Croatian Science Foundation under project 7549 'Millimeter and submillimeter observations of the solar chromosphere with ALMA'. The NSO is operated by the Association of Universities for Research in Astronomy, Inc., under cooperative agreement with the National Science Foundation. The work of A. Berlicki was partially supported by the programme 'Excellence Initiative - Research University' for years 2020-2026 for University of Wrocław, project no. BPIDUB.4610.15.2021.KP.B. E. S. Carlin acknowledges financial support from the Spanish Ministry of Science and Innovation (MICINN) through the Spanish State Research Agency, under Severo Ochoa Centres of Excellence Programme 2020-2023 (CEX2019-000920-S). L. Fletcher and N. Labrosse would like to acknowledge support from UK Research and Innovation Science and Technology Facilities Council grants ST/L006200/1 and ST/T000422/1. J. Fernandes acknowledges visiting facilities at Rosseland Centre for Solar Physics (University of Oslo), in 2019. P. Gömöry, A. Kučera and J. Rybák were supported by the Science Grant Agency project VEGA 2/0048/20. I. Kontogiannis is supported by KO 6283/2-1 of the Deutsche Forschungsgemeinschaft (DFG). C. J. Nelson is thankful to the Science and Technology Facilities Council (STFC), for support received through grant ST/T00021X/1, and ESA, for support as an ESA Research Fellow. K. Petrovay was supported by the Hungarian National Research, Development and Innovation Fund (grants no. NKFI K-128384 and TKP2021-NKTA-64). P. J. A. Simões acknowledges support from CNPq (contract 307612/2019-8). J. Trujillo Bueno acknowledges the funding received from the European Research Council (ERC) under de European Union's Horizon 2020 research and innovation programme (ERC Advanced Grant Agreement No. 742265). M. Verma is supported by VE 1112/1-1 of the Deutsche Forschungsgemeinschaft (DFG). L. Q. Zhang acknowledges that this work was supported by the National Natural Science Foundation of China (Grant No. 11727805, No. 1210030348). F. Zuccarello acknowledges that this work was supported by the Italian MIUR-PRIN grant 2017APKP7T and by the Università degli Studi di Catania (Piano per la Ricerca Università di Catania 2020-2022, Linea di intervento 2).

[1] Instituto de Astrofísica de Canarias, E-38205 La Laguna, Tenerife, Spain
[2] Departamento de Astrofísica, Universidad de La Laguna, E-38206 La Laguna, Tenerife, Spain
[3] Leibniz-Institut für Sonnenphysik (KIS), Schöneckstr. 6, 79104 Freiburg, Germany
[4] Instituto de Astrofísica de Andalucía (IAA-CSIC), Glorieta de la Astronomía s/n, 18008 Granada, Spain
[5] Astronomical Institute of the Czech Academy of Sciences, Fričova 298, 25165 Ondřejov, Czech Republic
[6] Institute for Solar Physics, Department of Astronomy, Stockholm University, AlbaNova University Center, 10691 Stockholm, Sweden
[7] Astronomical Institute of the Slovak Academy of Sciences, 05960 Tatranská Lomnica, Slovakia
[8] Astrophysics Research Centre, School of Mathematics and Physics, Queen's University Belfast, Belfast, BT7 1NN, Northern Ireland, U.K.
[9] European Space Agency (ESA), European Space Research and Technology Centre (ESTEC), Keplerlaan 1, 2201 AZ, Noordwijk, The Netherlands
[10] Institute for Astronomy, Astrophysics, Space Applications and Remote Sensing, National Observatory of Athens, GR-15236 Penteli, Greece
[11] Laboratoire de Physique des Plasmas (LPP), École Polytechnique, IP Paris, Sorbonne Université, CNRS, Observatoire de Paris, Université PSL, Université Paris Saclay, Paris, France
[12] Rosseland Centre for Solar Physics, University of Oslo, PO Box 1029, Blindern 0315, Oslo, Norway
[13] LESIA, Observatoire de Paris, Université PSL, Sorbonne Université, Université Paris Cité, CNRS, 5 place Jules Janssen, 92195 Meudon, France
[14] University of Applied Sciences and Arts of Southern Switzerland (SUPSI), Polo Universitario Lugano, Via Serafino Balestra 16, 6900 Lugano, Switzerland
[15] GEPI, Observatoire de Paris, PSL Research University, CNRS, 92195 Meudon, France
[16] INAF Osservatorio Astronomico di Trieste, Via G.B. Tiepolo 11, 34143 Trieste, Italy
[17] Leibniz-Institut für Astrophysik Potsdam (AIP), An der Sternwarte 16, 14482 Potsdam, Germany
[18] Instituto de Astrofísica e Ciências do Espaço, Universidade de Coimbra, OGAUC, Rua do Observatório s/n, 3040-004 Coimbra, Portugal
[19] PMOD/WRC, Dorfstrasse 33, 7260 Davos Dorf, Switzerland
[20] ETH-Zurich, Hönggerberg campus, HIT building, Zürich, Switzerland
[21] University of Applied Sciences Western Switzerland (HEIG-VD), 1, route de Cheseaux, CH-1401 Yverdon-les-Bains, Switzerland
[22] Department of Electrical Engineering, Pontificia Universidad Católica de Chile, 4860 Vicuña Mackenna, 7820436 Santiago, Chile.
[23] National Solar Observatory (NSO), 3665 Discovery Drive, Boulder, CO 80303, USA
[24] IRSOL – Istituto Ricerche Solari "Aldo e Cele Daccò", Università della Svizzera italiana (USI), 6605 Locarno-Monti, Switzerland

[25] Euler Institute, Università della Svizzera italiana (USI), CH-6900 Lugano, Switzerland
[26] UCL Mullard Space Science Laboratory, Holmbury St Mary, Dorking RH5 6NT, UK
[27] Royal Observatory of Belgium, 1180 Ukkel, Belgium
[28] University of Wrocław, Centre of Scientific Excellence - Solar and Stellar Activity, Kopernika 11, 51-622 Wrocław, Poland
[29] Astronomical Institute, University of Wrocław, ul. Kopernika 11, 51-622 Wrocław, Poland
[30] Physics Department, University of Rome Tor Vergata, I-00133 Rome, Italy
[31] Leiden Observatory, Leiden University, Niels Bohrweg 2, 2333 CA, Leiden, The Netherlands
[32] Hvar Observatory, Faculty of Geodesy, University of Zagreb, Kačićeva 26, 10000 Zagreb, Croatia
[33] IGAM/Institute of Physics, Karl-Franzens University Graz, Universitätsplatz 5/II,Graz, Austria
[34] Max-Planck-Institut für Sonnensystemforschung (MPS), Justus-von-Liebig-Weg 3, 37077 Göttingen, Germany
[35] Centre for Advanced Instrumentation, Department of Physics, Durham University, Durham, UK
[36] Department of Physics, University of Calabria, Ponte P. Bucci 31C, 87036 Rende, Italy
[37] Institute of Theoretical Astrophysics, University of Oslo, PO Box 1029, Blindern 0315, Oslo, Norway
[38] GRANTECAN S.A., C/Vía Láctea s/n (IAC) 38205 La Laguna, Tenerife, Spain
[39] INAF Osservatorio Astrofisico di Arcetri, Largo E. Fermi 5, 50125 Firenze, Italy
[40] INAF - Fund. Galileo Galilei, Rambla Jose Ana Fernández Perez 7, 38712 Breña Baja (La Palma), Canary Islands, Spain
[41] INAF - Istituto di Astrofisica e Planetologia Spaziali, 00133 Roma, Italy
[42] Université Paul Sabatier, Observatoire Midi-Pyrénées, IRAP, Cnrs, Cnes, 14 ave. E. Belin, F-31400 Toulouse
[43] Astrophysics Research Institute, Liverpool John Moores University, 146 Brownlow Hill, Liverpool L3 5RF, UK
[44] Armagh Observatory & Planetarium, College Hill, Armagh BT61 9DG, N. Ireland
[45] Solar Physics & Space Plasma Research Center (SP2RC), School of Mathematics and Statistics,University of Sheffield, Hicks Building, Hounsfield Road, S3 7RH, UK
[46] ELTE Eötvös Loránd University, Institute of Geography and Earth Sciences, Department of Astronomy, Budapest, Hungary
[47] Hungarian Solar Physics Foundation (HSPF), Petőfi tér 3., Gyula, H-5700, Hungary
[48] INAF Osservatorio Astronomico di Roma, via Frascati 33, 00078 Monte Porzio Catone, Italy
[49] Université Côte d'Azur, Observatoire de la Côte d'Azur, CNRS, Laboratoire Lagrange, France
[50] CITEUC - Center for Earth and Space Research of the University of Coimbra, Geophysical and Astronomical Observatory and Department of Mathematics of the University of Coimbra, Coimbra, Portugal
[51] Subdireccion general de Internacionalización de la Ciencia e Innovación. Ministerio de Ciencia e Innovación. Paseo de la Castellana 162, planta 16 impares, 28046 Madrid
[52] SUPA School of Physics and Astronomy, University of Glasgow, Glasgow G12 8QQ, UK
[53] CNRS IRL2009 - THEMIS, Via Lactea s/n, ES-38205 La Laguna, Canary Islands, Spain
[54] MAX IV Laboratory, Lund University, Box 188, SE-221 00 Lund, Sweden
[55] CNR-Istituto Nazionale di Ottica, Largo E. Fermi 6, I-50125 Firenze, Italy
[56] INAF - Osservatorio Astrofisico di Catania, Via S. Sofia 78, I–95123 Catania, Italy
[57] Utrecht University, Department of Physics & Astronomy Postbus 80000, 3508TA Utrecht, the Netherlands
[58] Department of Optics, Palacký University Olomouc, 17. Listopadu 12, Olomouc, Czech Republic
[59] University of Geneva, 7, route de Drize, 1227 Carouge, Switzerland
[60] Department of Computer Science, Aalto University, PO Box 15400, FI-00076 Aalto, Finland
[61] Dipartimento di Fisica e Astronomia, Università di Firenze, Florence, Italy
[62] Univ Lyon, Univ Lyon1, Ens de Lyon, CNRS, Centre de Recherche Astrophysique de Lyon UMR5574, F-69230, Saint-Genis-Laval, France
[63] Faculty of Communication Sciences, Università della Svizzera Italiana, Lugano, Switzerland
[64] University of California Observatories, 1156 High St., Santa Cruz, CA 95064, USA
[65] INAF Osservatorio Astronomico di Capodimonte, Salita Moiariello 16, 80131 Napoli
[66] Expert Analytics AS, Møllergata 8, 0179 Oslo, Norway
[67] University of L'Aquila, Dept. of Physical and Chemical Sciences, Via Vetoio 48, 67100 Coppito (AQ) , Italy
[68] Centre for mathematical Plasma Astrophysics, Dept. of Mathematics, KU Leuven, 3001 Leuven, Belgium
[69] Institute of Physics, University of Maria Curie-Skłodowska, Pl. M. Curie-Skłodowska 5, 20-031 Lublin, Poland
[70] Skolkovo Institute of Science and Technology, Bolshoy Boulevard 30, bld. 1, Moscow 121205, Russia
[71] Thüringer Landessternwarte, Sternwarte 5, 07778 Tautenburg, Germany
[72] INAF - Istituto di Astrofisica Spaziale e Fisica Cosmica Via Alfonso Corti 12, 20133 Milano, Italy
[73] Centro de Rádio Astronomia e Astrofísica Mackenzie, Escola de Engenharia, Universidade Presbiteriana Mackenzie, São Paulo, Brazil
[74] School of Space Research, Kyung Hee University, Yongin, Gyeonggi, 446-701, Republic of Korea
[75] ASI, Italian Space Agency, Via del Politecnico snc, 00133 Rome, Italy
[76] Institute for Particle Physics and Astrophysics, ETH Zurich, Switzerland
[77] National Astronomical Observatory of Japan, 2-21-1 Osawa, Mitaka, Tokyo 181-8588, Japan
[78] Consejo Superior de Investigaciones Científicas, Spain
[79] The Key Laboratory on Adaptive Optics, Chinese Academy of Sciences, Chengdu 610209, China
[80] Institute of Optics and Electronics, Chinese Academy of Sciences, Chengdu 610209, China
[81] Dipartimento di Fisica e Astronomia "Ettore Majorana", Università di Catania, Via S. Sofia 78, 95123 Catania, Italy